\newif\ifarxiv
\arxivtrue

\ifarxiv
    \documentclass{article}
    \usepackage[]{amsthm}
\else
    \documentclass[acmjacm]{acmtrans2m}
\fi

\usepackage[]{amsmath}
\usepackage[]{amsfonts}
\usepackage[]{amssymb}
\usepackage[ruled,section]{algorithm}
\usepackage{algorithmic}
\usepackage[]{url}
\usepackage[]{tabularx}
\usepackage[]{booktabs}
\usepackage[]{array}
\usepackage[]{multirow}

\newcommand{\thisCaptionSummary}[0]{}

\newcommand{\sequence}[3][1]{\ensuremath{#2_{#1}\dots #2_{#3}}}
\newcommand{\twoPartSequence}[2]
   {\ensuremath{\langle#1\rangle\langle#2\rangle}}
\newcommand{\singletonSequence}[1]{%
   {\ensuremath{\langle#1\rangle}}}

\newcommand{\longsum}[3][1]{\ensuremath{#2_{#1} + \dots + #2_{#3}}}
\newcommand{\prepend}[0]{\ensuremath{\cdot}}
\newcommand{\concatenate}[0]{} 

\newcommand{\sac}[0]{\ensuremath{\mathcal{A}}}
\newcommand{\nac}[0]{\ensuremath{A}}
\newcommand{\lexminA}[0]{\ensuremath{\mathrm{M}_\mathcal{A}}}
\newcommand{\lexsuccA}[0]{\ensuremath{\mathrm{S}_\mathcal{A}}}

\newcommand{\ntac}[0]{\ensuremath{T_\mathcal{A}}}
\newcommand{\stac}[0]{\ensuremath{\mathcal{T_A}}}
\newcommand{\nnac}[0]{\ensuremath{N_\mathcal{A}}}
\newcommand{\snac}[0]{\ensuremath{\mathcal{N_A}}}

\newcommand{\sdc}[0]{\ensuremath{\mathcal{D}}}
\newcommand{\ndc}[0]{\ensuremath{D}}
\newcommand{\sdcf}[0]{\ensuremath{\mathcal{D}^*}}
\newcommand{\ndcf}[0]{\ensuremath{D^*}}
\newcommand{\lexsuccD}[0]{\ensuremath{\mathrm{S}_\mathcal{D}}}

\newcommand{\nump}[0]{\ensuremath{p}}

\newcommand{\iverson}[1]{\ensuremath{[#1]}}
\newcommand{\cardinality}[1]{\ensuremath{|#1|}}
\newcommand{\suchthat}[0]{\ensuremath{\mid}}
\newcommand{\floor}[1]{\ensuremath{\lfloor #1 \rfloor}}

\newcommand{\equivalent}[0]{\iff}

\newenvironment{basecase}[1]{\paragraph*{Base case: #1}}{}
\newenvironment{inductivecase}{\paragraph*{Induction step}}{}

\newcommand{\stacksum}[3]{\sum_{\substack{#1\\#2}}#3}

\newcommand{\pageNumber}[1]{p.#1}
\newcommand{\chapterNumber}[1]{ch.#1}
\newcommand{\sectionNumber}[1]{\S#1}
\newcommand{\exerciseNumber}[1]{ex.#1}
\newcommand{\volumeNumber}[1]{Vol.#1}

\newcommand{\sequenceNumber}[1]{Seq.#1}

\newcommand{\sectionRef}[1]{Section~\ref{#1}}
\newcommand{\figureRef}[1]{Figure~\ref{#1}}
\newcommand{\tableRef}[1]{Table~\ref{#1}}
\newcommand{\equationRef}[1]{\eqref{#1}} 
\newcommand{\lemmaRef}[1]{Lemma~\ref{#1}}
\newcommand{\theoremRef}[1]{Theorem~\ref{#1}}
\newcommand{\corollaryRef}[1]{Corollary~\ref{#1}}

\newcommand{\algorithmRef}[1]{Algorithm~\ref{#1}}

\newcommand{\thesisPageRef}[1]{%
Kelleher~\cite[\pageNumber{#1}]{kelleher-encoding}}

\newcommand{\twoAuthor}[2]{#1 \& #2}
\newcommand{\threeAuthor}[3]{#1, #2 \& #3}

\newcommand{\scarequote}[1]{`#1'}
\newcommand{\inlinequote}[1]{``#1"}

\newcommand{\bigoh}[1]{\ensuremath{O(#1)}}

\ifarxiv
    \newtheorem{theorem}{Theorem}[section]
    \newtheorem{corollary}{Corollary}[section]
    \newtheorem{lemma}{Lemma}[section]
    \newtheorem{definition}{Definition}[section]
\else
    \newtheorem{theorem}{Theorem}[section]
    \newtheorem{corollary}[theorem]{Corollary}
    \newtheorem{lemma}[theorem]{Lemma}
    \newdef{definition}[theorem]{Definition}
    \newdef{remark}[theorem]{Remark}
\fi

\newcommand{\ahfmt}[1]{\textsc{#1}}

\newcommand{\recAscHandle}[1]{\recAsc(\ensuremath{#1})}
\newcommand{\recDescHandle}[1]{\recDesc(\ensuremath{#1})}
\newcommand{\ruleAscHandle}[1]{\ruleAsc(\ensuremath{#1})}
\newcommand{\ruleDescHandle}[1]{\ruleDesc(\ensuremath{#1})}
\newcommand{\accelAscHandle}[1]{\accelAsc(\ensuremath{#1})}
\newcommand{\accelDescHandle}[1]{\accelDesc(\ensuremath{#1})}

\newcommand{\recAsc}[0]{\ahfmt{RecAsc}}
\newcommand{\recDesc}[0]{\ahfmt{RecDesc}}
\newcommand{\ruleAsc}[0]{\ahfmt{RuleAsc}}
\newcommand{\ruleDesc}[0]{\ahfmt{RuleDesc}}
\newcommand{\accelAsc}[0]{\ahfmt{AccelAsc}}
\newcommand{\accelDesc}[0]{\ahfmt{AccelDesc}}

\newcommand{\trs}[1]{\ensuremath{R_{{A\ref{#1}}}}}
\newcommand{\tws}[1]{\ensuremath{W_{{A\ref{#1}}}}}

\newcommand{\ti}[1]{\ensuremath{I_{{A\ref{#1}}}}}
\newcommand{\ai}[1]{\ensuremath{\bar I_{{A\ref{#1}}}}}

\newcommand{\freq}[1]{\ensuremath{t_{\ref{#1}}}}

\newcommand{\ert}[1]{\ensuremath{E_{\ref{#1}}}}
\newcommand{\visit}[0]{\ensuremath{\mathbf{visit}}}

\newcommand{\readwriteformat}[1]{\textbf{\small{#1}}}

\newenvironment{lineFrequencyLemma}[4]
{\begin{lemma}\label{lem-freq-#1-#2} %
The number of times line~\ref{#1-#2} is %
executed during the execution of %
\algorithmRef{#1} is given by \ensuremath{\freq{#1-#2}(#3) = #4}.
\end{lemma}
\begin{proof}}
{\end{proof}}

\newenvironment{twoLineFrequencyLemma}[5]
{\begin{lemma}\label{lem-freq-#1-#2+freq-#1-#3} %
The number of times line~\ref{#1-#2} is %
executed during the execution of %
\algorithmRef{#1} is given by 
\ensuremath{\freq{#1-#2}(#4)  + \freq{#1-#3}(#4) = #5}.
\end{lemma}
\begin{proof}}
{\end{proof}}

\newenvironment{totalReadsTheorem}[4][\sirc(n,m)]
{\begin{theorem}\label{the-trs-#2}
\algorithmRef{#2} requires \ensuremath{\trs{#2}(#3) = #4} read
operations to generate the set \ensuremath{#1}.
\end{theorem}
\begin{proof}}
{\end{proof}}

\newenvironment{totalWritesTheorem}[4][\sirc(n,m)]
{\begin{theorem}\label{the-tws-#2}
\algorithmRef{#2} requires \ensuremath{\tws{#2}(#3) = #4}
write operations to generate the set \ensuremath{#1}, excluding
initialisation.
\end{theorem}
\begin{proof}}
{\end{proof}}

\newcommand{\Gollnitz}[0]{G{\"o}llnitz}

\newcommand{\Stojmenovic}[0]{Stojmenovi\'c}

\newcolumntype{R}{>{\raggedleft}X}
\newcolumntype{L}{>{\raggedright}X}

\newcommand{\gapacoteabstract}{
Integer partitions may be encoded as either ascending or descending compositions for 
the purposes of systematic generation. Many algorithms exist to generate all descending 
compositions, yet none have previously been published to generate all ascending compositions.
We develop three new algorithms to generate all ascending compositions 
and compare these with descending composition generators from the literature.
We analyse the new algorithms and provide new and more precise analyses for the descending 
composition generators. In each case, the ascending composition generation algorithm 
is substantially more efficient than its descending composition counterpart. We 
develop a new formula for the partition function $p(n)$ as part of our analysis 
of the lexicographic succession rule for ascending compositions.
}

\ifarxiv
    
\begin{document}
    \title{Generating All Partitions: A Comparison Of Two Encodings}
    \author{Jerome Kelleher
        \thanks{Institute of Evolutionary Biology, University of Edinburgh, King's Buildings,
    West Mains Road, EH9 3JT U.K. \texttt{jerome.kelleher@ed.ac.uk}}
        \and Barry O'Sullivan\thanks{Cork Constraint Computation Centre,
    Western Gateway Building, University College Cork, Ireland. \texttt{b.osullivan@cs.ucc.ie}}}
    \maketitle
    \begin{abstract}
    \gapacoteabstract
    \end{abstract}
\else
    \markboth{J.\ Kelleher and B.\ O'Sullivan}{Generating All Partitions}
    \title{Generating All Partitions: A Comparison Of Two Encodings}
    \author{Jerome Kelleher\\ University of Edinburgh 
        \and Barry O'Sullivan\\University College Cork}
    \begin{abstract}
    \gapacoteabstract
    \end{abstract}

    \begin{document}

    \begin{bottomstuff}
    Authors' addresses: J. Kelleher, Institute of Evolutionary Biology, University of Edinburgh, King's Buildings,
    West Mains Road, EH9 3JT U.K, email: \url{jerome.kelleher@ed.ac.uk}; B. O'Sullivan, Cork Constraint Computation Centre,
    Western Gateway Building, University College Cork, Ireland, email: \url{b.osullivan@cs.ucc.ie}.
    \end{bottomstuff}

    \maketitle
\fi
\section{Introduction}
A partition of a positive integer $n$ is an unordered collection
of positive integers whose sum is $n$. Partitions have been the subject of
extensive study for many years and the theory of partitions is a large and
diverse body of knowledge. Partitions are a fundamental mathematical
concept and have connections with 
number theory~\cite{andrews-theory},
elliptic modular functions~\cite[\pageNumber{224}]{schroeder-number},
Schur algebras and representation
theory~\cite[\pageNumber{13}]{martin-schur},
derivatives~\cite{yang-derivatives},
symmetric groups~\cite{comet-notations,bivins-characters},
Gaussian polynomials~\cite[\chapterNumber{7}]{andrews-integer} and 
much else~\cite{ahlgren-addition}. The theory of partitions also 
has many and varied applications~\cite{desesquelles-calculation,planat-thermal,%
kubasiak-fermi,tran-on,grossmann-from,actor-infinite,%
temperley-statistical}.  

Combinatorial generation algorithms allow us to systematically traverse 
all possibilities in some combinatorial universe, and have been the subject 
of sustained interest for many years~\cite{knuth-history}. 
Many algorithms are known to generate fundamental combinatorial objects;
for example, in 1977 Sedgewick reviewed more than thirty
permutation generation algorithms~\cite{sedgewick-permutation}. 
Many different orders have been proposed for generating combinatorial 
objects, the most common being lexicographic~\cite{kemp-generating} and minimal-change 
order~\cite{savage-survey}. The choice of encoding, the representation of 
the objects we are interested in as simpler structures,
is of critical importance to the efficiency
of combinatorial generation.

In this paper we demonstrate that by changing the encoding for partitions
from \emph{descending} compositions to \emph{ascending} compositions we  
obtain significantly more efficient generation algorithms. 
We develop three 
new algorithms under the most common generation idioms: recursion
(\sectionRef{sec-gap-recursive-algorithms}),
succession rules (\sectionRef{sec-gap-succession-rules}), 
and efficient sequential generation (\sectionRef{sec-gap-accelerated-algorithms}).
In each case we rigorously analyse the 
new algorithm and use this analysis to compare with a commensurable algorithm
from the literature,  for which  we provide a new and more precise analysis.
These analyses are 
performed using a novel application of Kemp's abstraction of counting read and write 
operations~\cite{kemp-generating} and this approach is validated in an empirical 
study (\sectionRef{sec-gap-accelerated-algorithms-comparison}). 
In all three cases the new ascending composition generation algorithm
is substantially more efficient than the algorithm from the literature.
As part of our study of partition generation algorithms we provide a new 
proof of a partition identity 
in \sectionRef{sec-gap-accelerated-algorithms-ascending-compositions-terminal}.
We also develop a new 
formula for the partition function $\nump(n)$, one of the most important 
functions in the theory of partitions~\cite{andrews-partitions}, in  
\sectionRef{sec-gap-succession-rules-pn}.

\subsection{Related Work}
A composition of a positive integer $n$ is an expression of $n$ as an
ordered sum of positive integers~\cite[\pageNumber{14}]{stanley-enumerative}, and 
a composition $\longsum{a}{k} = n$ can be represented by the
sequence $\sequence{a}{k}$. Since there is a unique way of expressing each 
partition of $n$ as composition of $n$ in either ascending or 
descending order\footnote{For our purposes the terms  \scarequote{ascending} and \scarequote{descending} are
synonymous with \scarequote{nondecreasing} and \scarequote{nonincreasing},
respectively.}, we can generate either the set of ascending or 
descending compositions of $n$ in order to obtain the set of 
partitions. More precisely, we can say that we are 
encoding partitions as either ascending or descending compositions 
for the purposes of systematic generation.  

Although partitions are fundamentally unordered 
they have come to be defined in more and more concrete terms 
\emph{as} descending compositions. This trend can be clearly
seen in the works of Sylvester~\cite{sylvester-constructive},  
MacMahon~\cite[\volumeNumber{II} \pageNumber{91}]{macmahon-combinatory} and finally 
Andrews~\cite[\pageNumber{1}]{andrews-theory}. Sylvester's \inlinequote{constructive
theory of partitions}, based on the idea of treating a partition as 
\inlinequote{definite thing}~\cite{sylvester-constructive}
(in contrast to Euler's algebraical identities~\cite{hardy-asymptotic}),
has been extremely successful~\cite{pak-partition}. As a result of this,
partitions are now often defined
as descending compositions~\cite[\pageNumber{1}]{andrews-theory}; thus, algorithms to generate all partitions
have naturally followed the prevailing definition and generated 
descending compositions.

It is widely accepted that the most efficient means of generating
descending compositions is in reverse lexicographic
order: see Andrews~\cite[\pageNumber{230}]{andrews-theory},
Knuth~\cite[\pageNumber{1}]{knuth-generating-all-partitions},
\twoAuthor{Nijenhuis}
{Wilf}~\cite[\pageNumber{65--68}]{nijenhuis-combinatorial},
\twoAuthor{Page}{Wilson}~\cite[\sectionNumber{5.5}]{page-introduction},
Skiena~\cite[\pageNumber{52}]{skiena-implementing}, 
\twoAuthor{Stanton}{White}~\cite[\pageNumber{13}]{stanton-constructive},
Wells~\cite[\pageNumber{150}]{wells-elements} or
\twoAuthor{Zoghbi}{\Stojmenovic}~\cite{zoghbi-fast}.
Several different representations (concrete data structures) have been 
used for generating descending compositions: namely 
the sequence~\cite{mckay-partitions}, 
multiplicity~\cite[\chapterNumber{9}]{nijenhuis-combinatorial}
and part-count~\cite{stockmal-generation} representations. 
Although the lexicographic succession rules for
descending compositions in the multiplicity or part-count representations
can be implemented looplessly~\cite{ehrlich-loopless}, they 
tend to be less efficient that their sequence representation 
counterparts~\cite[\exerciseNumber{5}]{knuth-generating-all-partitions}.
In an empirical analysis, 
\twoAuthor{Zoghbi}{\Stojmenovic}~\cite{zoghbi-fast} demonstrated that
their sequence
representation algorithms are significantly more efficient than all known
multiplicity and part-count representation algorithms. 

Algorithms to generate descending compositions in lexicographic order
have also been published. 
See Knuth~\cite[\pageNumber{147}]{knuth-stanford} and 
\twoAuthor{Zoghbi}{\Stojmenovic}~\cite{zoghbi-fast} 
for implementations using the sequence representation; 
\threeAuthor{Reingold}{Nievergelt}
{Deo}~\cite[\pageNumber{193}]{reingold-combinatorial}
and
\twoAuthor{Fenner}{Loizou}~\cite{fenner-analysis} for implementations 
using the multiplicity representation; 
and Klimko~\cite{klimko-algorithm} 
for an implementation using the  part-count representation.
\twoAuthor{Fenner}{Loizou}'s tree construction
operations~\cite{fenner-binary} can be used to generate descending compositions 
in several other orders.

Several algorithms are known to generate descending 
$k$-compositions in lexicographic~\cite{gupta-ranking-B,tomasi-two},
reverse lexicographic~\cite{riha-efficient},
and minimal-change~\cite{savage-gray} order.
Hindenburg's
eighteenth century algorithm~\cite[\pageNumber{106}]{dickson-history}
generates ascending $k$-compositions in
lexicographic order  and is regarded as the canonical method to generate partitions
into a fixed number of parts: see Knuth~\cite[\pageNumber{2}]{knuth-generating-all-partitions},
Andrews~\cite[\pageNumber{232}]{andrews-theory} or 
\threeAuthor{Reingold}{Nievergelt}
{Deo}~\cite[\pageNumber{191}]{reingold-combinatorial}. 
Algorithms due to 
Stockmal~\cite{stockmal-generation-constraints},
Lehmer~\cite[\pageNumber{26}]{lehmer-machine}, 
\threeAuthor{Narayana}{Mathsen}{Sarangi}~\cite{narayana-algorithm},
and Boyer~\cite{boyer-simple} also generate ascending $k$-compositions
in lexicographic order. 
Algorithms to generate \emph{all} ascending 
compositions, however, have not been considered.

\subsection{Notation}
In general we use the notation and conventions advocated by 
Knuth~\cite[\pageNumber{1}]{knuth-generating-all-n-tuples}, using the term 
\emph{visit} to refer to the process of making a complete object available to some 
consuming procedure. Thus, any combinatorial generation algorithm must 
visit every element in the combinatorial universe in question exactly once.
In discussing the efficiency of combinatorial generation, 
we say that an algorithm is \emph{constant amortised 
time}~\cite[\sectionNumber{1.7}]{ruskey-combinatorial}
if the average amount of time required to generate an object 
is bounded, from above, by some constant.

Ordinarily, we denote a sequence of integers as $\sequence{a}{k}$, which
denotes a sequence of $k$ integers indexed $a_1$, $a_2$, etc. When 
referring to short specific sequences it is convenient to enclose
each element using $\langle$ and $\rangle$. Thus, if we let
$\sequence{a}{k} = \twoPartSequence{3}{23}$, we have $k = 2$, $a_1 = 3$
and $a_2 = 23$. We will also use the idea of prepending a particular value
to the head of a sequence: thus, the notation
$3\prepend\singletonSequence{23}$ is the same sequence as given in the
preceding example. 

\begin{definition}\label{def-ascending-composition}
A sequence of positive integers $\sequence{a}{k}$ is an
\emph{ascending composition}
of the positive integer $n$ if $\longsum{a}{k} = n$ and $a_1 \leq \dots
\leq a_k$. 
\end{definition}
\begin{definition}
\label{def-ascending-composition-functions}
Let $\sac(n)$ be the set of all ascending compositions of $n$ for some $n
\geq 1$, and let  $\sac(n, m) \subseteq \sac(n)$ be defined for $1\leq m
\leq n$ as $\sac(n, m) = \{\sequence{a}{k}\suchthat \sequence{a}{k} \in
\sac(n)$ and $a_1 \geq m\}$. Also, let $\nac(n) = \cardinality{\sac(n)}$ 
and $\nac(n, m) = \cardinality{\sac(n, m)}$.
\end{definition}

\begin{definition}\label{def-descending-composition}
A sequence of positive integers $\sequence{d}{k}$ is a
\emph{descending composition}
of the positive integer $n$ if $\longsum{d}{k} = n$ and $d_1 \geq \dots
\geq d_k$. 
\end{definition}

\begin{definition}
\label{def-descending-composition-functions}
Let $\sdc(n)$ be the set of all descending compositions of $n$ for some $n
\geq 1$, and let  $\sdcf(n, m) \subseteq \sdc(n)$ be defined for $1\leq m
\leq n$ as $\sdcf(n, m) = \{\sequence{d}{k}\suchthat \sequence{d}{k} \in
\sdc(n)$ and $d_1 = m\}$. Also, let $\ndc(n) = \cardinality{\sdc(n)}$ 
and $\ndcf(n) = \cardinality{\sdcf(n, m)}$.
\end{definition}

There is an asymmetry between the function used to enumerate the ascending compositions
and the descending compositions: $\nac(n, m)$ counts the ascending compositions of $n$ 
where the first part is \emph{at least} $m$, whereas $\ndcf(n,m)$ counts the number 
of descending compositions of $n$ where the first part is \emph{exactly} $m$. This asymmetry
is necessary as we require $\nac(n, m)$ in our analysis of ascending composition 
generation algorithms and $\ndcf(n, m)$ is essential for the analysis of the 
recursive descending composition generation algorithm of 
\sectionRef{sec-gap-recursive-algorithms-descending-compositions}.

\section{Recursive Algorithms}\label{sec-gap-recursive-algorithms}
In this section we examine recursive algorithms to generate
ascending and descending compositions. Recursion is a popular
technique in combinatorial generation as it leads to elegant
and concise generation procedures~\cite{ruskey-combinatorial}. In
\sectionRef{sec-gap-recursive-algorithms-ascending-compositions} we
develop and analyse a simple constant amortised time recursive algorithm  to 
generate all ascending compositions of $n$.
Then, in
\sectionRef{sec-gap-recursive-algorithms-descending-compositions}
we study Ruskey's descending composition 
generator~\cite[\sectionNumber{4.8}]{ruskey-combinatorial}, and provide
a new analysis of this algorithm. We  compare these
algorithms in \sectionRef{sec-gap-recursive-algorithms-comparison} in
terms of the total number of recursive invocations required to generate
all $\nump(n)$ partitions of $n$.

\subsection{Ascending Compositions}
\label{sec-gap-recursive-algorithms-ascending-compositions}
The only recursive algorithm to generate all ascending compositions
available in the literature is de Moivre's method~\cite{deMoivre-method}.
In de Moivre's method we generate the
ascending compositions of $n$ by prepending $m$ to the (previously listed)
ascending compositions of $n - m$, for $m = 1, \dots,
n$~\cite[\pageNumber{20}]{knuth-history}.
Our new recursive algorithm to generate all ascending compositions of $n$ operates
on a similar principle, but does not require us to have large sets of partitions 
in memory.

We first note that we can generate all
ascending compositions of $n$, with smallest part at least $m$, by prepending $m$
to all ascending compositions of $n - m$.  We then  observe that $m$ 
can range from $1$ to $\floor{n / 2}$, since the 
smallest part in a partition of $n$ (with more than one part) cannot be less 
than $1$ or greater than 
$\floor{n / 2}$; and we complete the process by visiting the 
singleton composition $\singletonSequence{n}$. This provides sufficient 
information for us to derive a 
recursive generation procedure, \algorithmRef{alg-rec-asc}, in the idiom of 
\twoAuthor{Page}{Wilson}~\cite{page-introduction}.
This algorithm generates all ascending compositions of $n$
where the first part is at least $m$ in lexicographic order. See 
Kelleher~\cite[\sectionNumber{5.2.1}]{kelleher-encoding} 
for a complete discussion and proof of correctness of \algorithmRef{alg-rec-asc}. 

Following the standard practise for the analysis of recursive generation
algorithms, we count the number of recursive calls required to generate
the set of combinatorial objects in
question (e.g.\ Sawada~\cite{sawada-generating}). 
By counting the total
number of recursive invocations required, we obtain a bound on the total
time required, as each invocation, discounting the time spent in recursive
calls, requires constant time.  To establish that
\algorithmRef{alg-rec-asc} generates the set $\sac(n)$ in constant
amortised time we must count the total
number of invocations,  $\ti{alg-rec-asc}(n)$, and show that this value is proportional to
$\nump(n)$. 

\begin{theorem}\label{the-ti-alg-rec-asc}
For all positive integers $n$, $\ti{alg-rec-asc}(n) = \nump(n)$.
\end{theorem}
\begin{proof}
Each invocation of \algorithmRef{alg-rec-asc}  visits exactly one
composition (line 8). The invocation $\recAscHandle{n,m,1}$ 
correctly visits all $\nump(n)$
ascending compositions of $n$~\cite[\pageNumber{78}]{kelleher-encoding} and  
it immediately follows, therefore, that
there must be $\nump(n)$ invocations. Hence, $\ti{alg-rec-asc}(n) =
\nump(n)$.
\end{proof}

\begin{algorithm}[t]
\caption{$\recAscHandle{n,m,k}$}
\begin{algorithmic}[1]
\label{alg-rec-asc}
\REQUIRE $1 \leq m \leq n$
\STATE $x \leftarrow m$
\WHILE {$2x \leq n$}
    \STATE $a_k \leftarrow x$
    \STATE $\recAscHandle{n - x, x, k + 1}$
    \STATE $x \leftarrow x + 1$
\ENDWHILE
\STATE $a_k \leftarrow n$
\STATE $\visit$ $\sequence{a}{k}$
\end{algorithmic}
\end{algorithm}

\theoremRef{the-ti-alg-rec-asc} gives us an asymptotic measure of the total
computational effort required to generate all partitions of $n$ using
\algorithmRef{alg-rec-asc}. It is also useful 
to know the average amount of effort that this total
implies per partition. Therefore, we let
$\ai{alg-rec-asc}(n)$ denote the average number of invocations of
\recAsc\ required to generate an ascending composition
of $n$. We then trivially get
\begin{equation}\label{eqn-ai-alg-rec-asc}
\ai{alg-rec-asc}(n) = 1
\end{equation}
from \theoremRef{the-ti-alg-rec-asc}, and we can see that  
\algorithmRef{alg-rec-asc} is obviously constant amortised time.

In this subsection we have developed a new algorithm to generate all 
ascending compositions of $n$. This algorithm, although concise 
and simple, can be easily shown to be constant amortised time. In the 
next subsection we examine the most efficient known algorithm 
to generate descending compositions, which we subsequently 
compare to the ascending composition generator of this subsection. 


\subsection{Descending Compositions}
\label{sec-gap-recursive-algorithms-descending-compositions}
Two recursive algorithms are available to generate all descending compositions of $n$:
\twoAuthor{Page}{Wilson}'s~\cite[\sectionNumber{5.5}]{page-introduction}
generator
(variants of which have appeared in several texts, including
\twoAuthor{Kreher}{Stinson}~\cite[\pageNumber{68}]{kreher-combinatorial},
Skiena~\cite[\pageNumber{51}]{skiena-implementing} and 
\twoAuthor{Pemmaraju}{Skiena}~\cite[\pageNumber{136}]
{pemmaraju-computational})
and Ruskey's improvement 
thereof~\cite[\sectionNumber{4.8}]{ruskey-combinatorial}. 
Ruskey's algorithm, given in
\algorithmRef{alg-rec-desc}, generates all
descending compositions of $n$ in which the first (and largest) part is
\emph{exactly} $m$; thus $\recDescHandle{8,4,1}$ visits
the compositions $41111, 4211, 422, 431, 44$. 
\recDesc\ uses what Ruskey refers to as a
\scarequote{path
elimination technique}~\cite[\sectionNumber{4.3}]{ruskey-combinatorial} to
attain constant amortised time performance.

\begin{algorithm}[t]
\caption{$\recDescHandle{n,m,k}$}
\begin{algorithmic}[1]
\label{alg-rec-desc}
\REQUIRE $1 \leq m \leq n$ \textbf{and} $d_j = 1$ for $j > k$
\STATE $d_k \leftarrow m$
\IF {$n = m$ \textbf{or} $m = 1$}
    \STATE $\visit$ $\sequence{d}{k + n - m}$
\ELSE
    \FOR {$x \leftarrow 1$ \textbf{to} $\min(m, n - m)$}    
        \STATE $\recDescHandle{n - m, x ,k + 1}$
    \ENDFOR
    \STATE $d_k \leftarrow 1$
\label{alg-rec-desc-petbackup}
\ENDIF
\end{algorithmic}
\end{algorithm}

A slight complication arises when we wish to use
\recDesc\ to generate
\emph{all} descending compositions. As the algorithm generates all
descending compositions where the first part is exactly $m$, we must
iterate through all $j  \in \{1,\dots,n\}$ and invoke 
\recDescHandle{n, j, 1}. Following Ruskey's recommendations~\cite[\sectionNumber{4.3}]{ruskey-combinatorial},
we consider instead the
invocation $\recDescHandle{2n,n,1}$. This invocation will generate all
descending compositions of $2n$ where the first part is exactly $n$;
therefore the remaining parts will be a descending composition of $n$.
Thus, if we alter line 3 to ignore the first part in $d$ (i.e.\ $\visit$
$\sequence[2]{d}{k + n - m}$), we will visit all descending compositions of
$n$ in lexicographic order.


Ruskey's algorithm generates descending compositions where the largest
part is exactly $m$, and so we require a recurrence relation to count
objects of this type. Ruskey~\cite[\sectionNumber{4.8}]{ruskey-combinatorial}
provides a recurrence relation to compute $\ndcf(n, m)$, which we shall use
for our analysis. Thus, we define $\ndcf(n, n)= \ndcf(n, 1) = 1$,
and in general,
\begin{equation}\label{eqn-ndcf}
\ndcf(n, m) = \sum_{x = 1}^{\min(m, n - m)} \ndcf(n - m, x).
\end{equation}
Recurrence \equationRef{eqn-ndcf} is useful here because it is
the recurrence relation upon which \recDesc\ is
based.  Using this recurrence we can then easily count  
the number of invocations of \recDesc\ required to generate the descending
compositions of $n$. Let us define  
$\ti{alg-rec-desc}'(n, m)$ as the number of invocation of 
\recDesc\ required to generate all descending compositions of $n$
where the first part is exactly $m$.
Then,
$\ti{alg-rec-desc}'(n, n)=
\ti{alg-rec-desc}'(n, 1) = 1$, and 
\begin{equation}\label{eqn-tip-alg-rec-desc}
\ti{alg-rec-desc}'(n, m) = 1 + \sum_{x =
1}^{\min(m, n - m)} \ti{alg-rec-desc}'(n - m, x).
\end{equation}
Recurrence \equationRef{eqn-tip-alg-rec-desc}
computes the number of invocations of
\algorithmRef{alg-rec-desc} required to generate
all descending compositions of $n$ with first part exactly $m$, but
tells us little about the actual magnitude of this value. As a step
towards solving this recurrence in terms of the partition function
$\nump(n)$ we require the following lemma, in which we relate the 
$\ti{alg-rec-desc}'(n, m)$ numbers to the
$\ndcf(n, m)$ numbers.

\begin{lemma}
\label{lem-ti-alg-rec-desc-ndcf}
If $1 < m \leq n$ then
$\ti{alg-rec-desc}'(n, m) = \ndcf(n,
m) +
\ndcf(n - 1, m)$.
\end{lemma}
\begin{proof}
Proceed by strong induction on $n$.

\begin{basecase}{$n = 2$}
Suppose $1 < m \leq 2$; it follows immediately that $m = 2$. Thus, by
recurrence \equationRef{eqn-tip-alg-rec-desc} we
compute $\ti{alg-rec-desc}'(2, 2) = 1$ and by
recurrence \equationRef{eqn-ndcf} compute $\ndcf(2, 2) = 1$ and $\ndcf(1,
2) = 0$. Therefore, $\ti{alg-rec-desc}'(2, 2)
= \ndcf(2,2) + \ndcf(1, 2)$, and so the inductive basis holds.
\end{basecase}
\begin{inductivecase}
Suppose, for some positive integer $n$, $\ti{alg-rec-desc}'(n', m') =
\ndcf(n', m') + \ndcf(n' - 1, m')$ for all positive integers $1 < m' \leq
n' < n$. Then, suppose $m$ is an arbitrary positive integer such that  $1 < m
\leq n$. Now, suppose $m = n$. By \equationRef{eqn-tip-alg-rec-desc} we know
that  $\ti{alg-rec-desc}'(n, m) = 1$ since $m = m$.
Also,  $\ndcf(n, m) = 1$ as $m = n$, and $\ndcf(n - 1, m) = 0$ as $n - 1
\neq m$, $m \neq 1$ and $\min(m, n - m - 1) = -1$, ensuring that the sum
in \equationRef{eqn-ndcf} is empty. Therefore, $\ti{alg-rec-desc}'(n, m) =
\ndcf(n, m) + \ndcf(n - 1, m)$.

Suppose, on the other hand, that $1 < m < n$. We can see immediately that
$\min(m, n - m) \geq 1$, and so there must be at least one term in the sum
of \equationRef{eqn-tip-alg-rec-desc}. Extracting
this first term where $x = 1$
from \equationRef{eqn-tip-alg-rec-desc} we get
\[
\ti{alg-rec-desc}'(n, m) = 1 +
\ti{alg-rec-desc}'(n - m, 1) +
\sum_{x = 2}^{\min(m, n -
m)}\ti{alg-rec-desc}'(n - m, x),
\]
and furthermore, as $\ti{alg-rec-desc}'(n,
1) = 1$, we obtain
\begin{equation}\label{eqn-gritluck-donkey-punt}
\ti{alg-rec-desc}'(n, m) = 2 + \sum_{x =
2}^{\min(m, n - m)}\ti{alg-rec-desc}'(n -
m, x).
\end{equation}
We are assured that $1 < x \leq n - m$ by the upper and lower bounds of
the summation in \equationRef{eqn-gritluck-donkey-punt}, and so we can
apply the inductive hypothesis to get
\begin{align}
\nonumber
\ti{alg-rec-desc}'(n, m) & = 2 + \sum_{x =
2}^{\min(m, n - m)}(
 \ndcf(n - m, x) + \ndcf(n -m - 1, x)) \\
\nonumber
 &= 2 + \sum_{x = 2}^{\min(m, n - m)}\ndcf(n - m, x)
 +  \sum_{x = 2}^{\min(m, n - m)} \ndcf(n -m - 1, x).
\end{align}
By the definition of $\ndcf$ we know that $\ndcf(n, 1) = 1$, and so
$\ndcf(n - m, 1) + \ndcf(n -  m - 1, 1) = 2$. Replacing the leading $2$
above with this expression, and inserting the terms $\ndcf(n - m, 1)$
and $\ndcf(n -  m - 1, 1)$ into the appropriate summations we find that
\begin{equation}\label{eqn-almost-there-almost-there}
\ti{alg-rec-desc}'(n, m) = \sum_{x =
1}^{\min(m, n - m)}\ndcf(n - m, x)
 +  \sum_{x = 1}^{\min(m, n - m)} \ndcf(n -m - 1, x) \,.
\end{equation}
By \equationRef{eqn-ndcf} we know that 
the first term of \equationRef{eqn-almost-there-almost-there} 
is equal to the first term of  $\ti{alg-rec-desc}'(n, m) = \ndcf(n, m)
+ \ndcf(n - 1, m)$, it therefore remains to show that
\[
\ndcf(n - 1, m) = \sum_{x = 1}^{\min(m, n - m)} \ndcf(n -m - 1, x),
\] 
or equivalently, that
\begin{equation}\label{eqn-awkward-hoor}
\sum_{x = 1}^{\min(m, n - m - 1)} \ndcf(n -m - 1, x)
=
\sum_{x = 1}^{\min(m, n - m)} \ndcf(n -m - 1, x).
\end{equation}

Suppose $m \leq n - m - 1$. Then, $\min(m, n - m - 1) = \min(m, n - m)$, 
and so the left and right-hand sides of \equationRef{eqn-awkward-hoor} are equal. 
Suppose, alternatively, that $m > n - m - 1$. 
Hence, $\min(m, n - m - 1) = n - m - 1$ and $\min(m, n - m) = n - m$ and so we get
\[
\sum_{x = 1}^{n - m} \ndcf(n -m - 1, x)= 
\sum_{x = 1}^{n - m - 1} \ndcf(n -m - 1, x) + \ndcf(n -m - 1, n - m).
\]
Since $n - m - 1 < n - m$ we know that $\ndcf(n -m - 1, n - m) = 0$, and therefore
\equationRef{eqn-awkward-hoor} is verified.

Therefore, by \equationRef{eqn-almost-there-almost-there}  and 
\equationRef{eqn-awkward-hoor} we know that $\ti{alg-rec-desc}'(n, m)
 = \ndcf(n, m) + \ndcf(n - 1, m)$, as required.
\end{inductivecase}
\end{proof}

\lemmaRef{lem-ti-alg-rec-desc-ndcf} is a crucial step in our analysis of
\algorithmRef{alg-rec-desc} as it relates the number of invocations
required to generate a given set of descending compositions to the function
$\ndcf(n, m)$. Much is known about the $\ndcf(n, m)$ numbers, as
they count the partitions of $n$ where the largest part is $m$; thus, we
can then relate the number of invocations required to the partition
numbers, $\nump(n)$.  Therefore, let us formally define
$\ti{alg-rec-desc}(n)$ to be number of invocations of
\algorithmRef{alg-rec-desc} required to generate all $\nump(n)$ descending
compositions of $n$. We then get the following result. 

\begin{theorem}\label{the-ti-alg-rec-desc}
If $n > 1$ then $\ti{alg-rec-desc}(n) = \nump(n) + \nump(n - 1)$.
\end{theorem}
\begin{proof}
Suppose $n > 1$.  To generate all descending compositions of $n$ we
invoke $\recDescHandle{2n, n, 1}$ (see discussion above), and as $n > 1$
we can apply  \lemmaRef{lem-ti-alg-rec-desc-ndcf}, to obtain
$\ti{alg-rec-desc}'(2n, n) = \ndcf(2n, n) + \ndcf(2n - 1, n)$, and thus
$\ti{alg-rec-desc}(n) = \ndcf(2n, n) + \ndcf(2n - 1, n)$.  We know that
$\ndcf(2n,n) = \nump(n)$, as we can clearly obtain a descending composition
of $n$ from a descending composition of $2n$ where the first part is
exactly $n$ by removing that first part. Similarly, $\ndcf(2n - 1, n) =
\nump(n - 1)$, as we can remove the first part of size $n$ from any
descending composition of $2n - 1$ with first part equal to $n$, obtaining
a descending composition of $n - 1$. Thus, the descending
compositions counted by the functions $\ndcf(2n, n) = \nump(n)$ and 
$\ndcf(2n - 1, n) =  \nump(n - 1)$. Hence, $\ti{alg-rec-desc}(n) =
\nump(n) + \nump(n - 1)$, completing the proof.
\end{proof}

Note that in \theoremRef{the-ti-alg-rec-desc}, and in many of the
following analyses, we restrict our attention to values $n > 1$. This is to
avoid unnecessary complication of the relevant formulas in accounting for
the case where $n = 1$. In the above, if we compute $\ti{alg-rec-desc}(n) =
\nump(n) + \nump(n - 1)$ for $n = 1$, we arrive at the conclusion that the
number of invocations required is $2$, as $\nump(0) = 1$ by convention. 
In the interest of clarity we shall ignore such contingencies, as they do
not affect the general conclusions we draw.

Using \theoremRef{the-ti-alg-rec-desc} it is now straightforward to show
that \recDesc\ generates all descending compositions of
$n$ in constant amortised time. To show that the algorithm is constant
amortised time we must demonstrate that the average number of invocations
of the algorithm per object generated is bounded, from above, by some 
constant. To do this, let us
formally define $\ai{alg-rec-desc}(n)$ as the
average number of invocations of \recDesc\ required to
generate a descending composition of $n$. Clearly, as the total number
of invocations is $\ti{alg-rec-desc}(n)$ and the number of objects
generated is $\nump(n)$, we have $\ai{alg-rec-desc}(n) =
\ti{alg-rec-desc}(n) / \nump(n)$. 

Since $\ti{alg-rec-desc}(n) = \nump(n) + \nump(n - 1)$ by
\theoremRef{the-ti-alg-rec-desc}, we have $\ai{alg-rec-desc}(n) = 1 +
\nump(n) / \nump(n - 1)$. It is well known that $\nump(n) > \nump(n - 1)$
for all $n > 1$, and therefore $\nump(n - 1) / \nump(n) < 1$. From this
inequality we can then deduce that $\ai{alg-rec-desc}(n) < 2$, 
proving that 
\algorithmRef{alg-rec-desc} is constant amortised time.
It is useful to have a more precise asymptotic
expression for the average number of invocations required to generate a
descending composition using \recDesc, $\ai{alg-rec-desc}(n)$.
By the asymptotic estimate for
$\nump(n -
t)/\nump(n)$~\cite[\pageNumber{11}]{knuth-generating-all-partitions} we
then get $\ai{alg-rec-desc}(n) = 1 + e^{-C/\sqrt{n}}\left(
1 + \bigoh{n^{-1/6}} \right)$, with $C = \pi / \sqrt{6}$. Simplifying this
expression we get 
\begin{equation}\label{eqn-ai-alg-rec-desc}
\ai{alg-rec-desc}(n) =  1 +
e^{-\pi/\sqrt{6n}}\left(1 + \bigoh{n^{-1/6}} \right).
\end{equation}

In this subsection we have described and provided a new analysis
for the most efficient known recursive descending composition generation
algorithm, which is due to
Ruskey~\cite[\sectionNumber{4.8}]{ruskey-combinatorial}.
Ruskey demonstrates that
\recDesc\ is constant amortised
time by reasoning about the number of children each node in the computation
tree has, but does not derive the precise number of invocations involved.
In this section we rigorously counted the number of invocations required to
generate all descending compositions of $n$ using this algorithm, and
related the recurrence involved to the partition numbers. We then used an
asymptotic formula for $\nump(n)$ to derive the number of invocations
required to generate each partition, on average. In the next subsection we
use this analysis to compare Ruskey's descending composition generator
with our new ascending composition generator.

\subsection{Comparison}\label{sec-gap-recursive-algorithms-comparison}

Performing the comparison between the recursive algorithms to 
generate all ascending compositions and to generate all descending compositions 
of $n$ is a simple procedure. \recAsc\ requires
$\nump(n)$ invocations to generate all $\nump(n)$ partitions of $n$
whereas \recDesc\ requires $\nump(n) + \nump(n  - 1)$ invocations.
 The asymptotics of $\nump(n)$ show that, as $n$ becomes large, $\nump(n -
1)/\nump(n)$ approaches $1$. Thus, we can reasonably expect the descending
composition generator to require approximately twice as long as the
ascending composition generator to generate all partitions of $n$.

\begin{table}
\renewcommand{\thisCaptionSummary}{A comparison of recursive partition
generators.}
\caption[\thisCaptionSummary]{\thisCaptionSummary\ The ratio of the time
required by our ascending composition generation algorithm and Ruskey's
algorithm in the Java and C languages is
shown.\label{tab-recursive-algorithms-comparison}}
\begin{tabularx}{\textwidth}{llRRRRRR}
\toprule
\multicolumn{2}{r}{$n = $} & $61$ & $72$ & $77$ & $90$ & $95$ & $109$\tabularnewline 
\multicolumn{2}{r}{{\footnotesize $\nump(n)$ = }} & {\footnotesize$1.12\times 10^{6}$} & {\footnotesize$5.39\times 10^{6}$} & {\footnotesize$1.06\times 10^{7}$} & {\footnotesize$5.66\times 10^{7}$} & {\footnotesize$1.05\times 10^{8}$} & {\footnotesize$5.42\times 10^{8}$}\tabularnewline 
 \cmidrule(l{2mm}r{2mm}){1-8} 
\multicolumn{2}{r}{Java}	&$0.56$	&$0.56$	&$0.56$	&$0.55$	&$0.55$	&$0.55$\tabularnewline
\multicolumn{2}{r}{C}	&$0.40$	&$0.48$	&$0.49$	&$0.50$	&$0.50$	&$0.50$\tabularnewline
\multicolumn{2}{r}{Theoretical}	&$0.54$	&$0.54$	&$0.53$	&$0.53$	&$0.53$	&$0.53$\tabularnewline
\bottomrule\end{tabularx}
\end{table}

In \tableRef{tab-recursive-algorithms-comparison} we see a comparison 
of the actual time spent in generating partitions of $n$ using Ruskey's
algorithm, \algorithmRef{alg-rec-desc}, and our ascending composition
generator, \algorithmRef{alg-rec-asc}. In this table we report the time
spent by \algorithmRef{alg-rec-asc} in generating all ascending compositions of $n$,
divided by the time required by Ruskey's algorithm (we report these ratios as 
the actual durations are of little
interest). 
Several steps were taken in an effort to address Sedgewick's 
concerns about the empirical comparisons
of algorithms~\cite{sedgewick-permutation}. Direct and literal implementation of the algorithms concerned
were written in the C and Java languages and compiled in the simplest 
possible manner (i.e., without the use of compiler \scarequote{optimisations}).
Execution times were measured as accurately as possible and the minimum 
value over five runs used.
The C programs were compiled using GCC version 3.3.4 and the Java programs compiled and
run on the Java 2 
Standard Edition, version 1.4.2. All programs were executed on an
Intel Pentium 4 processor running Linux kernel
2.6.8. See \thesisPageRef{111--114} for 
a full discussion of the methodology adopted in making these observations. 

The 
values of $n$ are selected such that $n$ is the smallest integer where 
$\nump(n) > 1\times10^x$ and $\nump(n) > 5\times10^x$ for $6 \leq x \leq 8$. 
Orders of magnitude larger than these values proved to be infeasible on the 
experimental platform; similarly, the time elapsed in generating fewer
than a million partitions was too brief to measure accurately.
Along with the observed ratios of the time required by
\recAsc\ and \recDesc\ we also report
the theoretically predicted ratio of the running times: $\nump(n) /(
\nump(n) + \nump(n - 1))$. We can see from
\tableRef{tab-recursive-algorithms-comparison} that these theoretically
predicted ratios agree well with the empirical evidence. We can also see
that as $n$ becomes larger, Ruskey's algorithm is tending towards taking
twice as long as \recAsc\ to generate the same partitions.

\section{Succession Rules}\label{sec-gap-succession-rules}
In this section we consider algorithms of the form studied by Kemp in his
general treatment of the problem of generating combinatorial
objects~\cite{kemp-generating}. 
Kemp reduced the problem of generating combinatorial objects to the
generation of all words in a formal language $\mathcal{L}$, and developed
powerful general techniques to analyse such algorithms. Specifically, Kemp
studied \inlinequote{direct generation algorithms} that obey a simple two
step procedure: (1) scan the current word from right-to-left until we find
the end of the common prefix shared by the current word and its immediate
successor; and (2) attach the new suffix to the end of this shared prefix.
The cost of this process can be easily quantified by counting the number of
\scarequote{read} operations required in step (1), and the number of
\scarequote{write} operations in step (2). To determine the complexity of
generating a given language, we can count the number of these operations
incurred in the process of generating all words in the language.

The section proceeds as follows. In 
\sectionRef{sec-gap-succession-rules-ascending-compositions}
we derive a new succession rule for ascending compositions. We
then use this succession rule to develop a generation algorithm,
which we subsequently analyse. 
Then, in 
\sectionRef{sec-gap-succession-rules-descending-compositions} we 
examine the well-know succession rule for generating descending 
compositions in reverse lexicographic order, and analyse the resulting 
algorithm. Following this, \sectionRef{sec-gap-succession-rules-comparison}  
compares the two algorithms in terms of Kemp's read and write 
operations. Finally, in \sectionRef{sec-gap-succession-rules-pn} we 
develop a new formula for $\nump(n)$ using our analysis the succession
rule for ascending compositions.

\subsection{Ascending Compositions}
\label{sec-gap-succession-rules-ascending-compositions}
We are concerned here with developing a simple succession rule 
that will allow us to generate the lexicographic successor of a given 
ascending composition, and using this rule to develop a generation
algorithm.  To do this it is convenient to define the following notation.
\begin{definition}[Lexicographic Minimum]\label{def-lexminA}
For some positive integers $m \leq n$, the function $\lexminA(n, m)$
computes the lexicographically least element of the set $\sac(n, m)$.
\end{definition}
\begin{definition}[Lexicographic Successor]\label{def-lexsuccA}
For any $\sequence{a}{k} \in \sac(n) \setminus \singletonSequence{n}$
the function $\lexsuccA(\sequence{a}{k})$ computes the immediate
lexicographic successor of $\sequence{a}{k}$.
\end{definition}

The succession rule for ascending compositions is then stated simply.
We obtain the lexicographically least composition in  $\sac(n, m)$ by appending $m$ to
the lexicographically least composition in $\sac(n - m, m)$. If
$2m  > n$ then there are no compositions in $\sac(n, m)$ with
more than one part, leading us to conclude that there is only one possible
composition; and this must be the lexicographically least. This leads us 
to the following recurrence:
\begin{equation}\label{eqn-rec-lexminA}
\lexminA(n, m) = m\prepend \lexminA(n - m, m)
\end{equation}
where $\lexminA(n, m)  = \singletonSequence{n}$ if $2m > n$. 
See \thesisPageRef{84} for a proof of \equationRef{eqn-rec-lexminA}.  
We can also derive a nonrecursive succession rule for $\lexsuccA$,
which we develop in the following sequence of results.

\begin{lemma}\label{lem-lexminA} 
For all positive integers $m \leq n$, the lexicographically least element
of the set $\sac(n, m)$ is given by 
\begin{equation}\label{eqn-lexminA} 
\lexminA(n, m) = \overbrace{m\dots m}^{\mu}\concatenate
\singletonSequence{n - \mu m},
\end{equation}
where $\mu = \floor{ n / m } - 1$.
\end{lemma}
\begin{proof}
Proceed by strong induction on $n$.

\begin{basecase}{$n = 1$}
Since $1 \leq m \leq n$ and $n = 1$, then $m = 1$, and so $2m > n$. Then,
by \equationRef{eqn-rec-lexminA}, we know that $\lexminA(n, m) = 
\singletonSequence{n}$. Thus, as $\mu = 0$ when $n = 1$, 
\equationRef{eqn-lexminA} correctly computes $\lexminA(n, m)$ when $n =
1$.
\end{basecase}
\begin{inductivecase}
Suppose, for some positive integer $n$ that \equationRef{eqn-lexminA}
holds true for all positive integers $m'\leq n' < n$. Suppose $m$ is an
arbitrary positive integer such that $m \leq n$.
Suppose then that $2m > n$. By dividing both sides of this inequality
by $m$, we see that $n / m < 2$, and so $\floor{  n / m } \leq 1$.
Similarly, as $m \leq n$, it follows that $1 \leq n / m$, and so 
$1 \leq \floor{ n / m }$. Thus, $1 \leq  \floor{ n / m } \leq
1$, and so $\floor{ n / m } = 1$; hence $\mu = 0$. By
\equationRef{eqn-rec-lexminA} $\lexminA(n, m) = \singletonSequence{n}$, 
and as $\mu = 0$, zero copies of $m$ are concatenated with
$\singletonSequence{n - \mu m}$, and so \equationRef{eqn-lexminA} correctly
computes  $\lexminA(n, m)$.

Suppose then that $2m \leq n$. By the inductive hypothesis and
\equationRef{eqn-rec-lexminA} we have
\[
\lexminA(n, m) = m\prepend 
\overbrace{m\dots m}^{\mu'}
\concatenate\singletonSequence{n - m - \mu'm}
\]
Clearly, if $\mu = \mu' + 1$, then \equationRef{eqn-lexminA} correctly
computes the lexicographically least element of $\sac(n, m)$. We know that
$\mu' = \floor{ (n - m) / m } - 1$, which gives us $\mu' =
\floor{ n / m - 1 } - 1$. It follows that  $\mu' = \floor{ n / m
} - 2$, and, as $\mu = \floor{n / m} - 1$ from \equationRef{eqn-lexminA},
we have $\mu = \mu' + 1$, completing the proof.
\end{inductivecase}
\end{proof}

\begin{theorem}[Lexicographic Successor]\label{the-lexsuccA}
If $\sequence{a}{k} \in \sac(n) \setminus \{\singletonSequence{n} \}$
then
\begin{equation}\label{eqn-lexsuccA}
\lexsuccA(\sequence{a}{k}) =  \sequence{a}{k - 2} \concatenate
\overbrace{m\dots m}^{\mu}
\concatenate\singletonSequence{n' - \mu m}
\end{equation}
where $m = a_{k - 1} + 1$, $n' = a_{k - 1} + a_{k}$, and $\mu = \floor{n'
/ m} - 1$. 
\end{theorem}
\begin{proof}
Suppose $n$ is an arbitrary positive integer. Let $\sequence{a}{k}$ be an arbitrary
element of $\sac(n)\setminus \{\singletonSequence{n}\}$. 
Clearly, there is no positive integer $x$
such that $\sequence{a}{k - 1}\concatenate\singletonSequence{a_k + x} \in
\sac(n)$.  The initial part of $\lexminA(a_k + a_{k - 1}, a_{k - 1} + 1)$ is
the least possible value we can assign to $a_{k - 1}$; and the remaining
parts (if any) are the lexicographically least way to extend
$\sequence{a}{k - 1}$ to a complete ascending composition of $n$.
Therefore, $\lexsuccA(\sequence{a}{k}) =  \sequence{a}{k -
2} \concatenate \lexminA(a_{k - 1} + a_{k}, a_{k - 1} +1)$. Then, using
\lemmaRef{lem-lexminA} we get \equationRef{eqn-lexsuccA} as required.
\end{proof}

\begin{algorithm}[t]
\caption{$\ruleAscHandle{n}$}
\begin{algorithmic}[1]
\label{alg-rule-asc}
\REQUIRE $n > 0$
\STATE $k \leftarrow 2$ \label{alg-rule-asc-k=2} 
\STATE $a_{1} \leftarrow 0$ \label{alg-rule-asc-a1=0}
\STATE $a_{2} \leftarrow n$ \label{alg-rule-asc-a2=n}
\WHILE {$k \neq 1$}  \label{alg-rule-asc-k-neq-1}
    \STATE $y \leftarrow a_{k} - 1$ \label{alg-rule-asc-y=ak-1}
    \STATE $k \leftarrow k - 1$ \label{alg-rule-asc-k=k-1}
    \STATE $x \leftarrow a_{k} + 1$ \label{alg-rule-asc-x=ak+1}
    \WHILE {$x \leq y$}\label{alg-rule-asc-x-leq-y}
        \STATE $a_k \leftarrow x$ \label{alg-rule-asc-ak=x}        
        \STATE $y \leftarrow y - x$ 
        \STATE $k \leftarrow k + 1$ \label{alg-rule-asc-k=k+1}
     \ENDWHILE 
    \STATE $a_{k} \leftarrow x + y$ \label{alg-rule-asc-ak=x+y}    
    \STATE $\visit$ $\sequence{a}{k}$\label{alg-rule-asc-visit}
\ENDWHILE 
\end{algorithmic}
\end{algorithm}

The succession rule \equationRef{eqn-lexsuccA} is implemented in \ruleAsc\
(\algorithmRef{alg-rule-asc}). Each iteration of the main loop visits
exactly one composition, and the internal loop generates any sequences 
of parts required to find the lexicographic successor.  
We concentrate here on analysis of this algorithm;
see Kelleher~\cite[\sectionNumber{5.3.1}]{kelleher-encoding}
for a full discussion and proof of 
correctness.
The goal of our analysis is to derive a simple expression, in terms of the
number of partitions of $n$, for the total number of read and write
operations~\cite{kemp-generating} made in the process of generating all
ascending compositions of $n$. We do this by first determining the 
frequency of certain key instructions and using this information 
to determine the number of read and  write operations involved.

\begin{lineFrequencyLemma}{alg-rule-asc}{k=k-1}{n}{\nump(n)}
As \algorithmRef{alg-rule-asc} correctly visits all $\nump(n)$ ascending
compositions of $n$, we know that line~\ref{alg-rule-asc-visit} is executed
exactly $\nump(n)$ times. Clearly line~\ref{alg-rule-asc-k=k-1} is executed
precisely the same number of times as line~\ref{alg-rule-asc-visit}, and so
we have $\freq{alg-rule-asc-k=k-1}(n) = \nump(n)$, as required.
\end{lineFrequencyLemma}

\begin{lineFrequencyLemma}{alg-rule-asc}{k=k+1}{n}{\nump(n) - 1}
The variable $k$ is used to control termination of the algorithm. From
line~\ref{alg-rule-asc-k=2} we know that $k$ is initially $2$, and from
line~\ref{alg-rule-asc-k-neq-1} we know that the algorithm terminates when
$k = 1$. Furthermore, the value of $k$ is modified only on 
lines~\ref{alg-rule-asc-k=k-1} and \ref{alg-rule-asc-k=k+1}. By
\lemmaRef{lem-freq-alg-rule-asc-k=k-1} we know that $k$ is
decremented $\nump(n)$ times; it then follows immediately that $k$ must be
incremented $\nump(n) - 1$ times, and so we have
$\freq{alg-rule-asc-k=k+1}(n) = \nump(n) - 1$, as required.
\end{lineFrequencyLemma}

\newcommand{\trsAlgRuleAsc}{\ensuremath{2\nump(n)}}
\begin{totalReadsTheorem}[\sac(n)]{alg-rule-asc}{n}{\trsAlgRuleAsc}
Read operations are carried out on lines~\ref{alg-rule-asc-x=ak+1} and 
\ref{alg-rule-asc-y=ak-1}, which are executed $\nump(n)$ times each by
 \lemmaRef{lem-freq-alg-rule-asc-k=k-1}.
Thus, the total number of read operations is $\trs{alg-rule-asc}(n) =
\trsAlgRuleAsc$.
\end{totalReadsTheorem}

\newcommand{\twsAlgRuleAsc}{\ensuremath{2\nump(n) - 1}}
\begin{totalWritesTheorem}[\sac(n)]{alg-rule-asc}{n}{\twsAlgRuleAsc}
After initialisation, write operations are carried out in
\algorithmRef{alg-rule-asc} only on lines~\ref{alg-rule-asc-ak=x} and
\ref{alg-rule-asc-ak=x+y}. Line~\ref{alg-rule-asc-ak=x+y} is executed
$\nump(n)$ times by \lemmaRef{lem-freq-alg-rule-asc-k=k-1}. We can also see
that line~\ref{alg-rule-asc-ak=x} is executed
exactly as many times as line~\ref{alg-rule-asc-k=k+1}, and by 
\lemmaRef{lem-freq-alg-rule-asc-k=k+1} we know that this value is
$\nump(n) - 1$. Therefore, summing these contributions, we get
$\tws{alg-rule-asc}(n) = 2\nump(n) - 1$, completing the proof.
\end{totalWritesTheorem}

From \theoremRef{the-tws-alg-rule-asc} and
\theoremRef{the-trs-alg-rule-asc} it is easy to see that we require an
average of two read and two write operations per partition generated, as 
we required $2\nump(n)$ of both operations to generate all $\nump(n)$
partitions of $n$. Thus, for any value of
$n$ we are assured that the total time required to generate all partitions
of $n$ will be proportional to the number of partitions generated,
implying that the algorithm is constant amortised time. 

\subsection{Descending Compositions}
\label{sec-gap-succession-rules-descending-compositions}
Up to this point we have considered only algorithms that generate
compositions in lexicographic order. The majority of descending
composition generation algorithms, however, visit compositions in
\emph{reverse} lexicographic order (McKay~\cite{mckay-partitions}
refers to it as the \scarequote{natural order} for partitions). 
There are many different
presentations of the succession rule required to transform a
descending composition from this list into its immediate successor: see
Andrews~\cite[\pageNumber{230}]{andrews-theory},
Knuth~\cite[\pageNumber{1}]{knuth-generating-all-partitions},
\twoAuthor{Nijenhuis}
{Wilf}~\cite[\pageNumber{65--68}]{nijenhuis-combinatorial},
\twoAuthor{Page}{Wilson}~\cite[\sectionNumber{5.5}]{page-introduction},
Skiena~\cite[\pageNumber{52}]{skiena-implementing}, 
\twoAuthor{Stanton}
{White}~\cite[\pageNumber{13}]{stanton-constructive},
Wells~\cite[\pageNumber{150}]{wells-elements} or
\twoAuthor{Zoghbi}{\Stojmenovic}~\cite{zoghbi-fast}. 
No analysis of this succession rule in terms of the number of read and write
operations~\cite{kemp-generating} involved has been published, however, 
and in this
section we analyse a basic implementation of the rule (we study more
sophisticated techniques in
\sectionRef{sec-gap-accelerated-algorithms-descending-compositions}).

If we formally define $\lexsuccD(\sequence{d}{k})$ to be the immediate
lexicographic predecessor of a $\sequence{d}{k}\in \sdc(n) \setminus
1\dots1$, the succession rule can be formulated as follows. Given a
descending composition $\sequence{d}{k}$ where $d_1 \neq 1$, we obtain the
next composition in the ordering by applying the transformation
\begin{equation}\label{eqn-lexsuccD}
\lexsuccD(\sequence{d}{k})  = \sequence{d}{q - 1} \concatenate
\overbrace{m\dots m}^{\mu}
\concatenate\singletonSequence{n' - \mu m}
\end{equation}
where $q$ is the rightmost non-$1$ value (i.e., $d_j > 1$ for $1 \leq j
\leq q$ and $d_j = 1 $ for $q < j \leq k$), $m = d_q - 1$, $n' = d_q + k
- q$ and $\mu = \floor{n' / m} - \iverson{n' \bmod m = 0}$. This
presentation can readily be derived from the treatments cited in the
previous paragraph.

\begin{algorithm}[t]
\caption{\ruleDescHandle{n}}
\begin{algorithmic}[1]
\label{alg-rule-desc}
\REQUIRE $n > 0$
\STATE $d_1 \leftarrow n$               \label{alg-rule-desc-d1=n}
\STATE $k \leftarrow 1$                 \label{alg-rule-desc-k=1}
\STATE $\visit$ $d_1$                   \label{alg-rule-desc-visit1}
\WHILE {$k \neq n$}                     \label{alg-rule-desc-k-neq-n}
    \STATE $\ell \leftarrow k$          \label{alg-rule-desc-l=k}
    \STATE $m \leftarrow d_k$           \label{alg-rule-desc-m=dq-o}
    \WHILE {$m = 1$}                    \label{alg-rule-desc-m-eq-1}
        \STATE $k \leftarrow k - 1$     \label{alg-rule-desc-k=k-1}
        \STATE $m \leftarrow d_k$       \label{alg-rule-desc-m=dq-i}
    \ENDWHILE                               
    \STATE $n' \leftarrow m + \ell - k$ \label{alg-rule-desc-n'=m+l-k}
    \STATE $m \leftarrow m - 1$         \label{alg-rule-desc-m=m-1}
    \WHILE {$m < n'$}                   \label{alg-rule-desc-m<n'}
         \STATE $d_k \leftarrow m$      \label{alg-rule-desc-dk=m}
         \STATE $n' \leftarrow n' - m$  \label{alg-rule-desc-n'=n'-m}
         \STATE $k \leftarrow k + 1$    \label{alg-rule-desc-k=k+1}
    \ENDWHILE    
    \STATE $d_k \leftarrow n'$          \label{alg-rule-desc-dk=n'}
    \STATE $\visit$ $\sequence{d}{k}$   \label{alg-rule-desc-visit2} 
\ENDWHILE 
\end{algorithmic}
\end{algorithm}   

The succession rule \equationRef{eqn-lexsuccD} is implemented in
\ruleDesc\ (\algorithmRef{alg-rule-desc}), where each iteration of the main
loop implements a single application of the rule.  The internal loop of
lines~\ref{alg-rule-desc-m-eq-1}--9 implements a right-to-left scan for
the largest index $q$ such that $d_q > 1$, and the loop of lines 13--17 
inserts $\mu$ copies of $m$ into the array. We analyse the algorithm
by first determining the frequency of certain key statements, and 
using this information to derive the number of read and write operations 
needed to generate all descending compositions of $n$.

\begin{lineFrequencyLemma}{alg-rule-desc}{k=k-1}{n}
{1 - n + \sum_{x = 1}^{n - 1}\nump(x)}
As exactly one descending composition is visited per iteration of the
outer while loop, we know that upon reaching line $6$ there is a complete
descending composition of $n$ contained in $\sequence{d}{k}$. Furthermore,
as $d_1 \geq \dots \geq d_k$, we know that all parts of size $1$ are at
the end of the composition, and so it is clear that line $7$ will be
executed exactly once for each part of size $1$ in any given composition.
As we visit the compositions at the end of the loop and we terminate when
$k = n$ we will not reach line 5 when the composition in question
consists of $n$ copies of $1$ (as this is the lexicographically least, and
hence the last descending composition in reverse lexicographic order).
Thus, line 7 will be  executed exactly as many times as there are
parts of size $1$ in all partitions of $n$, minus the $n$ $1$s contained
in the last composition. It is well
known~\cite[\pageNumber{8}]{honsberger-mathematical} that the number of
$1$s in all partitions of $n$ is $1 + \nump(1) + \dots +
\nump(n - 1)$, and therefore we see that line 7 is executed exactly $1 - n
+ \sum_{x = 1}^{n - 1}\nump(x)$, as required.
\end{lineFrequencyLemma}

\begin{lineFrequencyLemma}{alg-rule-desc}{k=k+1}{n}
{\sum_{x = 1}^{n - 1}\nump(x)}
The variable $k$ is used to control termination of
\algorithmRef{alg-rule-desc}: the
algorithm begins with $k = 1$ and terminates when $k = n$. Examining
\algorithmRef{alg-rule-desc} we see that $k$ is modified on only two
lines: it is incremented on line~\ref{alg-rule-desc-k=k+1}  and decremented
on line~\ref{alg-rule-desc-k=k-1}. Thus, we must have $n - 1$ more
increment operations than decrements; by
\lemmaRef{lem-freq-alg-rule-desc-k=k-1} there are exactly  $1 - n + \sum_{x
= 1}^{n - 1}\nump(x)$ decrement operations, and so we see that line 14 is
executed  $\sum_{x = 1}^{n - 1}\nump(x)$ times, as required.
\end{lineFrequencyLemma}

\newcommand{\trsAlgRuleDesc}{\ensuremath{\sum_{x =
1}^{n}\nump(x) - n}}
\begin{totalReadsTheorem}[\sdc(n)]{alg-rule-desc}{n}{\trsAlgRuleDesc}
Read operations are performed on lines~\ref{alg-rule-desc-m=dq-o}
and~\ref{alg-rule-desc-m=dq-i} of \algorithmRef{alg-rule-desc}. By
\lemmaRef{lem-freq-alg-rule-desc-k=k-1} we know that
line~\ref{alg-rule-desc-k=k-1} is executed $1 - n + \sum_{x = 1}^{n -
1}\nump(x)$ times, and so line~\ref{alg-rule-desc-m=dq-i} is executed an
equal number of times. Clearly
line~\ref{alg-rule-desc-m=dq-o} is executed $\nump(n) - 1$ times, and so
we get a total of $\trs{alg-rule-desc}(n) = \trsAlgRuleDesc$, as required.
\end{totalReadsTheorem}

\newcommand{\twsAlgRuleDesc}{\ensuremath{\sum_{x = 1}^{n}\nump(x) - 1}}
\begin{totalWritesTheorem}[\sdc(n)]{alg-rule-desc}{n}{\twsAlgRuleDesc}
The only occasions in \algorithmRef{alg-rule-desc} where a value is
written to the array $d$ are lines~\ref{alg-rule-desc-dk=m} and
\ref{alg-rule-desc-dk=n'}.
By \lemmaRef{lem-freq-alg-rule-desc-k=k+1} we know that
line~\ref{alg-rule-desc-k=k+1} is executed exactly
$\sum_{x = 1}^{n - 1}\nump(x)$ times, and it is straightforward to see that
line~\ref{alg-rule-desc-dk=m} is executed precisely the same number of
times. As we visit exactly one composition per iteration of the outer while
loop, and all descending compositions except the composition
$\singletonSequence{n}$ are visited with this loop, we then see that
line~\ref{alg-rule-desc-dk=n'} is executed $\nump(n) - 1$ times in all.
Therefore, summing these contributions we get 
$ \tws{alg-rule-desc}(n) =  \sum_{x = 1}^{n - 1}\nump(x) + 
\nump(n) - 1  = \twsAlgRuleDesc$ as required.
\end{totalWritesTheorem}

Theorems~\ref{the-trs-alg-rule-desc} and~\ref{the-tws-alg-rule-desc}
derive the precise number of read and write operations required to
generate all descending compositions of $n$ using
\algorithmRef{alg-rule-desc}, and this completes our analysis of the
algorithm. We discuss the implications of these results in the next
subsection, where we compare the total number of read and write operations
required by \ruleAscHandle{n} and \ruleDescHandle{n}.

\subsection{Comparison}
\label{sec-gap-succession-rules-comparison}

In this section we developed two algorithms. The first algorithm we considered,
\ruleAsc\ (\algorithmRef{alg-rule-asc}), generates ascending compositions
of $n$; the second algorithm, \ruleDesc\ (\algorithmRef{alg-rule-desc}),
generates descending compositions of $n$. We analysed the total number of
read and write operations required by these algorithms to generate all
partitions of $n$ by iteratively applying the succession rule
involved. The totals  obtained, disregarding unimportant (i.e.\ $\bigoh{1}$ or
$\bigoh{n}$) trailing
terms, for the ascending composition generator are summarised as follows.
\begin{equation}\label{trs-tws-alg-rule-asc}
\trs{alg-rule-asc}(n) \approx 2\nump(n)
\quad \text{and} \quad
\tws{alg-rule-asc}(n) \approx  2\nump(n)
\end{equation}
That is, we require approximately $2\nump(n)$ operations of the form $x \leftarrow a_j$
and approximately $2\nump(n)$ operations of the form $a_j \leftarrow x$ to generate all
partitions of $n$ using the ascending composition generator. Turning then
to the descending composition generator, we obtained the following totals,
again removing insignificant trailing terms.
\begin{equation}\label{trs-tws-alg-rule-desc}
\trs{alg-rule-desc}(n) \approx \sum_{x = 1}^{n}\nump(x)
\quad \text{and} \quad
\tws{alg-rule-desc}(n) \approx \sum_{x = 1}^{n}\nump(x) 
\end{equation}
These totals would appear to indicate a large disparity between the
algorithms, but we must examine the asymptotics of $\sum_{x =
1}^{n}\nump(x)$ to determine whether this is significant. We shall
do this in terms of the average number of read and write operations per
partition which is implied by these totals.

We know the total number of read and write operations required to generate
all $\nump(n)$ partitions of $n$ using both algorithms. Thus, to determine
the expected number of read and write operations required to transform
the average partition into its immediate successor we must divide 
these totals by $\nump(n)$. In the case of the ascending
composition generation algorithms this is trivial, as both expressions are
of the form $2\nump(n)$, and so dividing by $\nump(n)$ plainly yields
the value $2$. Determining the average number of read and write
operations using the succession rule for descending compositions is more
difficult, however, as both expressions involve a factor of the form
$\sum_{x = 1}^{n}\nump(x)$. 

Using the asymptotic expressions for $\nump(n)$ we can get a qualitative
estimate of these
functions. Odlyzko~\cite[\pageNumber{1083}]{odlyzko-asymptotic} derived an
estimate for the value of sums of partition numbers which can be stated as
follows
\[
\sum_{x = 1}^{n}\nump(x) = \frac{ e^{\pi \sqrt{2n/3}}}{2\pi\sqrt{2n}}
\left(1 + \bigoh{n^{-1/6}}\right).
\]
Then, dividing this by the asymptotic expression for $\nump(n)$ we get
the following approximation 
\begin{equation}\label{eqn-sum-pn-over-pn}
\frac{1}{\nump(n)} \sum_{x = 1}^{n}\nump(x) \approx 1 +
\frac{\sqrt{6n}}{\pi},
\end{equation}
which, although crude, is sufficient for our purposes. The key feature of
\equationRef{eqn-sum-pn-over-pn} is that the value is not constant: it
is $\bigoh{\sqrt{n}}$. Using this approximation we obtain the
following values for the number of read and write operations 
expected to transform a random partition of $n$ into its successor.
\begin{equation*}
\begin{array}{lrr}
 & \readwriteformat{Reads} & \readwriteformat{Writes} \\
\hline
\text{Ascending} & 2 & 2 \\
\text{Descending}&  
1 + 0.78\sqrt{n} & 1 +  0.78\sqrt{n} \\
\hline
\end{array}
\end{equation*}

\newcommand{\totalFormat}[1]{\ensuremath{(\mathit{#1})}}
\begin{figure}
\begin{tabularx}{\textwidth}{LLRR}
    \setlength{\unitlength}{2mm}
\begin{picture}(10,77)(0,-77)
\put(0,0){\circle{0.5}}
\put(1,0){\circle{0.5}}
\put(10,-1){\circle{0.5}}
\put(11,-1){\circle{0.5}}
\put(9,-2){\circle{0.5}}
\put(10,-2){\circle{0.5}}
\put(8,-3){\circle{0.5}}
\put(9,-3){\circle{0.5}}
\put(8,-4){\circle{0.5}}
\put(9,-4){\circle{0.5}}
\put(7,-5){\circle{0.5}}
\put(8,-5){\circle{0.5}}
\put(7,-6){\circle{0.5}}
\put(8,-6){\circle{0.5}}
\put(6,-7){\circle{0.5}}
\put(7,-7){\circle{0.5}}
\put(7,-8){\circle{0.5}}
\put(8,-8){\circle{0.5}}
\put(6,-9){\circle{0.5}}
\put(7,-9){\circle{0.5}}
\put(6,-10){\circle{0.5}}
\put(7,-10){\circle{0.5}}
\put(5,-11){\circle{0.5}}
\put(6,-11){\circle{0.5}}
\put(6,-12){\circle{0.5}}
\put(7,-12){\circle{0.5}}
\put(5,-13){\circle{0.5}}
\put(6,-13){\circle{0.5}}
\put(5,-14){\circle{0.5}}
\put(6,-14){\circle{0.5}}
\put(4,-15){\circle{0.5}}
\put(5,-15){\circle{0.5}}
\put(6,-16){\circle{0.5}}
\put(7,-16){\circle{0.5}}
\put(5,-17){\circle{0.5}}
\put(6,-17){\circle{0.5}}
\put(5,-18){\circle{0.5}}
\put(6,-18){\circle{0.5}}
\put(4,-19){\circle{0.5}}
\put(5,-19){\circle{0.5}}
\put(4,-20){\circle{0.5}}
\put(5,-20){\circle{0.5}}
\put(4,-21){\circle{0.5}}
\put(5,-21){\circle{0.5}}
\put(3,-22){\circle{0.5}}
\put(4,-22){\circle{0.5}}
\put(5,-23){\circle{0.5}}
\put(6,-23){\circle{0.5}}
\put(4,-24){\circle{0.5}}
\put(5,-24){\circle{0.5}}
\put(4,-25){\circle{0.5}}
\put(5,-25){\circle{0.5}}
\put(3,-26){\circle{0.5}}
\put(4,-26){\circle{0.5}}
\put(4,-27){\circle{0.5}}
\put(5,-27){\circle{0.5}}
\put(3,-28){\circle{0.5}}
\put(4,-28){\circle{0.5}}
\put(3,-29){\circle{0.5}}
\put(4,-29){\circle{0.5}}
\put(2,-30){\circle{0.5}}
\put(3,-30){\circle{0.5}}
\put(5,-31){\circle{0.5}}
\put(6,-31){\circle{0.5}}
\put(4,-32){\circle{0.5}}
\put(5,-32){\circle{0.5}}
\put(4,-33){\circle{0.5}}
\put(5,-33){\circle{0.5}}
\put(3,-34){\circle{0.5}}
\put(4,-34){\circle{0.5}}
\put(3,-35){\circle{0.5}}
\put(4,-35){\circle{0.5}}
\put(3,-36){\circle{0.5}}
\put(4,-36){\circle{0.5}}
\put(2,-37){\circle{0.5}}
\put(3,-37){\circle{0.5}}
\put(3,-38){\circle{0.5}}
\put(4,-38){\circle{0.5}}
\put(2,-39){\circle{0.5}}
\put(3,-39){\circle{0.5}}
\put(2,-40){\circle{0.5}}
\put(3,-40){\circle{0.5}}
\put(2,-41){\circle{0.5}}
\put(3,-41){\circle{0.5}}
\put(1,-42){\circle{0.5}}
\put(2,-42){\circle{0.5}}
\put(4,-43){\circle{0.5}}
\put(5,-43){\circle{0.5}}
\put(3,-44){\circle{0.5}}
\put(4,-44){\circle{0.5}}
\put(3,-45){\circle{0.5}}
\put(4,-45){\circle{0.5}}
\put(2,-46){\circle{0.5}}
\put(3,-46){\circle{0.5}}
\put(3,-47){\circle{0.5}}
\put(4,-47){\circle{0.5}}
\put(2,-48){\circle{0.5}}
\put(3,-48){\circle{0.5}}
\put(2,-49){\circle{0.5}}
\put(3,-49){\circle{0.5}}
\put(1,-50){\circle{0.5}}
\put(2,-50){\circle{0.5}}
\put(2,-51){\circle{0.5}}
\put(3,-51){\circle{0.5}}
\put(2,-52){\circle{0.5}}
\put(3,-52){\circle{0.5}}
\put(1,-53){\circle{0.5}}
\put(2,-53){\circle{0.5}}
\put(1,-54){\circle{0.5}}
\put(2,-54){\circle{0.5}}
\put(1,-55){\circle{0.5}}
\put(2,-55){\circle{0.5}}
\put(0,-56){\circle{0.5}}
\put(1,-56){\circle{0.5}}
\put(4,-57){\circle{0.5}}
\put(5,-57){\circle{0.5}}
\put(3,-58){\circle{0.5}}
\put(4,-58){\circle{0.5}}
\put(3,-59){\circle{0.5}}
\put(4,-59){\circle{0.5}}
\put(2,-60){\circle{0.5}}
\put(3,-60){\circle{0.5}}
\put(2,-61){\circle{0.5}}
\put(3,-61){\circle{0.5}}
\put(2,-62){\circle{0.5}}
\put(3,-62){\circle{0.5}}
\put(1,-63){\circle{0.5}}
\put(2,-63){\circle{0.5}}
\put(2,-64){\circle{0.5}}
\put(3,-64){\circle{0.5}}
\put(1,-65){\circle{0.5}}
\put(2,-65){\circle{0.5}}
\put(1,-66){\circle{0.5}}
\put(2,-66){\circle{0.5}}
\put(1,-67){\circle{0.5}}
\put(2,-67){\circle{0.5}}
\put(0,-68){\circle{0.5}}
\put(1,-68){\circle{0.5}}
\put(2,-69){\circle{0.5}}
\put(3,-69){\circle{0.5}}
\put(1,-70){\circle{0.5}}
\put(2,-70){\circle{0.5}}
\put(1,-71){\circle{0.5}}
\put(2,-71){\circle{0.5}}
\put(0,-72){\circle{0.5}}
\put(1,-72){\circle{0.5}}
\put(1,-73){\circle{0.5}}
\put(2,-73){\circle{0.5}}
\put(0,-74){\circle{0.5}}
\put(1,-74){\circle{0.5}}
\put(0,-75){\circle{0.5}}
\put(1,-75){\circle{0.5}}
\put(0,-76){\circle{0.5}}
\put(1,-76){\circle{0.5}}
\multiput(-0.5,0.5)(0,-77){2}{\line(1,0){12}}
\multiput(-0.5,0.5)(12,0){2}{\line(0,-1){77}}
\end{picture}
 & 
    \setlength{\unitlength}{2mm}
\begin{picture}(10,77)(0,-77)
\put(0,0){\circle*{0.5}}
\put(1,0){\circle*{0.5}}
\put(0,0){\circle*{0.5}}
\put(1,0){\circle*{0.5}}
\put(2,0){\circle*{0.5}}
\put(3,0){\circle*{0.5}}
\put(4,0){\circle*{0.5}}
\put(5,0){\circle*{0.5}}
\put(6,0){\circle*{0.5}}
\put(7,0){\circle*{0.5}}
\put(8,0){\circle*{0.5}}
\put(9,0){\circle*{0.5}}
\put(10,0){\circle*{0.5}}
\put(11,0){\circle*{0.5}}
\put(10,-1){\circle*{0.5}}
\put(9,-2){\circle*{0.5}}
\put(8,-3){\circle*{0.5}}
\put(9,-3){\circle*{0.5}}
\put(8,-4){\circle*{0.5}}
\put(7,-5){\circle*{0.5}}
\put(8,-5){\circle*{0.5}}
\put(7,-6){\circle*{0.5}}
\put(6,-7){\circle*{0.5}}
\put(7,-7){\circle*{0.5}}
\put(8,-7){\circle*{0.5}}
\put(7,-8){\circle*{0.5}}
\put(6,-9){\circle*{0.5}}
\put(7,-9){\circle*{0.5}}
\put(6,-10){\circle*{0.5}}
\put(5,-11){\circle*{0.5}}
\put(6,-11){\circle*{0.5}}
\put(7,-11){\circle*{0.5}}
\put(6,-12){\circle*{0.5}}
\put(5,-13){\circle*{0.5}}
\put(6,-13){\circle*{0.5}}
\put(5,-14){\circle*{0.5}}
\put(4,-15){\circle*{0.5}}
\put(5,-15){\circle*{0.5}}
\put(6,-15){\circle*{0.5}}
\put(7,-15){\circle*{0.5}}
\put(6,-16){\circle*{0.5}}
\put(5,-17){\circle*{0.5}}
\put(6,-17){\circle*{0.5}}
\put(5,-18){\circle*{0.5}}
\put(4,-19){\circle*{0.5}}
\put(5,-19){\circle*{0.5}}
\put(4,-20){\circle*{0.5}}
\put(5,-20){\circle*{0.5}}
\put(4,-21){\circle*{0.5}}
\put(3,-22){\circle*{0.5}}
\put(4,-22){\circle*{0.5}}
\put(5,-22){\circle*{0.5}}
\put(6,-22){\circle*{0.5}}
\put(5,-23){\circle*{0.5}}
\put(4,-24){\circle*{0.5}}
\put(5,-24){\circle*{0.5}}
\put(4,-25){\circle*{0.5}}
\put(3,-26){\circle*{0.5}}
\put(4,-26){\circle*{0.5}}
\put(5,-26){\circle*{0.5}}
\put(4,-27){\circle*{0.5}}
\put(3,-28){\circle*{0.5}}
\put(4,-28){\circle*{0.5}}
\put(3,-29){\circle*{0.5}}
\put(2,-30){\circle*{0.5}}
\put(3,-30){\circle*{0.5}}
\put(4,-30){\circle*{0.5}}
\put(5,-30){\circle*{0.5}}
\put(6,-30){\circle*{0.5}}
\put(5,-31){\circle*{0.5}}
\put(4,-32){\circle*{0.5}}
\put(5,-32){\circle*{0.5}}
\put(4,-33){\circle*{0.5}}
\put(3,-34){\circle*{0.5}}
\put(4,-34){\circle*{0.5}}
\put(3,-35){\circle*{0.5}}
\put(4,-35){\circle*{0.5}}
\put(3,-36){\circle*{0.5}}
\put(2,-37){\circle*{0.5}}
\put(3,-37){\circle*{0.5}}
\put(4,-37){\circle*{0.5}}
\put(3,-38){\circle*{0.5}}
\put(2,-39){\circle*{0.5}}
\put(3,-39){\circle*{0.5}}
\put(2,-40){\circle*{0.5}}
\put(3,-40){\circle*{0.5}}
\put(2,-41){\circle*{0.5}}
\put(1,-42){\circle*{0.5}}
\put(2,-42){\circle*{0.5}}
\put(3,-42){\circle*{0.5}}
\put(4,-42){\circle*{0.5}}
\put(5,-42){\circle*{0.5}}
\put(4,-43){\circle*{0.5}}
\put(3,-44){\circle*{0.5}}
\put(4,-44){\circle*{0.5}}
\put(3,-45){\circle*{0.5}}
\put(2,-46){\circle*{0.5}}
\put(3,-46){\circle*{0.5}}
\put(4,-46){\circle*{0.5}}
\put(3,-47){\circle*{0.5}}
\put(2,-48){\circle*{0.5}}
\put(3,-48){\circle*{0.5}}
\put(2,-49){\circle*{0.5}}
\put(1,-50){\circle*{0.5}}
\put(2,-50){\circle*{0.5}}
\put(3,-50){\circle*{0.5}}
\put(2,-51){\circle*{0.5}}
\put(3,-51){\circle*{0.5}}
\put(2,-52){\circle*{0.5}}
\put(1,-53){\circle*{0.5}}
\put(2,-53){\circle*{0.5}}
\put(1,-54){\circle*{0.5}}
\put(2,-54){\circle*{0.5}}
\put(1,-55){\circle*{0.5}}
\put(0,-56){\circle*{0.5}}
\put(1,-56){\circle*{0.5}}
\put(2,-56){\circle*{0.5}}
\put(3,-56){\circle*{0.5}}
\put(4,-56){\circle*{0.5}}
\put(5,-56){\circle*{0.5}}
\put(4,-57){\circle*{0.5}}
\put(3,-58){\circle*{0.5}}
\put(4,-58){\circle*{0.5}}
\put(3,-59){\circle*{0.5}}
\put(2,-60){\circle*{0.5}}
\put(3,-60){\circle*{0.5}}
\put(2,-61){\circle*{0.5}}
\put(3,-61){\circle*{0.5}}
\put(2,-62){\circle*{0.5}}
\put(1,-63){\circle*{0.5}}
\put(2,-63){\circle*{0.5}}
\put(3,-63){\circle*{0.5}}
\put(2,-64){\circle*{0.5}}
\put(1,-65){\circle*{0.5}}
\put(2,-65){\circle*{0.5}}
\put(1,-66){\circle*{0.5}}
\put(2,-66){\circle*{0.5}}
\put(1,-67){\circle*{0.5}}
\put(0,-68){\circle*{0.5}}
\put(1,-68){\circle*{0.5}}
\put(2,-68){\circle*{0.5}}
\put(3,-68){\circle*{0.5}}
\put(2,-69){\circle*{0.5}}
\put(1,-70){\circle*{0.5}}
\put(2,-70){\circle*{0.5}}
\put(1,-71){\circle*{0.5}}
\put(0,-72){\circle*{0.5}}
\put(1,-72){\circle*{0.5}}
\put(2,-72){\circle*{0.5}}
\put(1,-73){\circle*{0.5}}
\put(0,-74){\circle*{0.5}}
\put(1,-74){\circle*{0.5}}
\put(0,-75){\circle*{0.5}}
\put(1,-75){\circle*{0.5}}
\put(0,-76){\circle*{0.5}}
\multiput(-0.5,0.5)(0,-77){2}{\line(1,0){12}}
\multiput(-0.5,0.5)(12,0){2}{\line(0,-1){77}}
\end{picture}
 &
    \setlength{\unitlength}{2mm}
\begin{picture}(10,77)(0,-77)
\put(0,-1){\circle{0.5}}
\put(1,-2){\circle{0.5}}
\put(0,-2){\circle{0.5}}
\put(1,-3){\circle{0.5}}
\put(2,-4){\circle{0.5}}
\put(1,-4){\circle{0.5}}
\put(0,-4){\circle{0.5}}
\put(1,-5){\circle{0.5}}
\put(2,-6){\circle{0.5}}
\put(1,-6){\circle{0.5}}
\put(3,-7){\circle{0.5}}
\put(2,-7){\circle{0.5}}
\put(1,-7){\circle{0.5}}
\put(0,-7){\circle{0.5}}
\put(1,-8){\circle{0.5}}
\put(2,-9){\circle{0.5}}
\put(1,-9){\circle{0.5}}
\put(2,-10){\circle{0.5}}
\put(3,-11){\circle{0.5}}
\put(2,-11){\circle{0.5}}
\put(1,-11){\circle{0.5}}
\put(4,-12){\circle{0.5}}
\put(3,-12){\circle{0.5}}
\put(2,-12){\circle{0.5}}
\put(1,-12){\circle{0.5}}
\put(0,-12){\circle{0.5}}
\put(1,-13){\circle{0.5}}
\put(2,-14){\circle{0.5}}
\put(1,-14){\circle{0.5}}
\put(2,-15){\circle{0.5}}
\put(3,-16){\circle{0.5}}
\put(2,-16){\circle{0.5}}
\put(1,-16){\circle{0.5}}
\put(3,-17){\circle{0.5}}
\put(2,-17){\circle{0.5}}
\put(4,-18){\circle{0.5}}
\put(3,-18){\circle{0.5}}
\put(2,-18){\circle{0.5}}
\put(1,-18){\circle{0.5}}
\put(5,-19){\circle{0.5}}
\put(4,-19){\circle{0.5}}
\put(3,-19){\circle{0.5}}
\put(2,-19){\circle{0.5}}
\put(1,-19){\circle{0.5}}
\put(0,-19){\circle{0.5}}
\put(1,-20){\circle{0.5}}
\put(2,-21){\circle{0.5}}
\put(1,-21){\circle{0.5}}
\put(2,-22){\circle{0.5}}
\put(3,-23){\circle{0.5}}
\put(2,-23){\circle{0.5}}
\put(1,-23){\circle{0.5}}
\put(2,-24){\circle{0.5}}
\put(3,-25){\circle{0.5}}
\put(2,-25){\circle{0.5}}
\put(4,-26){\circle{0.5}}
\put(3,-26){\circle{0.5}}
\put(2,-26){\circle{0.5}}
\put(1,-26){\circle{0.5}}
\put(3,-27){\circle{0.5}}
\put(4,-28){\circle{0.5}}
\put(3,-28){\circle{0.5}}
\put(2,-28){\circle{0.5}}
\put(5,-29){\circle{0.5}}
\put(4,-29){\circle{0.5}}
\put(3,-29){\circle{0.5}}
\put(2,-29){\circle{0.5}}
\put(1,-29){\circle{0.5}}
\put(6,-30){\circle{0.5}}
\put(5,-30){\circle{0.5}}
\put(4,-30){\circle{0.5}}
\put(3,-30){\circle{0.5}}
\put(2,-30){\circle{0.5}}
\put(1,-30){\circle{0.5}}
\put(0,-30){\circle{0.5}}
\put(2,-31){\circle{0.5}}
\put(3,-32){\circle{0.5}}
\put(2,-32){\circle{0.5}}
\put(1,-32){\circle{0.5}}
\put(2,-33){\circle{0.5}}
\put(3,-34){\circle{0.5}}
\put(2,-34){\circle{0.5}}
\put(4,-35){\circle{0.5}}
\put(3,-35){\circle{0.5}}
\put(2,-35){\circle{0.5}}
\put(1,-35){\circle{0.5}}
\put(3,-36){\circle{0.5}}
\put(2,-36){\circle{0.5}}
\put(3,-37){\circle{0.5}}
\put(4,-38){\circle{0.5}}
\put(3,-38){\circle{0.5}}
\put(2,-38){\circle{0.5}}
\put(5,-39){\circle{0.5}}
\put(4,-39){\circle{0.5}}
\put(3,-39){\circle{0.5}}
\put(2,-39){\circle{0.5}}
\put(1,-39){\circle{0.5}}
\put(4,-40){\circle{0.5}}
\put(3,-40){\circle{0.5}}
\put(5,-41){\circle{0.5}}
\put(4,-41){\circle{0.5}}
\put(3,-41){\circle{0.5}}
\put(2,-41){\circle{0.5}}
\put(6,-42){\circle{0.5}}
\put(5,-42){\circle{0.5}}
\put(4,-42){\circle{0.5}}
\put(3,-42){\circle{0.5}}
\put(2,-42){\circle{0.5}}
\put(1,-42){\circle{0.5}}
\put(7,-43){\circle{0.5}}
\put(6,-43){\circle{0.5}}
\put(5,-43){\circle{0.5}}
\put(4,-43){\circle{0.5}}
\put(3,-43){\circle{0.5}}
\put(2,-43){\circle{0.5}}
\put(1,-43){\circle{0.5}}
\put(0,-43){\circle{0.5}}
\put(2,-44){\circle{0.5}}
\put(3,-45){\circle{0.5}}
\put(2,-45){\circle{0.5}}
\put(3,-46){\circle{0.5}}
\put(4,-47){\circle{0.5}}
\put(3,-47){\circle{0.5}}
\put(2,-47){\circle{0.5}}
\put(5,-48){\circle{0.5}}
\put(4,-48){\circle{0.5}}
\put(3,-48){\circle{0.5}}
\put(2,-48){\circle{0.5}}
\put(1,-48){\circle{0.5}}
\put(3,-49){\circle{0.5}}
\put(4,-50){\circle{0.5}}
\put(3,-50){\circle{0.5}}
\put(2,-50){\circle{0.5}}
\put(4,-51){\circle{0.5}}
\put(3,-51){\circle{0.5}}
\put(5,-52){\circle{0.5}}
\put(4,-52){\circle{0.5}}
\put(3,-52){\circle{0.5}}
\put(2,-52){\circle{0.5}}
\put(6,-53){\circle{0.5}}
\put(5,-53){\circle{0.5}}
\put(4,-53){\circle{0.5}}
\put(3,-53){\circle{0.5}}
\put(2,-53){\circle{0.5}}
\put(1,-53){\circle{0.5}}
\put(4,-54){\circle{0.5}}
\put(5,-55){\circle{0.5}}
\put(4,-55){\circle{0.5}}
\put(3,-55){\circle{0.5}}
\put(6,-56){\circle{0.5}}
\put(5,-56){\circle{0.5}}
\put(4,-56){\circle{0.5}}
\put(3,-56){\circle{0.5}}
\put(2,-56){\circle{0.5}}
\put(7,-57){\circle{0.5}}
\put(6,-57){\circle{0.5}}
\put(5,-57){\circle{0.5}}
\put(4,-57){\circle{0.5}}
\put(3,-57){\circle{0.5}}
\put(2,-57){\circle{0.5}}
\put(1,-57){\circle{0.5}}
\put(8,-58){\circle{0.5}}
\put(7,-58){\circle{0.5}}
\put(6,-58){\circle{0.5}}
\put(5,-58){\circle{0.5}}
\put(4,-58){\circle{0.5}}
\put(3,-58){\circle{0.5}}
\put(2,-58){\circle{0.5}}
\put(1,-58){\circle{0.5}}
\put(0,-58){\circle{0.5}}
\put(3,-59){\circle{0.5}}
\put(4,-60){\circle{0.5}}
\put(3,-60){\circle{0.5}}
\put(5,-61){\circle{0.5}}
\put(4,-61){\circle{0.5}}
\put(3,-61){\circle{0.5}}
\put(2,-61){\circle{0.5}}
\put(4,-62){\circle{0.5}}
\put(5,-63){\circle{0.5}}
\put(4,-63){\circle{0.5}}
\put(3,-63){\circle{0.5}}
\put(6,-64){\circle{0.5}}
\put(5,-64){\circle{0.5}}
\put(4,-64){\circle{0.5}}
\put(3,-64){\circle{0.5}}
\put(2,-64){\circle{0.5}}
\put(7,-65){\circle{0.5}}
\put(6,-65){\circle{0.5}}
\put(5,-65){\circle{0.5}}
\put(4,-65){\circle{0.5}}
\put(3,-65){\circle{0.5}}
\put(2,-65){\circle{0.5}}
\put(1,-65){\circle{0.5}}
\put(5,-66){\circle{0.5}}
\put(4,-66){\circle{0.5}}
\put(6,-67){\circle{0.5}}
\put(5,-67){\circle{0.5}}
\put(4,-67){\circle{0.5}}
\put(3,-67){\circle{0.5}}
\put(7,-68){\circle{0.5}}
\put(6,-68){\circle{0.5}}
\put(5,-68){\circle{0.5}}
\put(4,-68){\circle{0.5}}
\put(3,-68){\circle{0.5}}
\put(2,-68){\circle{0.5}}
\put(8,-69){\circle{0.5}}
\put(7,-69){\circle{0.5}}
\put(6,-69){\circle{0.5}}
\put(5,-69){\circle{0.5}}
\put(4,-69){\circle{0.5}}
\put(3,-69){\circle{0.5}}
\put(2,-69){\circle{0.5}}
\put(1,-69){\circle{0.5}}
\put(9,-70){\circle{0.5}}
\put(8,-70){\circle{0.5}}
\put(7,-70){\circle{0.5}}
\put(6,-70){\circle{0.5}}
\put(5,-70){\circle{0.5}}
\put(4,-70){\circle{0.5}}
\put(3,-70){\circle{0.5}}
\put(2,-70){\circle{0.5}}
\put(1,-70){\circle{0.5}}
\put(0,-70){\circle{0.5}}
\put(5,-71){\circle{0.5}}
\put(6,-72){\circle{0.5}}
\put(5,-72){\circle{0.5}}
\put(4,-72){\circle{0.5}}
\put(7,-73){\circle{0.5}}
\put(6,-73){\circle{0.5}}
\put(5,-73){\circle{0.5}}
\put(4,-73){\circle{0.5}}
\put(3,-73){\circle{0.5}}
\put(8,-74){\circle{0.5}}
\put(7,-74){\circle{0.5}}
\put(6,-74){\circle{0.5}}
\put(5,-74){\circle{0.5}}
\put(4,-74){\circle{0.5}}
\put(3,-74){\circle{0.5}}
\put(2,-74){\circle{0.5}}
\put(9,-75){\circle{0.5}}
\put(8,-75){\circle{0.5}}
\put(7,-75){\circle{0.5}}
\put(6,-75){\circle{0.5}}
\put(5,-75){\circle{0.5}}
\put(4,-75){\circle{0.5}}
\put(3,-75){\circle{0.5}}
\put(2,-75){\circle{0.5}}
\put(1,-75){\circle{0.5}}
\put(10,-76){\circle{0.5}}
\put(9,-76){\circle{0.5}}
\put(8,-76){\circle{0.5}}
\put(7,-76){\circle{0.5}}
\put(6,-76){\circle{0.5}}
\put(5,-76){\circle{0.5}}
\put(4,-76){\circle{0.5}}
\put(3,-76){\circle{0.5}}
\put(2,-76){\circle{0.5}}
\put(1,-76){\circle{0.5}}
\put(0,-76){\circle{0.5}}
\multiput(-0.5,0.5)(0,-77){2}{\line(1,0){12}}
\multiput(-0.5,0.5)(12,0){2}{\line(0,-1){77}}
\end{picture}
 &
    \setlength{\unitlength}{2mm}
\begin{picture}(10,77)(0,-77)
\put(0,0){\circle*{0.5}}
\put(0,-1){\circle*{0.5}}
\put(1,-1){\circle*{0.5}}
\put(0,-2){\circle*{0.5}}
\put(1,-2){\circle*{0.5}}
\put(1,-3){\circle*{0.5}}
\put(2,-3){\circle*{0.5}}
\put(0,-4){\circle*{0.5}}
\put(1,-4){\circle*{0.5}}
\put(1,-5){\circle*{0.5}}
\put(2,-5){\circle*{0.5}}
\put(1,-6){\circle*{0.5}}
\put(2,-6){\circle*{0.5}}
\put(3,-6){\circle*{0.5}}
\put(0,-7){\circle*{0.5}}
\put(1,-7){\circle*{0.5}}
\put(1,-8){\circle*{0.5}}
\put(2,-8){\circle*{0.5}}
\put(1,-9){\circle*{0.5}}
\put(2,-9){\circle*{0.5}}
\put(2,-10){\circle*{0.5}}
\put(3,-10){\circle*{0.5}}
\put(1,-11){\circle*{0.5}}
\put(2,-11){\circle*{0.5}}
\put(3,-11){\circle*{0.5}}
\put(4,-11){\circle*{0.5}}
\put(0,-12){\circle*{0.5}}
\put(1,-12){\circle*{0.5}}
\put(1,-13){\circle*{0.5}}
\put(2,-13){\circle*{0.5}}
\put(1,-14){\circle*{0.5}}
\put(2,-14){\circle*{0.5}}
\put(2,-15){\circle*{0.5}}
\put(3,-15){\circle*{0.5}}
\put(1,-16){\circle*{0.5}}
\put(2,-16){\circle*{0.5}}
\put(3,-16){\circle*{0.5}}
\put(2,-17){\circle*{0.5}}
\put(3,-17){\circle*{0.5}}
\put(4,-17){\circle*{0.5}}
\put(1,-18){\circle*{0.5}}
\put(2,-18){\circle*{0.5}}
\put(3,-18){\circle*{0.5}}
\put(4,-18){\circle*{0.5}}
\put(5,-18){\circle*{0.5}}
\put(0,-19){\circle*{0.5}}
\put(1,-19){\circle*{0.5}}
\put(1,-20){\circle*{0.5}}
\put(2,-20){\circle*{0.5}}
\put(1,-21){\circle*{0.5}}
\put(2,-21){\circle*{0.5}}
\put(2,-22){\circle*{0.5}}
\put(3,-22){\circle*{0.5}}
\put(1,-23){\circle*{0.5}}
\put(2,-23){\circle*{0.5}}
\put(2,-24){\circle*{0.5}}
\put(3,-24){\circle*{0.5}}
\put(2,-25){\circle*{0.5}}
\put(3,-25){\circle*{0.5}}
\put(4,-25){\circle*{0.5}}
\put(1,-26){\circle*{0.5}}
\put(2,-26){\circle*{0.5}}
\put(3,-26){\circle*{0.5}}
\put(3,-27){\circle*{0.5}}
\put(4,-27){\circle*{0.5}}
\put(2,-28){\circle*{0.5}}
\put(3,-28){\circle*{0.5}}
\put(4,-28){\circle*{0.5}}
\put(5,-28){\circle*{0.5}}
\put(1,-29){\circle*{0.5}}
\put(2,-29){\circle*{0.5}}
\put(3,-29){\circle*{0.5}}
\put(4,-29){\circle*{0.5}}
\put(5,-29){\circle*{0.5}}
\put(6,-29){\circle*{0.5}}
\put(0,-30){\circle*{0.5}}
\put(1,-30){\circle*{0.5}}
\put(2,-30){\circle*{0.5}}
\put(2,-31){\circle*{0.5}}
\put(3,-31){\circle*{0.5}}
\put(1,-32){\circle*{0.5}}
\put(2,-32){\circle*{0.5}}
\put(2,-33){\circle*{0.5}}
\put(3,-33){\circle*{0.5}}
\put(2,-34){\circle*{0.5}}
\put(3,-34){\circle*{0.5}}
\put(4,-34){\circle*{0.5}}
\put(1,-35){\circle*{0.5}}
\put(2,-35){\circle*{0.5}}
\put(3,-35){\circle*{0.5}}
\put(2,-36){\circle*{0.5}}
\put(3,-36){\circle*{0.5}}
\put(3,-37){\circle*{0.5}}
\put(4,-37){\circle*{0.5}}
\put(2,-38){\circle*{0.5}}
\put(3,-38){\circle*{0.5}}
\put(4,-38){\circle*{0.5}}
\put(5,-38){\circle*{0.5}}
\put(1,-39){\circle*{0.5}}
\put(2,-39){\circle*{0.5}}
\put(3,-39){\circle*{0.5}}
\put(4,-39){\circle*{0.5}}
\put(3,-40){\circle*{0.5}}
\put(4,-40){\circle*{0.5}}
\put(5,-40){\circle*{0.5}}
\put(2,-41){\circle*{0.5}}
\put(3,-41){\circle*{0.5}}
\put(4,-41){\circle*{0.5}}
\put(5,-41){\circle*{0.5}}
\put(6,-41){\circle*{0.5}}
\put(1,-42){\circle*{0.5}}
\put(2,-42){\circle*{0.5}}
\put(3,-42){\circle*{0.5}}
\put(4,-42){\circle*{0.5}}
\put(5,-42){\circle*{0.5}}
\put(6,-42){\circle*{0.5}}
\put(7,-42){\circle*{0.5}}
\put(0,-43){\circle*{0.5}}
\put(1,-43){\circle*{0.5}}
\put(2,-43){\circle*{0.5}}
\put(2,-44){\circle*{0.5}}
\put(3,-44){\circle*{0.5}}
\put(2,-45){\circle*{0.5}}
\put(3,-45){\circle*{0.5}}
\put(3,-46){\circle*{0.5}}
\put(4,-46){\circle*{0.5}}
\put(2,-47){\circle*{0.5}}
\put(3,-47){\circle*{0.5}}
\put(4,-47){\circle*{0.5}}
\put(5,-47){\circle*{0.5}}
\put(1,-48){\circle*{0.5}}
\put(2,-48){\circle*{0.5}}
\put(3,-48){\circle*{0.5}}
\put(3,-49){\circle*{0.5}}
\put(4,-49){\circle*{0.5}}
\put(2,-50){\circle*{0.5}}
\put(3,-50){\circle*{0.5}}
\put(4,-50){\circle*{0.5}}
\put(3,-51){\circle*{0.5}}
\put(4,-51){\circle*{0.5}}
\put(5,-51){\circle*{0.5}}
\put(2,-52){\circle*{0.5}}
\put(3,-52){\circle*{0.5}}
\put(4,-52){\circle*{0.5}}
\put(5,-52){\circle*{0.5}}
\put(6,-52){\circle*{0.5}}
\put(1,-53){\circle*{0.5}}
\put(2,-53){\circle*{0.5}}
\put(3,-53){\circle*{0.5}}
\put(4,-53){\circle*{0.5}}
\put(4,-54){\circle*{0.5}}
\put(5,-54){\circle*{0.5}}
\put(3,-55){\circle*{0.5}}
\put(4,-55){\circle*{0.5}}
\put(5,-55){\circle*{0.5}}
\put(6,-55){\circle*{0.5}}
\put(2,-56){\circle*{0.5}}
\put(3,-56){\circle*{0.5}}
\put(4,-56){\circle*{0.5}}
\put(5,-56){\circle*{0.5}}
\put(6,-56){\circle*{0.5}}
\put(7,-56){\circle*{0.5}}
\put(1,-57){\circle*{0.5}}
\put(2,-57){\circle*{0.5}}
\put(3,-57){\circle*{0.5}}
\put(4,-57){\circle*{0.5}}
\put(5,-57){\circle*{0.5}}
\put(6,-57){\circle*{0.5}}
\put(7,-57){\circle*{0.5}}
\put(8,-57){\circle*{0.5}}
\put(0,-58){\circle*{0.5}}
\put(1,-58){\circle*{0.5}}
\put(2,-58){\circle*{0.5}}
\put(3,-58){\circle*{0.5}}
\put(3,-59){\circle*{0.5}}
\put(4,-59){\circle*{0.5}}
\put(3,-60){\circle*{0.5}}
\put(4,-60){\circle*{0.5}}
\put(5,-60){\circle*{0.5}}
\put(2,-61){\circle*{0.5}}
\put(3,-61){\circle*{0.5}}
\put(4,-61){\circle*{0.5}}
\put(4,-62){\circle*{0.5}}
\put(5,-62){\circle*{0.5}}
\put(3,-63){\circle*{0.5}}
\put(4,-63){\circle*{0.5}}
\put(5,-63){\circle*{0.5}}
\put(6,-63){\circle*{0.5}}
\put(2,-64){\circle*{0.5}}
\put(3,-64){\circle*{0.5}}
\put(4,-64){\circle*{0.5}}
\put(5,-64){\circle*{0.5}}
\put(6,-64){\circle*{0.5}}
\put(7,-64){\circle*{0.5}}
\put(1,-65){\circle*{0.5}}
\put(2,-65){\circle*{0.5}}
\put(3,-65){\circle*{0.5}}
\put(4,-65){\circle*{0.5}}
\put(5,-65){\circle*{0.5}}
\put(4,-66){\circle*{0.5}}
\put(5,-66){\circle*{0.5}}
\put(6,-66){\circle*{0.5}}
\put(3,-67){\circle*{0.5}}
\put(4,-67){\circle*{0.5}}
\put(5,-67){\circle*{0.5}}
\put(6,-67){\circle*{0.5}}
\put(7,-67){\circle*{0.5}}
\put(2,-68){\circle*{0.5}}
\put(3,-68){\circle*{0.5}}
\put(4,-68){\circle*{0.5}}
\put(5,-68){\circle*{0.5}}
\put(6,-68){\circle*{0.5}}
\put(7,-68){\circle*{0.5}}
\put(8,-68){\circle*{0.5}}
\put(1,-69){\circle*{0.5}}
\put(2,-69){\circle*{0.5}}
\put(3,-69){\circle*{0.5}}
\put(4,-69){\circle*{0.5}}
\put(5,-69){\circle*{0.5}}
\put(6,-69){\circle*{0.5}}
\put(7,-69){\circle*{0.5}}
\put(8,-69){\circle*{0.5}}
\put(9,-69){\circle*{0.5}}
\put(0,-70){\circle*{0.5}}
\put(1,-70){\circle*{0.5}}
\put(2,-70){\circle*{0.5}}
\put(3,-70){\circle*{0.5}}
\put(4,-70){\circle*{0.5}}
\put(5,-70){\circle*{0.5}}
\put(5,-71){\circle*{0.5}}
\put(6,-71){\circle*{0.5}}
\put(4,-72){\circle*{0.5}}
\put(5,-72){\circle*{0.5}}
\put(6,-72){\circle*{0.5}}
\put(7,-72){\circle*{0.5}}
\put(3,-73){\circle*{0.5}}
\put(4,-73){\circle*{0.5}}
\put(5,-73){\circle*{0.5}}
\put(6,-73){\circle*{0.5}}
\put(7,-73){\circle*{0.5}}
\put(8,-73){\circle*{0.5}}
\put(2,-74){\circle*{0.5}}
\put(3,-74){\circle*{0.5}}
\put(4,-74){\circle*{0.5}}
\put(5,-74){\circle*{0.5}}
\put(6,-74){\circle*{0.5}}
\put(7,-74){\circle*{0.5}}
\put(8,-74){\circle*{0.5}}
\put(9,-74){\circle*{0.5}}
\put(1,-75){\circle*{0.5}}
\put(2,-75){\circle*{0.5}}
\put(3,-75){\circle*{0.5}}
\put(4,-75){\circle*{0.5}}
\put(5,-75){\circle*{0.5}}
\put(6,-75){\circle*{0.5}}
\put(7,-75){\circle*{0.5}}
\put(8,-75){\circle*{0.5}}
\put(9,-75){\circle*{0.5}}
\put(10,-75){\circle*{0.5}}
\put(0,-76){\circle*{0.5}}
\put(1,-76){\circle*{0.5}}
\put(2,-76){\circle*{0.5}}
\put(3,-76){\circle*{0.5}}
\put(4,-76){\circle*{0.5}}
\put(5,-76){\circle*{0.5}}
\put(6,-76){\circle*{0.5}}
\put(7,-76){\circle*{0.5}}
\put(8,-76){\circle*{0.5}}
\put(9,-76){\circle*{0.5}}
\put(10,-76){\circle*{0.5}}
\put(11,-76){\circle*{0.5}}
\multiput(-0.5,0.5)(0,-77){2}{\line(1,0){12}}
\multiput(-0.5,0.5)(12,0){2}{\line(0,-1){77}}
\end{picture}
  
\tabularnewline
\readwriteformat{Read} \totalFormat{154} &
\readwriteformat{Write} \totalFormat{154} &
\totalFormat{271} \readwriteformat{Read} & 
\totalFormat{271} \readwriteformat{Write} 
\tabularnewline
\multicolumn{1}{l}{\ruleAscHandle{12}} & 
\multicolumn{2}{c}{$\nump(12) = 77$} &
\multicolumn{1}{r}{\ruleDescHandle{12}}
\end{tabularx}
\renewcommand{\thisCaptionSummary}[0]{Read and write tapes for the direct
implementations of succession rules to generate ascending and descending
compositions.}
\caption[\thisCaptionSummary]{\thisCaptionSummary\ On the left we have
the read and write tapes for the ascending composition generator,
\algorithmRef{alg-rule-asc}; on the right, then, are the corresponding
tapes for the descending composition generator,
\algorithmRef{alg-rule-desc}. In both cases, the traces correspond to the
read and write operations carried out in generating all partitions of
$12$.\label{fig-succession-rule-tapes}}
\end{figure}

We can see the qualitative difference between the algorithms by
examining their read and write tapes in
\figureRef{fig-succession-rule-tapes}. The tapes in question are generated
by imagining that read and write heads mark a tape each time one of these
operations is made. The horizontal position of each
head is determined by the index of the array element involved. The tape is
advanced one unit each time a composition is visited, and so we can see the
number of read and write operations required for each individual partition
generated. Regarding \figureRef{fig-succession-rule-tapes} then, and
examining the read tape for \ruleAsc, we can see that
every partition requires exactly $2$ reads; in contrast, the read tape for 
\ruleDesc\ shows a maximum of $n - 1$ read operations
per partition, and this oscillates rapidly as we move along the tape.
Similarly, the write tape for 
\ruleAsc\ shows that we sometimes need to make a long
sequence of write operations to make the transition in question, but that
these are compensated for --- as our analysis has shown --- by the
occasions where we need only one write. The behaviour of the write head in
\ruleDesc\ is very similar to that of its read head, and
we again see many transitions where a large number of writes are required.

The difference between \ruleAsc\ and \ruleDesc\ is not due to some  
algorithmic nuance; rather, it reflects of a structural property of
the objects in question. The total suffix length~\cite{kemp-generating} of
descending compositions is much greater than that of ascending
compositions, because in many descending composition the suffix consists
of the sequence of $1$s; and we known that the total number of $1$s in all
partitions of $n$ is $\sum_{x = 1}^{n - 1}\nump(x)$. In this well-defined
way, it is more efficient to generate all ascending compositions than it 
is to generate all descending compositions.

\subsection{A new formula for $p(n)$}
\label{sec-gap-succession-rules-pn}
Although not strictly relevant to our analyses of ascending and descending
composition generation algorithms, another result follows directly from the
analysis of \algorithmRef{alg-rule-asc}.  If we compare the lexicographic succession
rule \equationRef{eqn-lexsuccA} and \algorithmRef{alg-rule-asc} carefully, we realise that the
$\mu$ copies of $m$ must be inserted into the array within the inner loop
of lines~8--12; and our analysis has given us the precise number of times
that this happens. Therefore, we know that the sum of $\mu$ values
over all ascending compositions of $n$ (except the last composition,
$\singletonSequence{n}$), must equal the number of write operations made in
the inner loop. Using this observation we then get the following theorem. 

\begin{theorem}\label{the-ak-pn}
For all $n \geq 1$
\begin{equation}\label{eqn-ak-pn}
\nump(n) = \frac{1}{2}\left(
1 + n + 
\stacksum{\sequence{a}{k} \in}{\sac(n) \setminus\{\singletonSequence{n}\}}
{\left\lfloor \frac{a_{k - 1} + a_k}{a_{k - 1} + 1}
\right\rfloor}
\right)
\end{equation}
\end{theorem}
\begin{proof}
We know from \lemmaRef{lem-freq-alg-rule-asc-k=k+1} that the total number
of write operations made by \algorithmRef{alg-rule-asc} in the
inner loop of lines~8--12 is given by $\nump(n) - 1$.
\algorithmRef{alg-rule-asc} applies the lexicographic succession rule above
to all elements of $\sac(n) \setminus \{\singletonSequence{n}\}$, as well
as one extra composition, which we refer to as the
\scarequote{initialisation composition}. The initialisation composition is
not in the set $\sac(n)$ as $a_1 = 0$, and so we
must discount the number of writes incurred by 
applying the succession rule to this composition. The composition visited
immediately after $0n$ is $1\dots1$, and so $n - 1$ copies of $1$ must have
been inserted into the array in the
inner loop during this transition. Therefore, the total number of
writes made within the inner loop in applying the succession rule to
all elements of $\sac(n)
\setminus \{\singletonSequence{n}\}$ is given by $\nump(n) - 1 -(n - 1)
= \nump(n) - n$. Therefore, from this result and the succession rule of 
\theoremRef{the-lexsuccA} we get
\[
\nump(n) - n = 
\stacksum{\sequence{a}{k} \in}{\sac(n) \setminus\{\singletonSequence{n}\}}
{\left( \left\lfloor \frac{a_{k - 1} + a_k}{a_{k - 1} + 1}
\right\rfloor - 1\right)}, 
\]
from which it is easy to derive \equationRef{eqn-ak-pn}, completing the proof.
\end{proof}

We can simplify \equationRef{eqn-ak-pn} if we suppose that
all $\sequence{a}{k}\in \sac(n)$ are prefixed by a value $0$. More
formally, a direct consequence of \theoremRef{the-ak-pn} is that
\begin{equation}\label{eqn-ak-pn-simplified}
\nump(n) = \frac{1}{2}
\left( 1 + 
\sum_{\sequence{a}{k} \in\sac'(n)} 
\left\lfloor \frac{a_{k - 1} + a_k}{a_{k - 1} + 1}
\right\rfloor \right),
\end{equation}
where $\sac'(n) = \{0\prepend\sequence{a}{k} \suchthat \sequence{a}{k} \in
\sac(n)\}$.  Fundamentally, what \theoremRef{the-ak-pn}
shows us is that if we let $y$ be the largest part and $x$ the second 
largest part in an arbitrary partition of $n$, we can count the partitions
of $n$ by summing $\floor{(x + y) / (x + 1)}$ over all partitions of $n$.

The partition function $\nump(n)$ is one of the most important functions in 
the theory of partitions and has been studied for several centuries~\cite{andrews-partitions}.
The asymptotic~\cite{hardy-asymptotic} and arithmetic~\cite{ahlgren-addition} 
properties of $\nump(n)$ have been very thoroughly examined. While 
\equationRef{eqn-ak-pn-simplified} is clearly not an efficient means of computing
$\nump(n)$, it may provide some new insight into this celebrated function.

\section{Accelerated Algorithms}\label{sec-gap-accelerated-algorithms}
In this section we examine algorithms that use structural properties of
the sets of ascending and descending compositions to reduce the number of
read and write operations required. The algorithms presented are the
most efficient known examples of ascending and descending composition
generators, ensuring that we have a fair comparison of the algorithms
arising from the two candidate encodings for partitions. In
\sectionRef{sec-gap-accelerated-algorithms-ascending-compositions} we
develop a new ascending composition generator that requires 
fewer read operations than \ruleAsc. 
Then, in
\sectionRef{sec-gap-accelerated-algorithms-descending-compositions} we
study the most efficient known descending composition generation
algorithm, due to \twoAuthor{Zoghbi}{\Stojmenovic}~\cite{zoghbi-fast},
which
requires far fewer read and write operations than \ruleDesc.
In \sectionRef{sec-gap-accelerated-algorithms-comparison}, we compare these
two algorithms to determine which of the two is more efficient.

\subsection{Ascending Compositions}
\label{sec-gap-accelerated-algorithms-ascending-compositions}
In this subsection we improve on
\ruleAsc\ (\algorithmRef{alg-rule-asc}) by applying the theory of
\scarequote{terminal} and \scarequote{nonterminal} compositions.
To enable us to fully analyse the resulting algorithm we require an
expression to enumerate terminal ascending compositions in terms of 
$\nump(n)$. In the opening
part of this subsection we develop the theory of terminal and
nonterminal compositions. A byproduct of
this analysis is a new proof for a partition identity on the number
of partitions where the largest part is less than twice the second largest
part. After developing this necessary theory, we move on to the
description of the algorithm itself, and its subsequent analysis.

\subsubsection{Terminal and Nonterminal Compositions}
\label{sec-gap-accelerated-algorithms-ascending-compositions-terminal}
The algorithm that we shall examine shortly uses some structure within the 
set of ascending compositions to make many transitions very efficient. 
This structure is based on the ideas of \scarequote{terminal} and 
\scarequote{nonterminal} compositions. We now define these concepts 
and derive some basic enumerative results to aid us in our analysis.

\begin{definition}[Terminal Ascending Composition] 
\label{def-terminal-ascending-compositions}
For some positive integer $n$, an ascending composition
$\sequence{a}{k}\in\sac(n)$ is \emph{terminal}
if $k = 1$ or $2a_{k - 1} \leq a_k$. Let $\stac(n, m)$ denote
the set of terminal compositions in $\sac(n, m)$, and $\ntac(n, m)$ denote
the cardinality of this set (i.e.\ $\ntac(n, m) =
\cardinality{\stac(n, m)}$).
\end{definition}

\begin{definition}[Nonterminal Ascending Composition]
\label{def-nonterminal-ascending-compositions}
For some positive integer $n$, $\sequence{a}{k} \in \sac(n)$ is
\emph{nonterminal} if $k > 1$ and $2a_{k - 1} > a_k$. Let
$\snac(n, m)$ denote the set of nonterminal compositions in $\sac(n, m)$,
and let $\nnac(n, m)$ denote the cardinality of this set (i.e.\ $\nnac(n,
m) = \cardinality{\snac(n, m)}$).
\end{definition}

If we let  $\nac(n,m)$ denote the number of ascending compositions
of $n$ where the initial part is at least $m$ it can be 
shown~\cite[\chapterNumber{3}]{kelleher-encoding} that 

\begin{equation}\label{eqn-nac-iter}
\nac(n, m) = 1 + \sum_{x = m}^{\floor{n / 2}} \nac(n - x, x)
\end{equation}
holds for all positive integers $m\leq n$.
We require a similar recurrence to enumerate the terminal ascending
compositions, and so we let  $\ntac(n, m)$ denote the
number of terminal compositions in the set $\sac(n, m)$. The terminal
ascending compositions are a subset of the ascending compositions, and the
construction rule implied is the same: the number of terminal ascending
compositions of $n$ where the initial part is \emph{exactly} $m$ is equal
to the number of terminal compositions of $n - m$ with initial part
\emph{at least} $m$. The only difference, then, between the recurrences
for ascending compositions and terminal ascending compositions 
occurs in the boundary conditions. The recurrence can be stated as follows: 
for all positive integers $m\leq n$, $\ntac(n, m)$ satisfies 
\begin{equation}\label{eqn-ntac-iter}
\ntac(n, m) = 1 + \sum_{x = m}^{\floor{n / 3}} \ntac(n - x, x).
\end{equation}
See Kelleher~\cite[\pageNumber{160--161}]{kelleher-encoding} for the proofs 
of recurrences~\equationRef{eqn-nac-iter} and~\equationRef{eqn-ntac-iter}.

 Before we
move onto the main result, where we prove that $\ntac(n,m) = \nac(n,
m) - \nac(n - 2, m)$, we require some auxiliary results which
simplify the proof of this assertion. In \lemmaRef{lem-x-leq-floor-n--x-m}
we prove an equivalence between logical statements of a particular form
involving the  floor function, which is useful in
\lemmaRef{lem-ascending-tail-ones}; the latter lemma then provides the
main inductive step in our proof of the central theorem of this section.
In the interest of brevity, we limit our proofs to values of $n > 3$,
since $n \leq 3$ can be easily demonstrated and would unnecessarily
complicate the proofs.

\begin{lemma}\label{lem-x-leq-floor-n--x-m}
If $x$, $m$ and $n$ are positive integers then $x \leq \floor{(n - x) / m}
\equivalent x \leq \floor{n / (m + 1)}$.
\end{lemma}
\begin{proof}
Suppose $x$, $m$ and $n$ are positive integers. 
Suppose $x \leq \floor{(n - x) / m}$. Thus, $x \leq (n - x) / m$, and so 
$x \leq n / (m + 1)$. Then, as $\floor{n / (m + 1)} \leq n / (m + 1)$ and
$x$ is an integer, we know that $x \leq \floor{n / (m + 1)}$, and so $x
\leq \floor{(n - x) / m} \implies x \leq \floor{n / (m + 1)}$. 

Suppose that $x \leq \floor{n / (m + 1)}$. Then, $x \leq n / (m +
1)$, and so $x \leq (n - x)/m$. Once again, as $x$ is an integer it is
apparent that $x \leq  \floor{(n - x)/m} \leq  (n - x)/m$, and so $x \leq
\floor{n / (m + 1)} \implies x \leq \floor{(n - x)/m}$. Therefore, as  $x
\leq \floor{(n - x) / m} \implies x \leq \floor{n / (m + 1)}$ and $x \leq
\floor{n / (m + 1)} \implies x \leq \floor{(n - x)/m}$ we see that 
$x \leq \floor{(n - x) / m} \equivalent x \leq \floor{n / (m + 1)}$, as
required.
\end{proof}

\begin{lemma}\label{lem-ascending-tail-ones}
For all positive integers $n > 3$
\begin{equation}\label{eqn-ascending-tail-ones}
\sum_{x = \floor{n / 3} + 1}^{\floor{n / 2}} \nac(n - x, x) = 1 + 
\sum_{x = \floor{n / 3} + 1}^{\floor{(n - 2) / 2}} \nac(n - 2 - x, x).
\end{equation}
\end{lemma}
\begin{proof}
Suppose $n > 3$ and $1 \leq m \leq n$, and consider the left-hand side of
\equationRef{eqn-ascending-tail-ones}. We know that $\nac(n, m) = 1$ if $m
> \floor{n / 2}$, as the summation in recurrence \equationRef{eqn-nac-iter}
will be empty. By the contrapositive of \lemmaRef{lem-x-leq-floor-n--x-m}
we know that $x > \floor{(n - x) / 2} \equivalent x > \floor{n / 3}$, and
we therefore know that each term in the summation of the left-hand side
of \equationRef{eqn-ascending-tail-ones} is equal to $1$. Thus, we see
that 
\begin{equation}\label{eqn-ascending-tail-ones-lhs}
\sum_{x = \floor{n / 3} + 1}^{\floor{n / 2}} \nac(n - x, x) = 
\floor{n / 2} -  \floor{n / 3} - 1.
\end{equation}

Similarly, as $x > \floor{n / 3} \implies x > \floor{(n - x) / 2}$, it
clearly follows that $x > \floor{n / 3} \implies x > \floor{(n - x) / 2} -
1$, or $x > \floor{n / 3} \implies x > \floor{(n - 2 - x) /
2}$. Thus, each term in the summation on the right-hand side of
\equationRef{eqn-ascending-tail-ones} must also equal $1$, and so we get
\begin{align}
\nonumber
1 +  \sum_{x = \floor{n / 3} + 1}^{\floor{(n - 2) / 2}} \nac(n - 2 - x, x)
& = 1 + \floor{(n - 2) / 2} - \floor{n / 3} - 1\\
\label{eqn-ascending-tail-ones-rhs}
& = \floor{n / 2} -  \floor{n / 3} - 1.
\end{align}
Therefore, as \equationRef{eqn-ascending-tail-ones-lhs} and
\equationRef{eqn-ascending-tail-ones-rhs} show that the left-hand and
right-hand side of \equationRef{eqn-ascending-tail-ones} are equal, the
proof is complete.
\end{proof}

\begin{theorem}\label{the-ntac}
If $n \geq 3$, then $\ntac(n, m) = \nac(n, m) - \nac(n - 2, m)$ for all
$1 \leq m \leq \floor{n / 2}$.
\end{theorem}
\begin{proof}
Proceed by strong induction on $n$.

\begin{basecase}{$n = 3$} As $1 \leq m \leq \floor{n / 2}$ and $n =
3$, we know that $m = 1$. Computing $\ntac(3,1)$, we get $1 + \ntac(2, 1)
= 2$. We also find $\nac(3, 1) = 3$ and $\nac(1, 1) = 1$, and so the base
case of the induction holds.
\end{basecase}

\begin{inductivecase}
Suppose  $\ntac(n', m) = \nac(n', m) - \nac(n' - 2, m)$ when 
$1 \leq m \leq \floor{n' / 2}$, for all $3 < n' < n$, and some integer $n$.
Then, as $x \leq \floor{(n - x) / 2} \equivalent x \leq \floor{n /
3}$, by \lemmaRef{lem-x-leq-floor-n--x-m}, we can apply this inductive
hypothesis to each term 
$\ntac(n - x, x)$ in \equationRef{eqn-ntac-iter}, giving us
\begin{align}
\ntac(n, m) & = 1 + \sum_{x = m}^{\floor{n / 3}}(\nac(n - x, x) - 
\nac(n - 2 - x, x)) \nonumber \\
 &= 1 + \sum_{x = m}^{\floor{n / 3}}\nac(n - x, x) - 
\sum_{x = m}^{\floor{n / 3}}\nac(n - 2 - x, x).
\label{mervin-muffly}
\end{align}
By \lemmaRef{lem-ascending-tail-ones} we know that 
\[ \sum_{x = \floor{n / 3}
+ 1}^{\floor{n / 2}} \nac(n - x, x) - 
\sum_{x = \floor{n / 3} + 1}^{\floor{(n - 2) / 2}} \nac(n - 2 - x, x) - 1 =
0,\]
 and so we can add the left-hand side of this equation to the right-hand
side of \equationRef{mervin-muffly}, to get
\begin{multline*}
 \ntac(n, m) = 1 + \sum_{x = m}^{\floor{n / 3}}\nac(n - x, x) - 
\sum_{x = m}^{\floor{n / 3}}\nac(n - 2 - x, x) \\
+ \sum_{x = \floor{n / 3}
+ 1}^{\floor{n / 2}} \nac(n - x, x) - 
\sum_{x = \floor{n / 3} + 1}^{\floor{(n - 2) / 2}} \nac(n - 2 - x, x) - 1.
\end{multline*}
Then, gathering the terms $\nac(n - x, x)$ and $\nac(n - 2 - x, x)$ into 
the appropriate summations we get 
\[
\ntac(n, m)  = 1 + \sum_{x = m}^{\floor{n / 2}}\nac(n - x, x) - 1
-  \sum_{x = m}^{\floor{(n - 2)/ 2}}\nac(n - 2 - x, x),
\]
which by \equationRef{eqn-ntac-iter} gives us $\ntac(n, m) = \nac(n, m) -
\nac(n - 2, m)$, as required.
\end{inductivecase}
\end{proof}

For the purposes of our analysis it is useful to know the total number of
terminal and nonterminal compositions of $n$, and it is worthwhile
formalising the results here for reference. Therefore, letting $\ntac(n) =
\ntac(n, 1)$ and $\nnac(n) = \nnac(n,1)$, we get the following corollaries
defined in terms of the partition function $\nump(n)$.

\begin{corollary}\label{cor-ntac}
For all positive  integers $n$, $\ntac(n) =\nump(n) - \nump(n - 2)$.
\end{corollary}
\begin{proof}
As $\ntac(n) = \ntac(n,1)$ and $\nac(n,1) = \nump(n)$, proof is immediate
by \theoremRef{the-ntac} for all $n \geq 3$. Since $\nump(n) = 0$ for all
$n < 0$ and $\nump(0) = 1$, we can readily verify that $\ntac(2) =
\ntac(1) = 1$, as required.
\end{proof}

\begin{corollary}\label{cor-nnac}
If $n$ is a positive integer then $\nnac(n) =\nump(n - 2)$.
\end{corollary}
\begin{proof}
An ascending composition is either terminal or nonterminal. As the total
number of ascending compositions of $n$ is given by $\nump(n)$, we get 
$\nnac(n) = \nump(n) - (\nump(n) - \nump(n - 2)) =  \nump(n - 2)$,
as required.
\end{proof}

Corollaries~\ref{cor-ntac} and~\ref{cor-nnac} prove a nontrivial
structural property of the set of all ascending compositions, and can be
phrased in more conventional partition theoretic language. Consider an
arbitrary partition of $n$, and let $y$ be the largest part in this
partition. We then let $x$ be the second largest part ($x \leq y$).
\corollaryRef{cor-nnac} then shows that the number of partitions of $n$
where $2x > y$ is equal to the number of partitions of $n - 2$. This
result is known, and has been reported by
Adams-Watters~\cite[\sequenceNumber{A027336}]{sloane-online}. The
preceding treatment, however, would appear to be the first published proof
of the identity.

\subsubsection{Algorithm}
Having derived some theoretical results about the terminal and nonterminal
ascending compositions of $n$, we are now in a position to exploit those 
properties in a generation algorithm.  
In the direct implementation of the lexicographic succession rule for 
ascending compositions, \ruleAsc,  we generate the
successor of $\sequence{a}{k}$ by computing the lexicographically least
element of the set $\sac(a_{k - 1} + a_{k}, a_{k - 1} + 1)$, and
visit the resulting composition. The
algorithm operates by implementing exactly one transition per iteration of
the main loop. The accelerated algorithm, \accelAsc, developed 
here operates on a slightly different principle: we compute the
lexicographically least composition of $\sac(a_{k - 1} +
a_{k}, a_{k - 1} + 1)$, as before, but we now keep a watchful eye to see
if the resulting composition is nonterminal. If it is, we can
compute the lexicographic successor simply by incrementing  $a_{k - 1}$ and 
decrementing $a_{k}$.  Otherwise,
we revert to the standard means of computing the lexicographic
successor. By analysing this algorithm, we shall see that this approach 
provides significant gains. We concentrate on the analysis of 
\algorithmRef{alg-accel-asc} here --- see Kelleher~\cite[\sectionNumber{4.4.2}]%
{kelleher-encoding} for further discussion and proof of correctness.

\begin{algorithm}[t]
\caption{$\accelAscHandle{n}$}
\begin{algorithmic}[1]
\label{alg-accel-asc}
\REQUIRE $n \geq 1$  
\STATE $k \leftarrow 2$                     \label{alg-accel-asc-k=2}
\STATE $a_{1} \leftarrow 0$                 \label{alg-accel-asc-a1=0}
\STATE $y \leftarrow n - 1$                 \label{alg-accel-asc-y=n-1}
\WHILE {$k \neq 1$}                         \label{alg-accel-asc-k-neq-1} 
    \STATE $k \leftarrow k - 1$             \label{alg-accel-asc-k=k-1}
    \STATE $x \leftarrow a_{k} + 1$         \label{alg-accel-asc-x=ak+1}
    \WHILE {$2x \leq y$}                    \label{alg-accel-asc-2x-leq-y}
        \STATE $a_k \leftarrow x$           \label{alg-accel-asc-ak=x-e}
        \STATE $y \leftarrow y - x$         \label{alg-accel-asc-y=y-x}
        \STATE $k \leftarrow k + 1$         \label{alg-accel-asc-k=k+1}
    \ENDWHILE  
    \STATE $\ell \leftarrow k + 1$          \label{alg-accel-asc-l=k+1}
    \WHILE {$x \leq y$}                     \label{alg-accel-asc-x-leq-y}
        \STATE $a_k \leftarrow x$           \label{alg-accel-asc-ak=x-v} 
        \STATE $a_\ell \leftarrow y$        \label{alg-accel-asc-al=y} 
        \STATE $\visit$ $\sequence{a}{\ell}$\label{alg-accel-asc-visit-v} 
        \STATE $x \leftarrow x + 1$         \label{alg-accel-asc-x=x+1} 
        \STATE $y \leftarrow y - 1$         \label{alg-accel-asc-y=y-1}
    \ENDWHILE
    \STATE $y \leftarrow y + x - 1$         \label{alg-accel-asc-y=y+x-1}
    \STATE $a_{k} \leftarrow y + 1$         \label{alg-accel-asc-ak=y+1}
    \STATE $\visit$ $\sequence{a}{k}$       \label{alg-accel-asc-visit-o}
\ENDWHILE 
\end{algorithmic}
\end{algorithm}

\begin{lineFrequencyLemma}{alg-accel-asc}{visit-v}{n}{\nump(n - 2)}
Compositions visited on line~\ref{alg-accel-asc-visit-v} must be
nonterminal because upon reaching line~\ref{alg-accel-asc-l=k+1}, the
condition $2x > y$ must hold. As $x$ and $y$ are the second-last and
last parts, respectively, of the composition visited on
line~\ref{alg-accel-asc-visit-v},
then this composition must be nonterminal by definition. Subsequent
operations on $x$ and $y$ within this loop do not alter the property that
$2x > y$, and so all compositions visited on
line~\ref{alg-accel-asc-visit-v} must be nonterminal. 

Furthermore, we also know that all compositions visited on
line~\ref{alg-accel-asc-visit-o} must be terminal. To demonstrate this
fact, we note that if $\sequence{a}{k}$ is the last composition
visited before we arrive at line~\ref{alg-accel-asc-y=y+x-1}, the
composition visited on line~\ref{alg-accel-asc-visit-o} must be
$\sequence{a}{k - 2}\concatenate\singletonSequence{a_{k - 1} + a_k}$.
Therefore, to demonstrate that this composition is terminal, we must show
that $2a_{k - 2} \leq a_{k - 1} + a_k$. We know that $a_{k - 2} \leq a_{k -
1} \leq a_k$. It follows that $2a_{k - 2} \leq 2a_{k - 1}$, and also that
$2a_{k - 1} \leq a_{k - 1} + a_k$. Combining these two inequalities, we see
that  $2a_{k - 2} \leq 2a_{k - 1} \leq a_{k - 1} + a_k$, and so $2a_{k - 2}
\leq a_{k - 1} + a_k$. Thus all compositions visited on
line~\ref{alg-accel-asc-visit-o} must be terminal.

Then, as \algorithmRef{alg-accel-asc} correctly visits all $\nump(n)$
ascending compositions of $n$~\cite[\pageNumber{105}]{kelleher-encoding}, 
since all compositions 
visited on line~\ref{alg-accel-asc-visit-o} are terminal and as all
compositions visited on line~\ref{alg-accel-asc-visit-v} are nonterminal,
we know that all nonterminal compositions of $n$ must be visited on
line~\ref{alg-accel-asc-visit-v}. By \corollaryRef{cor-nnac} there are
$\nump(n - 2)$ nonterminal compositions of $n$, and hence
$\freq{alg-accel-asc-visit-v} = \nump(n - 2)$, as required.
\end{lineFrequencyLemma}

\begin{lineFrequencyLemma}{alg-accel-asc}{k=k-1}{n}
{\nump(n) - \nump(n - 2)}
By \lemmaRef{lem-freq-alg-accel-asc-visit-v} we know that the visit
statement on line~\ref{alg-accel-asc-visit-v} is executed $\nump(n - 2)$
times. As \algorithmRef{alg-accel-asc} correctly visits all $\nump(n)$
ascending compositions of $n$, then the remaining $\nump(n) - \nump(n - 2)$
compositions must be visited on line~\ref{alg-accel-asc-visit-o}. Clearly
then, line~\ref{alg-accel-asc-visit-o} (and hence
line~\ref{alg-accel-asc-k=k-1}) is executed $\nump(n) - \nump(n - 2)$
times. Therefore, $\freq{alg-accel-asc-k=k-1} = \nump(n) - \nump(n - 2)$,
as required. 
\end{lineFrequencyLemma}

\begin{lineFrequencyLemma}{alg-accel-asc}{k=k+1}{n}
{\nump(n) - \nump(n - 2) - 1}
The variable $k$ is assigned the value $2$ upon initialisation, and the
algorithm terminates when $k = 1$. As the variable is only updated via
increment (line~\ref{alg-accel-asc-k=k+1}) and decrement
(line~\ref{alg-accel-asc-k=k-1}) operations, we know that there must be
one more decrement operation than increments. By
\lemmaRef{lem-freq-alg-accel-asc-k=k-1} we know that there are $\nump(n) -
\nump(n - 2)$ decrements, and so there must be $\nump(n) - \nump(n - 2) -
1$ increments on the variable. Therefore, $\freq{alg-accel-asc-k=k+1} =
\nump(n) - \nump(n - 2) - 1$.
\end{lineFrequencyLemma}

\newcommand{\trsAlgAccelAsc}{\ensuremath{\nump(n) - \nump(n - 2)}}
\begin{totalReadsTheorem}[\sac(n)]{alg-accel-asc}{n}{\trsAlgAccelAsc}
Only one read operation occurs \algorithmRef{alg-accel-asc}, and
this is done on line~\ref{alg-accel-asc-x=ak+1}. By
\lemmaRef{lem-freq-alg-accel-asc-k=k-1} we know that
line~\ref{alg-accel-asc-k=k-1} is executed $\nump(n) - \nump(n - 2)$
times, and it immediately follows that line~\ref{alg-accel-asc-x=ak+1} is
executed the same number of times. Therefore,  $\trs{alg-accel-asc}(n) =
\trsAlgAccelAsc$, as required.
\end{totalReadsTheorem}

\newcommand{\twsAlgAccelAsc}{\ensuremath{2\nump(n) - 1}}
\begin{totalWritesTheorem}[\sac(n)]{alg-accel-asc}{n}{\twsAlgAccelAsc}
Write operations are performed on lines~\ref{alg-accel-asc-ak=x-e},
\ref{alg-accel-asc-ak=x-v}, 
 \ref{alg-accel-asc-al=y} and~\ref{alg-accel-asc-ak=y+1}.
\lemmaRef{lem-freq-alg-accel-asc-k=k-1}
shows that line~\ref{alg-accel-asc-ak=y+1} is executed $\nump(n) - \nump(n
- 2)$ times. From \lemmaRef{lem-freq-alg-accel-asc-k=k+1} we know that
line~\ref{alg-accel-asc-ak=x-e} is executed $\nump(n) - \nump( n  - 2) - 1$
times. Then, by \lemmaRef{lem-freq-alg-accel-asc-visit-v} we know that
lines~\ref{alg-accel-asc-ak=x-v} and~\ref{alg-accel-asc-al=y} are
executed $\nump(n - 2)$ times each. Summing these contributions we get
$\tws{alg-accel-asc}(n) = \nump(n) - \nump(n - 2) + \nump(n) - \nump( n
 - 2) - 1 + 2\nump(n - 2) = \twsAlgAccelAsc$, as required.
\end{totalWritesTheorem}

Theorems~\ref{the-trs-alg-accel-asc} and~\ref{the-tws-alg-accel-asc}
derive the precise number of read and write operations required to
generate all partitions of $n$ using
\algorithmRef{alg-accel-asc}. This algorithm is a considerable improvement
over our basic implementation of the succession rule,
\algorithmRef{alg-rule-asc}, in two ways.  Firstly, by keeping $\nump(n - 2)$ of
the visit operations within the loop of lines 13--19,  we significantly
reduce the average cost of a write operation. Thus, although we do not
appreciably reduce the total number of write operations involved, we
ensure that $2\nump(n - 2)$ of those writes are executed at the cost of
an increment and decrement on a local variable and the cost of a
$\leq$ comparison of two local variables --- in short, \emph{very}
cheaply. 

The second improvement is that we dramatically reduce the total number
of read operations involved. Recall that \ruleAsc\ required $2\nump(n)$
read operations to generate all ascending
compositions of $n$; \theoremRef{the-trs-alg-accel-asc} shows that
\accelAsc\ requires only $\nump(n) - \nump(n -2)$ read
operations. We also reduced the number of read operations
by a factor of $2$ by maintaining the value of $y$ between iterations of
the main while loop, but this could equally be applied to 
\ruleAsc, and is only a minor improvement at any rate.
The real gain here is obtained from exploiting the block-based nature of
the set of ascending compositions, as we do not need to perform any read
operations once we have begun iterating through the nonterminal
compositions within a block.

\subsection{Descending Compositions}
\label{sec-gap-accelerated-algorithms-descending-compositions}
In \sectionRef{sec-gap-succession-rules-descending-compositions} we derived
a direct implementation of the succession rule for
descending compositions. We then analysed the cost of using this direct
implementation to generate all descending compositions of $n$, and found
that it implied an average of $\bigoh{\sqrt{n}}$  read and write
operations per partition. There are, however,
several constant amortised time algorithms to generate descending compositions, and in
this section we study the most efficient example.

There is one basic problem with the direct implementation of the succession
rule for descending compositions (\ruleDesc): most of the read and write
operations it makes are redundant. To begin with, the read
operations incurred by \ruleDesc\ in scanning the
current composition to find the rightmost non-$1$ value are unnecessary.
As McKay~\cite{mckay-partitions} noted, we can easily keep track of the
index of the largest non-$1$ value between iterations, and thereby
eliminate the right-to-left scan  altogether. The means by
which we can avoid the majority of the write operations is a
little more subtle, and was first noted by \twoAuthor{Zoghbi}
{\Stojmenovic}~\cite{zoghbi-fast}.
For instance, consider the transition
\begin{equation}\label{eqn-example-descending-transition}
3321111 \rightarrow 33111111.
\end{equation}
\ruleDesc\  implements the transition from $3321111$ to $33111111$
by finding the prefix $33$ and writing six copies of $1$ after it,
oblivious to the fact that $4$ of the array
indices \emph{already} contain $1$. Thus, a more reasonable approach is to
make a special case in the succession rule so that if $d_q = 2$, we simply
set $d_q \leftarrow 1$ and append $1$ to the end of the composition. This
observation proves to be sufficient to remove the worst excesses of
\ruleDesc, as $1$s are by far the most numerous part in
the partitions of $n$.

\twoAuthor{Zoghbi}{\Stojmenovic}'s algorithm implements both of these
ideas, and
makes one further innovation to reduce the number of
write operations required. By initialising the array to hold $n$ copies of
$1$, we know that any index $> k$ must contain the value $1$, and so we
can save another write operation in the special case of $d_q = 2$ outlined
above. Thus, \twoAuthor{Zoghbi}{\Stojmenovic}'s algorithm is the most
efficient
example, and consequently it is the algorithm that  we shall use for our
comparative analysis. Knuth developed a similar 
algorithm~\cite[\pageNumber{2}]{knuth-generating-all-partitions}: he also
noted the necessity of keeping track of the value of $q$ between
iterations, and also implemented the special case for $d_q$ outlined
above. Knuth's algorithm, however, does not contain the further
improvement included by \twoAuthor{Zoghbi}{\Stojmenovic}\ (i.e.\
initialising the
array to $1\dots1$ and avoiding the second write operation in the $d_q =
2$ special case), and therefore requires strictly more write operations
than \twoAuthor{Zoghbi}{\Stojmenovic}'s.  \twoAuthor{Zoghbi}
{\Stojmenovic}'s algorithm also
 consistently outperforms Knuth's algorithm in empirical tests.

\begin{algorithm}[!t]
\caption{$\accelDescHandle{n}$} 
\label{alg-accel-desc}
\begin{algorithmic}[1]
\label{alg}
\REQUIRE $n \geq 1$
\STATE $k \leftarrow 1$ \label{alg-accel-desc-k=1}
\STATE $q \leftarrow 1$ \label{alg-accel-desc-q=1}
\STATE $\sequence[2]{d}{n} \leftarrow 1\dots1$
    \label{alg-accel-desc-d2...dn=1..1}
\STATE $d_1 \leftarrow n$ \label{alg-accel-desc-d1=n}
\STATE $\visit$ $d_1$     \label{alg-accel-desc-visit-d1}
\WHILE {$q \neq 0$}       \label{alg-accel-desc-q-neq-0}
\IF {$d_q = 2$}           \label{alg-accel-desc-dq-eq-2}
    \STATE $k \leftarrow k + 1$ \label{alg-accel-desc-k=k+1}
    \STATE $d_q \leftarrow  1$  \label{alg-accel-desc-dq=1}
    \STATE $q \leftarrow q - 1$ \label{alg-accel-desc-q=q-1}
\ELSE
    \STATE $m \leftarrow d_q - 1$    \label{alg-accel-desc-m=dq-1}
    \STATE $n' \leftarrow k - q + 1$ \label{alg-accel-desc-n'=k-q+1}
    \STATE $d_q \leftarrow m$        \label{alg-accel-desc-dq=m} 
    \WHILE {$n' \geq m$}             \label{alg-accel-desc-n'-geq-m} 
        \STATE $q \leftarrow q + 1$  \label{alg-accel-desc-q=q+1-e} 
        \STATE $d_q \leftarrow m$    \label{alg-accel-desc-dq=m-e} 
        \STATE $n' \leftarrow n' - m$\label{alg-accel-desc-n'=n'-m}
    \ENDWHILE
    \IF {$n' = 0$} \label{alg-accel-desc-n'=0}
        \STATE $k = q$ \label{alg-accel-desc-k=q}
    \ELSE
        \STATE $k \leftarrow q + 1$ \label{alg-accel-desc-k=q+1}
        \IF {$n' > 1$}              \label{alg-accel-desc-n'>1}  
            \STATE $q \leftarrow q + 1$ \label{alg-accel-desc-q=q+1-p}
            \STATE $d_q \leftarrow n'$  \label{alg-accel-desc-dq=n'}
        \ENDIF
    \ENDIF
\ENDIF
\STATE $\visit$ $\sequence{d}{k}$ \label{alg-accel-desc-visit}
\ENDWHILE
\end{algorithmic}
\end{algorithm}    

\twoAuthor{Zoghbi}{\Stojmenovic}'s algorithm is presented
in \algorithmRef{alg-accel-desc}, which we shall also refer to as
\accelDesc. Each iteration of the main loop
implements a single transition, and two cases are identified for
performing the transition. In the conditional block of
lines~\ref{alg-accel-desc-k=k+1}--\ref{alg-accel-desc-q=q-1} we implement
the special case for $d_q = 2$: we can see that the length of the
composition is incremented, $d_q$ is assigned to $1$ and the value of $q$
is updated to point to the new rightmost non-$1$ part. The general case
is dealt with in the block of lines~11--29; the approach is 
much the same as that of \ruleDesc, except in this case we have the
additional
complexity of maintaining the value of $q$ between iterations.

\begin{lineFrequencyLemma}{alg-accel-desc}{q=q-1}{n}{\nump(n - 2)}
The variable $q$ points to the smallest non-$1$ value in
$\sequence{d}{k}$, and we have a complete descending composition in the
array each time we reach line~\ref{alg-accel-desc-dq-eq-2}. Therefore,
line~\ref{alg-accel-desc-q=q-1} will be executed once for every descending
composition of $n$ which contains at least one $2$; and it is well
known that this is $\nump(n - 2)$.
Therefore, $\freq{alg-accel-desc-q=q-1}(n) = \nump(n -
2)$, as required.
\end{lineFrequencyLemma}

\begin{twoLineFrequencyLemma}{alg-accel-desc}
{q=q+1-e}{q=q+1-p}{n}{\nump(n - 2) - 1}
The variable $q$ controls the termination of the algorithm. It is
initialised to $1$ on line~\ref{alg-accel-desc-q=1}, and the algorithm
terminates when $q=0$. We modify $q$ via increment
operations on lines~\ref{alg-accel-desc-q=q+1-e}
and~\ref{alg-accel-desc-q=q+1-p}, and decrement operations on
line~\ref{alg-accel-desc-q=q-1} only. Therefore, there must be one more
decrement operation than increments on $q$. By
\lemmaRef{lem-freq-alg-accel-desc-q=q-1} there are $\nump(n - 2)$
decrements performed on $q$, and there must therefore be $\nump(n - 2) - 1$
increments. Therefore, $\freq{alg-accel-desc-q=q+1-e}(n) +
\freq{alg-accel-desc-q=q+1-p}(n) = \nump(n - 2) - 1$, as required.
\end{twoLineFrequencyLemma}

\newcommand{\trsAlgAccelDesc}{\ensuremath{2\nump(n) - \nump(n - 2) - 2}}
\begin{totalReadsTheorem}[\sdc(n)]{alg-accel-desc}{n}{\trsAlgAccelDesc}
Read operations are performed on lines~\ref{alg-accel-desc-dq-eq-2}
and~\ref{alg-accel-desc-m=dq-1} of \algorithmRef{alg-accel-desc}. Clearly,
as all but the composition $\singletonSequence{n}$ are visited on
line~\ref{alg-accel-desc-visit}, line~\ref{alg-accel-desc-dq-eq-2} is
executed $\nump(n) - 1$ times. Then, as a consequence of
\lemmaRef{lem-freq-alg-accel-desc-q=q-1}, we know that
line~\ref{alg-accel-desc-m=dq-1} is executed $\nump(n) - \nump(n -
2) - 1$ times. Therefore, the total number of read operations is given by 
$\trs{alg-accel-desc}(n) = \trsAlgAccelDesc$, as required.
\end{totalReadsTheorem}

\newcommand{\twsAlgAccelDesc}{\ensuremath{\nump(n) + \nump(n - 2) - 2}}
\begin{totalWritesTheorem}[\sdc(n)]{alg-accel-desc}{n}{\twsAlgAccelDesc}
After initialisation, write operations are performed
on lines~\ref{alg-accel-desc-dq=1},
\ref{alg-accel-desc-dq=m}, \ref{alg-accel-desc-dq=m-e}
and~\ref{alg-accel-desc-dq=n'} of \algorithmRef{alg-accel-desc}.
Line~\ref{alg-accel-desc-dq=1} contributes $\nump(n - 2)$ writes by
\lemmaRef{lem-freq-alg-accel-desc-q=q-1}; and similarly,
line~\ref{alg-accel-desc-dq=m} is executed $\nump(n) - \nump(n - 2) - 1$
times. By
\lemmaRef{lem-freq-alg-accel-desc-q=q+1-e+freq-alg-accel-desc-q=q+1-p} we
know that the total number of write operations incurred by
lines~\ref{alg-accel-desc-dq=m-e} and~\ref{alg-accel-desc-dq=n'} is
$\nump(n - 2) - 1$. Therefore, summing these contributions we get 
$ \tws{alg-accel-desc}(n) =  \twsAlgAccelDesc$, as required.
\end{totalWritesTheorem}

Theorems~\ref{the-trs-alg-accel-desc} and~\ref{the-tws-alg-accel-desc}
show that \twoAuthor{Zoghbi}{\Stojmenovic}'s algorithm is a vast
improvement on
\ruleDesc. Recall that \ruleDescHandle{n} requires roughly 
$\sum_{x = 1}^{n}\nump(x)$ read and $\sum_{x = 1}^{n}\nump(x)$ write
operations; and we have seen that \accelDescHandle{n} requires only
$2\nump(n) - \nump(n - 2)$ read and $\nump(n) + \nump(n - 2)$ write
operations. 

\twoAuthor{Zoghbi}{\Stojmenovic}~\cite{zoghbi-fast} also provided an
analysis of \accelDesc, and proved that it generates partitions in
constant amortised time. We briefly summarise this analysis to
provide some perspective on the approach we have taken. \twoAuthor{Zoghbi}
{\Stojmenovic}\ begin their analysis by demonstrating that $\ndc(n, m) \geq
n^2/12$ for all $m > 2$, where $\ndc(n, m)$ enumerates the descending
compositions of $n$ in which the initial part is \emph{no more} than $m$.
They use this result to reason that, for each $d_q > 2$ encountered, the
total number of iterations of the internal \textbf{while} loop is $<2c$,
for
some constant $c$. Thus, since the number of iterations of the internal
loop is constant whenever $d_q \geq 3$ (the case for $d_q = 2$ obviously
requires constant time), the algorithm generates descending compositions in
constant amortised time.

The preceding paragraph is not a rigorous argument proving
that \accelDesc\ is constant amortised time. It is intended only to
illustrate the difference in the approach that we have taken in this
section to \twoAuthor{Zoghbi}{\Stojmenovic}'s analysis, and perhaps
highlight some
of the advantages of using Kemp's abstract model of counting read and
write operations~\cite{kemp-generating}. By using Kemp's model we were
able to ignore irrelevant details regarding the algorithm's
implementation, and concentrate instead on the algorithm's
\emph{effect}: reading and writing parts in compositions. 

\subsection{Comparison}\label{sec-gap-accelerated-algorithms-comparison}

Considering \accelAsc\ (\algorithmRef{alg-accel-asc}) first, we derived
the following numbers of read and write operations required to generate all
ascending compositions of $n$, ignoring inconsequential trailing terms.
\begin{equation}\label{trs-tws-alg-accel-asc}
\trs{alg-accel-asc}(n) \approx \trsAlgAccelAsc 
\quad \text{and} \quad
\tws{alg-accel-asc}(n) \approx  2\nump(n)
\end{equation}
We can see that the total number of write operations is $2\nump(n)$; i.e.,
the total number of write operations is twice the total number of
partitions generated. On the other hand, the total number of read
operations required is only $\nump(n) - \nump(n - 2)$, which, as we shall
see presently, is asymptotically negligible in comparison to $\nump(n)$.
The number of read operations is small because we only require one read
operation per iteration of the outer loop. Once we have stored $a_{k -
1}$ in a local variable, we can then
extend the composition as necessary and visit all of the
following nonterminal compositions without needing to perform a read
operation. Thus, it is the write operations that dominate the cost of
generation with this algorithm and, as we noted earlier, the average
cost of a write operation in this algorithm is quite small.

For the descending composition generator, \accelDesc\ 
(\algorithmRef{alg-accel-desc}), the following read and write totals were
derived (we ignore the insignificant trailing terms in both cases).
\begin{equation}\label{trs-tws-alg-accel-desc}
\trs{alg-accel-desc}(n) \approx 2\nump(n) - \nump(n - 2) \quad \text{and}
\quad
\tws{alg-accel-desc}(n) \approx  \nump(n) + \nump(n - 2)
\end{equation}
The total number of write operations required by this algorithm 
to generate all partitions of $n$ is $\nump(n) + \nump(n - 2)$. 
Although this value is strictly less than the
write total for \accelAsc, the difference is not
asymptotically significant as $\nump(n - 2) / \nump(n)$ tends towards $1$
as $n$ becomes large. Therefore, we should not expect any
appreciable difference between the performances of the two algorithms in
terms of the number of write operations involved. There is, however, an
asymptotically significant difference in the number of read operations
performed by the algorithms.

The total number of read operations required by 
\accelDesc\ is $2\nump(n) - \nump(n - 2)$. This
expression is complicated by an algorithmic consideration, where it proved
to be more efficient to perform $\nump(n) - \nump(n - 2)$ extra read
operations than to save the relevant value in a local variable.
Essentially, \accelDesc\ needs to perform one read
operation for every iteration of the external loop, to determine the value
of $d_q$. If $d_q = 2$ we execute the special case and quickly generate
the next descending composition; otherwise, we apply the general case. We
cannot keep the value of $d_q$ locally because the value of $q$ changes
constantly, and so we do not spend significant periods of time operating on
the same array indices, as we do in \accelAsc. Thus, we must read the value
of $d_q$ for every transition, and we can therefore simplify by saying that
\accelDescHandle{n} requires $\nump(n)$ read operations. 

In the interest of  the fairest possible comparison between
ascending and descending compositions generation algorithms, let us
therefore simplify, and assume that any descending composition
generation algorithm utilising the same properties as
\accelDesc\ requires $\nump(n)$ read operations. We know
from \equationRef{trs-tws-alg-accel-asc} that our ascending composition
generation algorithm required only $\nump(n) - \nump(n - 2)$ reads. We can
therefore expect that an ascending composition generator will require
$\nump(n - 2)$ \emph{less} read operations than a descending composition
generator similar to \accelDesc. Other things being equal, we should expect
a significant difference between the total time required to generate all
partitions using an ascending composition generation algorithm and a
commensurable descending composition generator.

We can gain a qualitative idea of the differences involved if we examine
the average numbers of read and write operations using the asymptotic
values of $\nump(n)$. Again, to determine the average number of read and
write operations required per partition generated we must divide the
totals involved by $\nump(n)$. We stated earlier that the value of
$\nump(n) - \nump(n - 2)$ is asymptotically negligible compared to
$\nump(n)$; we can quantify this statement using the asymptotic formulas
for $\nump(n)$.
Knuth~\cite[\pageNumber{11}]{knuth-generating-all-partitions} provides an 
approximation of $\nump(n - 2) / \nump(n)$, which can be expressed as
follows: 
\newcommand{\pnOverPnMinusTwo}{\ensuremath{e^{-2\pi/\sqrt{6n}}}}
\begin{equation}
\frac{\nump(n - 2)}{\nump(n)} \approx \frac{1}{e^{2\pi/\sqrt{6n}}}.
\end{equation}
Using this approximation, we obtain the following
estimates for the average number of read and write operations required to
generate each ascending and descending composition of $n$.
\begin{equation*}
\begin{array}{llr}
 & \readwriteformat{Reads} & \readwriteformat{Writes} \\
\hline
\text{Ascending} & 1 - \pnOverPnMinusTwo & 2 \\
\text{Descending}&  
 1 & 1 +\pnOverPnMinusTwo \\
\hline
\end{array}
\end{equation*}
Suppose we wished to generate all partitions of $1000$. Then, using the
best known descending composition generation algorithm we would expect to
make $1$ read and $1.92$ write operations per partition generated. On the
other hand, if we used \accelAsc, we would expect to
make only $0.08$ read and $2$ write operations per partition.

\begin{figure}
\begin{tabularx}{\textwidth}{LLRR}
    \setlength{\unitlength}{2mm}
\begin{picture}(10,77)(0,-77)
\put(0,0){\circle{0.5}}
\put(9,-2){\circle{0.5}}
\put(8,-3){\circle{0.5}}
\put(7,-5){\circle{0.5}}
\put(6,-7){\circle{0.5}}
\put(6,-9){\circle{0.5}}
\put(5,-11){\circle{0.5}}
\put(5,-13){\circle{0.5}}
\put(4,-15){\circle{0.5}}
\put(5,-17){\circle{0.5}}
\put(4,-19){\circle{0.5}}
\put(3,-22){\circle{0.5}}
\put(4,-24){\circle{0.5}}
\put(3,-26){\circle{0.5}}
\put(3,-28){\circle{0.5}}
\put(2,-30){\circle{0.5}}
\put(4,-32){\circle{0.5}}
\put(3,-34){\circle{0.5}}
\put(2,-37){\circle{0.5}}
\put(2,-39){\circle{0.5}}
\put(1,-42){\circle{0.5}}
\put(3,-44){\circle{0.5}}
\put(2,-46){\circle{0.5}}
\put(2,-48){\circle{0.5}}
\put(1,-50){\circle{0.5}}
\put(1,-53){\circle{0.5}}
\put(0,-56){\circle{0.5}}
\put(3,-58){\circle{0.5}}
\put(2,-60){\circle{0.5}}
\put(1,-63){\circle{0.5}}
\put(1,-65){\circle{0.5}}
\put(0,-68){\circle{0.5}}
\put(1,-70){\circle{0.5}}
\put(0,-72){\circle{0.5}}
\put(0,-74){\circle{0.5}}
\multiput(-0.5,0.5)(0,-77){2}{\line(1,0){12}}
\multiput(-0.5,0.5)(12,0){2}{\line(0,-1){77}}
\end{picture}
 & 
    \setlength{\unitlength}{2mm}
\begin{picture}(10,77)(0,-77)
\put(0,0){\circle*{0.5}}
\put(0,0){\circle*{0.5}}
\put(1,0){\circle*{0.5}}
\put(2,0){\circle*{0.5}}
\put(3,0){\circle*{0.5}}
\put(4,0){\circle*{0.5}}
\put(5,0){\circle*{0.5}}
\put(6,0){\circle*{0.5}}
\put(7,0){\circle*{0.5}}
\put(8,0){\circle*{0.5}}
\put(9,0){\circle*{0.5}}
\put(10,0){\circle*{0.5}}
\put(11,0){\circle*{0.5}}
\put(10,-1){\circle*{0.5}}
\put(9,-2){\circle*{0.5}}
\put(8,-3){\circle*{0.5}}
\put(9,-3){\circle*{0.5}}
\put(8,-4){\circle*{0.5}}
\put(7,-5){\circle*{0.5}}
\put(8,-5){\circle*{0.5}}
\put(7,-6){\circle*{0.5}}
\put(6,-7){\circle*{0.5}}
\put(7,-7){\circle*{0.5}}
\put(8,-7){\circle*{0.5}}
\put(7,-8){\circle*{0.5}}
\put(6,-9){\circle*{0.5}}
\put(7,-9){\circle*{0.5}}
\put(6,-10){\circle*{0.5}}
\put(5,-11){\circle*{0.5}}
\put(6,-11){\circle*{0.5}}
\put(7,-11){\circle*{0.5}}
\put(6,-12){\circle*{0.5}}
\put(5,-13){\circle*{0.5}}
\put(6,-13){\circle*{0.5}}
\put(5,-14){\circle*{0.5}}
\put(4,-15){\circle*{0.5}}
\put(5,-15){\circle*{0.5}}
\put(6,-15){\circle*{0.5}}
\put(7,-15){\circle*{0.5}}
\put(6,-16){\circle*{0.5}}
\put(5,-17){\circle*{0.5}}
\put(6,-17){\circle*{0.5}}
\put(5,-18){\circle*{0.5}}
\put(4,-19){\circle*{0.5}}
\put(5,-19){\circle*{0.5}}
\put(4,-20){\circle*{0.5}}
\put(5,-20){\circle*{0.5}}
\put(4,-21){\circle*{0.5}}
\put(3,-22){\circle*{0.5}}
\put(4,-22){\circle*{0.5}}
\put(5,-22){\circle*{0.5}}
\put(6,-22){\circle*{0.5}}
\put(5,-23){\circle*{0.5}}
\put(4,-24){\circle*{0.5}}
\put(5,-24){\circle*{0.5}}
\put(4,-25){\circle*{0.5}}
\put(3,-26){\circle*{0.5}}
\put(4,-26){\circle*{0.5}}
\put(5,-26){\circle*{0.5}}
\put(4,-27){\circle*{0.5}}
\put(3,-28){\circle*{0.5}}
\put(4,-28){\circle*{0.5}}
\put(3,-29){\circle*{0.5}}
\put(2,-30){\circle*{0.5}}
\put(3,-30){\circle*{0.5}}
\put(4,-30){\circle*{0.5}}
\put(5,-30){\circle*{0.5}}
\put(6,-30){\circle*{0.5}}
\put(5,-31){\circle*{0.5}}
\put(4,-32){\circle*{0.5}}
\put(5,-32){\circle*{0.5}}
\put(4,-33){\circle*{0.5}}
\put(3,-34){\circle*{0.5}}
\put(4,-34){\circle*{0.5}}
\put(3,-35){\circle*{0.5}}
\put(4,-35){\circle*{0.5}}
\put(3,-36){\circle*{0.5}}
\put(2,-37){\circle*{0.5}}
\put(3,-37){\circle*{0.5}}
\put(4,-37){\circle*{0.5}}
\put(3,-38){\circle*{0.5}}
\put(2,-39){\circle*{0.5}}
\put(3,-39){\circle*{0.5}}
\put(2,-40){\circle*{0.5}}
\put(3,-40){\circle*{0.5}}
\put(2,-41){\circle*{0.5}}
\put(1,-42){\circle*{0.5}}
\put(2,-42){\circle*{0.5}}
\put(3,-42){\circle*{0.5}}
\put(4,-42){\circle*{0.5}}
\put(5,-42){\circle*{0.5}}
\put(4,-43){\circle*{0.5}}
\put(3,-44){\circle*{0.5}}
\put(4,-44){\circle*{0.5}}
\put(3,-45){\circle*{0.5}}
\put(2,-46){\circle*{0.5}}
\put(3,-46){\circle*{0.5}}
\put(4,-46){\circle*{0.5}}
\put(3,-47){\circle*{0.5}}
\put(2,-48){\circle*{0.5}}
\put(3,-48){\circle*{0.5}}
\put(2,-49){\circle*{0.5}}
\put(1,-50){\circle*{0.5}}
\put(2,-50){\circle*{0.5}}
\put(3,-50){\circle*{0.5}}
\put(2,-51){\circle*{0.5}}
\put(3,-51){\circle*{0.5}}
\put(2,-52){\circle*{0.5}}
\put(1,-53){\circle*{0.5}}
\put(2,-53){\circle*{0.5}}
\put(1,-54){\circle*{0.5}}
\put(2,-54){\circle*{0.5}}
\put(1,-55){\circle*{0.5}}
\put(0,-56){\circle*{0.5}}
\put(1,-56){\circle*{0.5}}
\put(2,-56){\circle*{0.5}}
\put(3,-56){\circle*{0.5}}
\put(4,-56){\circle*{0.5}}
\put(5,-56){\circle*{0.5}}
\put(4,-57){\circle*{0.5}}
\put(3,-58){\circle*{0.5}}
\put(4,-58){\circle*{0.5}}
\put(3,-59){\circle*{0.5}}
\put(2,-60){\circle*{0.5}}
\put(3,-60){\circle*{0.5}}
\put(2,-61){\circle*{0.5}}
\put(3,-61){\circle*{0.5}}
\put(2,-62){\circle*{0.5}}
\put(1,-63){\circle*{0.5}}
\put(2,-63){\circle*{0.5}}
\put(3,-63){\circle*{0.5}}
\put(2,-64){\circle*{0.5}}
\put(1,-65){\circle*{0.5}}
\put(2,-65){\circle*{0.5}}
\put(1,-66){\circle*{0.5}}
\put(2,-66){\circle*{0.5}}
\put(1,-67){\circle*{0.5}}
\put(0,-68){\circle*{0.5}}
\put(1,-68){\circle*{0.5}}
\put(2,-68){\circle*{0.5}}
\put(3,-68){\circle*{0.5}}
\put(2,-69){\circle*{0.5}}
\put(1,-70){\circle*{0.5}}
\put(2,-70){\circle*{0.5}}
\put(1,-71){\circle*{0.5}}
\put(0,-72){\circle*{0.5}}
\put(1,-72){\circle*{0.5}}
\put(2,-72){\circle*{0.5}}
\put(1,-73){\circle*{0.5}}
\put(0,-74){\circle*{0.5}}
\put(1,-74){\circle*{0.5}}
\put(0,-75){\circle*{0.5}}
\put(1,-75){\circle*{0.5}}
\put(0,-76){\circle*{0.5}}
\multiput(-0.5,0.5)(0,-77){2}{\line(1,0){12}}
\multiput(-0.5,0.5)(12,0){2}{\line(0,-1){77}}
\end{picture}
 &
    \setlength{\unitlength}{2mm}
\begin{picture}(10,77)(0,-77)
\put(0,-1){\circle{0.5}}
\put(0,-1){\circle{0.5}}
\put(0,-2){\circle{0.5}}
\put(0,-2){\circle{0.5}}
\put(1,-3){\circle{0.5}}
\put(0,-4){\circle{0.5}}
\put(0,-4){\circle{0.5}}
\put(1,-5){\circle{0.5}}
\put(1,-5){\circle{0.5}}
\put(1,-6){\circle{0.5}}
\put(0,-7){\circle{0.5}}
\put(0,-7){\circle{0.5}}
\put(1,-8){\circle{0.5}}
\put(1,-8){\circle{0.5}}
\put(1,-9){\circle{0.5}}
\put(1,-9){\circle{0.5}}
\put(2,-10){\circle{0.5}}
\put(1,-11){\circle{0.5}}
\put(0,-12){\circle{0.5}}
\put(0,-12){\circle{0.5}}
\put(1,-13){\circle{0.5}}
\put(1,-13){\circle{0.5}}
\put(1,-14){\circle{0.5}}
\put(1,-14){\circle{0.5}}
\put(2,-15){\circle{0.5}}
\put(1,-16){\circle{0.5}}
\put(1,-16){\circle{0.5}}
\put(2,-17){\circle{0.5}}
\put(1,-18){\circle{0.5}}
\put(0,-19){\circle{0.5}}
\put(0,-19){\circle{0.5}}
\put(1,-20){\circle{0.5}}
\put(1,-20){\circle{0.5}}
\put(1,-21){\circle{0.5}}
\put(1,-21){\circle{0.5}}
\put(2,-22){\circle{0.5}}
\put(1,-23){\circle{0.5}}
\put(1,-23){\circle{0.5}}
\put(2,-24){\circle{0.5}}
\put(2,-24){\circle{0.5}}
\put(2,-25){\circle{0.5}}
\put(1,-26){\circle{0.5}}
\put(1,-26){\circle{0.5}}
\put(3,-27){\circle{0.5}}
\put(2,-28){\circle{0.5}}
\put(1,-29){\circle{0.5}}
\put(0,-30){\circle{0.5}}
\put(0,-30){\circle{0.5}}
\put(2,-31){\circle{0.5}}
\put(1,-32){\circle{0.5}}
\put(1,-32){\circle{0.5}}
\put(2,-33){\circle{0.5}}
\put(2,-33){\circle{0.5}}
\put(2,-34){\circle{0.5}}
\put(1,-35){\circle{0.5}}
\put(1,-35){\circle{0.5}}
\put(2,-36){\circle{0.5}}
\put(2,-36){\circle{0.5}}
\put(3,-37){\circle{0.5}}
\put(2,-38){\circle{0.5}}
\put(1,-39){\circle{0.5}}
\put(1,-39){\circle{0.5}}
\put(3,-40){\circle{0.5}}
\put(2,-41){\circle{0.5}}
\put(1,-42){\circle{0.5}}
\put(0,-43){\circle{0.5}}
\put(0,-43){\circle{0.5}}
\put(2,-44){\circle{0.5}}
\put(2,-44){\circle{0.5}}
\put(2,-45){\circle{0.5}}
\put(2,-45){\circle{0.5}}
\put(3,-46){\circle{0.5}}
\put(2,-47){\circle{0.5}}
\put(1,-48){\circle{0.5}}
\put(1,-48){\circle{0.5}}
\put(3,-49){\circle{0.5}}
\put(2,-50){\circle{0.5}}
\put(2,-50){\circle{0.5}}
\put(3,-51){\circle{0.5}}
\put(2,-52){\circle{0.5}}
\put(1,-53){\circle{0.5}}
\put(1,-53){\circle{0.5}}
\put(4,-54){\circle{0.5}}
\put(3,-55){\circle{0.5}}
\put(2,-56){\circle{0.5}}
\put(1,-57){\circle{0.5}}
\put(0,-58){\circle{0.5}}
\put(0,-58){\circle{0.5}}
\put(3,-59){\circle{0.5}}
\put(3,-59){\circle{0.5}}
\put(3,-60){\circle{0.5}}
\put(2,-61){\circle{0.5}}
\put(2,-61){\circle{0.5}}
\put(4,-62){\circle{0.5}}
\put(3,-63){\circle{0.5}}
\put(2,-64){\circle{0.5}}
\put(1,-65){\circle{0.5}}
\put(1,-65){\circle{0.5}}
\put(4,-66){\circle{0.5}}
\put(3,-67){\circle{0.5}}
\put(2,-68){\circle{0.5}}
\put(1,-69){\circle{0.5}}
\put(0,-70){\circle{0.5}}
\put(0,-70){\circle{0.5}}
\put(5,-71){\circle{0.5}}
\put(4,-72){\circle{0.5}}
\put(3,-73){\circle{0.5}}
\put(2,-74){\circle{0.5}}
\put(1,-75){\circle{0.5}}
\put(0,-76){\circle{0.5}}
\multiput(-0.5,0.5)(0,-77){2}{\line(1,0){12}}
\multiput(-0.5,0.5)(12,0){2}{\line(0,-1){77}}
\end{picture}
 &
    \setlength{\unitlength}{2mm}
\begin{picture}(10,77)(0,-77)
\put(1,0){\circle*{0.5}}
\put(2,0){\circle*{0.5}}
\put(3,0){\circle*{0.5}}
\put(4,0){\circle*{0.5}}
\put(5,0){\circle*{0.5}}
\put(6,0){\circle*{0.5}}
\put(7,0){\circle*{0.5}}
\put(8,0){\circle*{0.5}}
\put(9,0){\circle*{0.5}}
\put(10,0){\circle*{0.5}}
\put(11,0){\circle*{0.5}}
\put(0,0){\circle*{0.5}}
\put(0,-1){\circle*{0.5}}
\put(0,-2){\circle*{0.5}}
\put(1,-2){\circle*{0.5}}
\put(1,-3){\circle*{0.5}}
\put(0,-4){\circle*{0.5}}
\put(1,-4){\circle*{0.5}}
\put(1,-5){\circle*{0.5}}
\put(1,-6){\circle*{0.5}}
\put(0,-7){\circle*{0.5}}
\put(1,-7){\circle*{0.5}}
\put(1,-8){\circle*{0.5}}
\put(1,-9){\circle*{0.5}}
\put(2,-9){\circle*{0.5}}
\put(2,-10){\circle*{0.5}}
\put(1,-11){\circle*{0.5}}
\put(0,-12){\circle*{0.5}}
\put(1,-12){\circle*{0.5}}
\put(1,-13){\circle*{0.5}}
\put(1,-14){\circle*{0.5}}
\put(2,-14){\circle*{0.5}}
\put(2,-15){\circle*{0.5}}
\put(1,-16){\circle*{0.5}}
\put(2,-16){\circle*{0.5}}
\put(2,-17){\circle*{0.5}}
\put(1,-18){\circle*{0.5}}
\put(0,-19){\circle*{0.5}}
\put(1,-19){\circle*{0.5}}
\put(1,-20){\circle*{0.5}}
\put(1,-21){\circle*{0.5}}
\put(2,-21){\circle*{0.5}}
\put(2,-22){\circle*{0.5}}
\put(1,-23){\circle*{0.5}}
\put(2,-23){\circle*{0.5}}
\put(2,-24){\circle*{0.5}}
\put(2,-25){\circle*{0.5}}
\put(1,-26){\circle*{0.5}}
\put(2,-26){\circle*{0.5}}
\put(3,-26){\circle*{0.5}}
\put(3,-27){\circle*{0.5}}
\put(2,-28){\circle*{0.5}}
\put(1,-29){\circle*{0.5}}
\put(0,-30){\circle*{0.5}}
\put(1,-30){\circle*{0.5}}
\put(2,-30){\circle*{0.5}}
\put(2,-31){\circle*{0.5}}
\put(1,-32){\circle*{0.5}}
\put(2,-32){\circle*{0.5}}
\put(2,-33){\circle*{0.5}}
\put(2,-34){\circle*{0.5}}
\put(1,-35){\circle*{0.5}}
\put(2,-35){\circle*{0.5}}
\put(2,-36){\circle*{0.5}}
\put(3,-36){\circle*{0.5}}
\put(3,-37){\circle*{0.5}}
\put(2,-38){\circle*{0.5}}
\put(1,-39){\circle*{0.5}}
\put(2,-39){\circle*{0.5}}
\put(3,-39){\circle*{0.5}}
\put(3,-40){\circle*{0.5}}
\put(2,-41){\circle*{0.5}}
\put(1,-42){\circle*{0.5}}
\put(0,-43){\circle*{0.5}}
\put(1,-43){\circle*{0.5}}
\put(2,-43){\circle*{0.5}}
\put(2,-44){\circle*{0.5}}
\put(2,-45){\circle*{0.5}}
\put(3,-45){\circle*{0.5}}
\put(3,-46){\circle*{0.5}}
\put(2,-47){\circle*{0.5}}
\put(1,-48){\circle*{0.5}}
\put(2,-48){\circle*{0.5}}
\put(3,-48){\circle*{0.5}}
\put(3,-49){\circle*{0.5}}
\put(2,-50){\circle*{0.5}}
\put(3,-50){\circle*{0.5}}
\put(3,-51){\circle*{0.5}}
\put(2,-52){\circle*{0.5}}
\put(1,-53){\circle*{0.5}}
\put(2,-53){\circle*{0.5}}
\put(3,-53){\circle*{0.5}}
\put(4,-53){\circle*{0.5}}
\put(4,-54){\circle*{0.5}}
\put(3,-55){\circle*{0.5}}
\put(2,-56){\circle*{0.5}}
\put(1,-57){\circle*{0.5}}
\put(0,-58){\circle*{0.5}}
\put(1,-58){\circle*{0.5}}
\put(2,-58){\circle*{0.5}}
\put(3,-58){\circle*{0.5}}
\put(3,-59){\circle*{0.5}}
\put(3,-60){\circle*{0.5}}
\put(2,-61){\circle*{0.5}}
\put(3,-61){\circle*{0.5}}
\put(4,-61){\circle*{0.5}}
\put(4,-62){\circle*{0.5}}
\put(3,-63){\circle*{0.5}}
\put(2,-64){\circle*{0.5}}
\put(1,-65){\circle*{0.5}}
\put(2,-65){\circle*{0.5}}
\put(3,-65){\circle*{0.5}}
\put(4,-65){\circle*{0.5}}
\put(4,-66){\circle*{0.5}}
\put(3,-67){\circle*{0.5}}
\put(2,-68){\circle*{0.5}}
\put(1,-69){\circle*{0.5}}
\put(0,-70){\circle*{0.5}}
\put(1,-70){\circle*{0.5}}
\put(2,-70){\circle*{0.5}}
\put(3,-70){\circle*{0.5}}
\put(4,-70){\circle*{0.5}}
\put(5,-70){\circle*{0.5}}
\put(5,-71){\circle*{0.5}}
\put(4,-72){\circle*{0.5}}
\put(3,-73){\circle*{0.5}}
\put(2,-74){\circle*{0.5}}
\put(1,-75){\circle*{0.5}}
\put(0,-76){\circle*{0.5}}
\multiput(-0.5,0.5)(0,-77){2}{\line(1,0){12}}
\multiput(-0.5,0.5)(12,0){2}{\line(0,-1){77}}
\end{picture}
  
\tabularnewline
\readwriteformat{Read} \totalFormat{35} &
\readwriteformat{Write} \totalFormat{154}  &
\totalFormat{77} \readwriteformat{Read}  & 
\totalFormat{119} \readwriteformat{Write} 
\tabularnewline
\multicolumn{1}{l}{\accelAscHandle{12}} & 
\multicolumn{2}{c}{$\nump(12) = 77$} &
\multicolumn{1}{r}{\accelDescHandle{12}}
\end{tabularx}
\renewcommand{\thisCaptionSummary}[0]{Read and write tapes for the
accelerated algorithms to generate ascending and descending
compositions.}
\caption[\thisCaptionSummary]{\thisCaptionSummary\ On the left we have
the read and write tapes for the ascending composition generator,
\algorithmRef{alg-accel-asc}; on the right, then, are the corresponding
tapes for the descending composition generator,
\algorithmRef{alg-accel-desc}. In both cases, the traces correspond to the
read and write operations carried out in generating all partitions of
$12$.\label{fig-accel-tapes}}
\end{figure}

The qualitative behaviour of 
\accelAsc\ and \accelDesc\ can be seen from their read and write
tapes
(\figureRef{fig-accel-tapes}). Comparing the write tapes for
the algorithms, we can see that the total number of write operations is
roughly equal in both algorithms, although they follow an altogether
different spatial pattern. The read tapes for the algorithms, however,
demonstrate the essential difference between the algorithms:
\accelDesc\ makes one read operation for every partition
generated, while the read operations for \accelAsc\ are
sparsely distributed across the tape.

We have derived expressions to count the total number of
read and write operations required to generate all partitions of $n$ using
\accelAsc\  and \accelDesc. We can now use these expressions to 
make some quantitative predictions about the relative efficiencies 
of the algorithms. If we assume that
the cost of read and write operations are equal, we can then derive a
prediction for the ratio of the total time elapsed using both algorithms.
Therefore, let $\ert{alg-accel-asc}(n)$ be the expected total running
time of \accelAscHandle{n}, and similarly define 
$\ert{alg-accel-desc}(n)$ for \accelDescHandle{n}. We can then
predict that the ratio of the running times should be equal to the ratio
of their total read and write counts. Thus, using the values of
\equationRef{trs-tws-alg-accel-asc} and
\equationRef{trs-tws-alg-accel-desc}, we get 
\begin{equation}
\frac{\ert{alg-accel-asc}(n)}{\ert{alg-accel-desc}(n)} = \frac
{ 3\nump(n) - \nump(n - 2)} {3\nump(n)}.
\end{equation}
Consequently, we expect that the total amount of time required to generate
all ascending compositions of $n$ should be a factor of $\nump(n - 2) /
3\nump(n)$ less than that required to generate all descending compositions
of $n$.  To test this hypothesis we measured the
total elapsed time required to generate all partitions of $n$ using
\accelAsc\ and \accelDesc, using the methodology outlined in 
\sectionRef{sec-gap-recursive-algorithms-comparison}.
We report the ratio of these times in \tableRef{tab-model-verification},
for both the C and Java implementations of the algorithms.

\tableRef{tab-model-verification} supports our qualitative predictions
well. The theoretical analysis of ascending and descending composition
generation algorithms in this section suggests that the ascending
composition generator should require significantly less time to generate
all partitions of $n$ than its descending composition counterpart; and
the data of \tableRef{tab-model-verification} supports this prediction. In
the Java implementations, the ascending composition generator
requires $15\%$ less time to generate all partitions of $100$ than the
descending composition generation algorithm; in the C version, the
difference is around $23\%$. These differences increase as the value of $n$
increases: when $n = 135$, we see that 
\accelAsc\ requires $18\%$ and $26\%$ less time
than \accelDesc\ in the C and Java implementations, respectively.

\begin{table}
\renewcommand{\thisCaptionSummary}{Empirical analysis of accelerated
ascending and descending composition generation algorithms.}
\caption[\thisCaptionSummary]{\thisCaptionSummary\ The ratio of the time
required to generate all partitions of $n$ using
\accelAsc\ and \accelDesc\ is given:
measured ratios for implementations in the Java and C languages as well as
the theoretically predicted ratio are shown.
\label{tab-model-verification}}
\begin{tabularx}{\textwidth}{p{15mm}LRRR}
\toprule
$n$ & $\nump(n)$ & Java & C & Theoretical\tabularnewline 
 \cmidrule(l{2mm}r{2mm}){1-5} 
$100$ & $1.91 \times 10^{8}$ & 0.85 & 0.77 & 0.74\tabularnewline
$105$ & $3.42 \times 10^{8}$ & 0.85 & 0.77 & 0.74\tabularnewline
$110$ & $6.07 \times 10^{8}$ & 0.84 & 0.75 & 0.74\tabularnewline
$115$ & $1.06 \times 10^{9}$ & 0.84 & 0.75 & 0.73\tabularnewline
$120$ & $1.84 \times 10^{9}$ & 0.83 & 0.75 & 0.73\tabularnewline
$125$ & $3.16 \times 10^{9}$ & 0.83 & 0.74 & 0.73\tabularnewline
$130$ & $5.37 \times 10^{9}$ & 0.83 & 0.74 & 0.73\tabularnewline
$135$ & $9.04 \times 10^{9}$ & 0.82 & 0.74 & 0.73\tabularnewline
\cmidrule(l{1cm}){3-4}& \multicolumn{2}{r}{
$r_{\text{Java}} = 0.9891$} & $r_{\text{C}} = 0.9321$&\tabularnewline\bottomrule\end{tabularx}
\end{table}

We also made a quantitative prediction about the ratio of the time
required to
generate all partitions of $n$ using \accelAsc\ and 
\accelDesc. Using the theoretical analysis, where we
counted the total number of read and write operations required by these
algorithms, we can predict the expected ratio of the time required by both
algorithms. This ratio is also reported in
\tableRef{tab-model-verification}, and we can see that it is consistent
with the measured ratios for the Java and C implementations of the
algorithms. In the case of the Java implementation, the theoretically
predicted ratios are too optimistic, suggesting that the model of counting
only read and write operations is a little overly simplistic in this case.
The correspondence between the measured and predicted ratios in the C
implementation is much closer, as we can see from
\tableRef{tab-model-verification}. In both cases there is a 
strong positive correlation between the predicted and measured
ratios.
\section{Conclusion} \label{sec-conclusion}
In this paper we have systematically compared algorithms to generate all ascending 
and descending compositions, two possible encodings for integer partitions. In 
\sectionRef{sec-gap-recursive-algorithms} we compared two recursive algorithms: our
new ascending composition generator, and Ruskey's descending composition generator.
By analysing these algorithms we were able to show that although both algorithms are 
constant amortised time, the descending composition generator requires approximately
twice as long to generate all partitions of $n$. In \sectionRef{sec-gap-succession-rules}
we compared two generators in Kemp's idiom: succession rules that require no state to 
be maintained between transitions. We developed a new succession rule for ascending 
compositions in lexicographic order, and implemented the well known succession rule 
for descending compositions in reverse lexicographic order. The analyses of these 
algorithms showed that the ascending composition generator required constant 
time, on average, to make each transition; whereas the descending composition 
generator required $\bigoh{\sqrt{n}}$ time. \sectionRef{sec-gap-accelerated-algorithms}
then compared the most efficient known algorithms to generate all ascending and 
descending compositions. We developed a new generation algorithm for the ascending 
compositions by utilising structure within the set of ascending compositions. 
We also analysed \twoAuthor{Zoghbi}{\Stojmenovic}'s algorithm and compared 
these two algorithms theoretically and empirically. As a result of this analysis,
we showed that the ascending composition generator requires roughly three quarters of 
the time required by the descending composition generator. These three 
comparisons of algorithms show that ascending compositions are a superior
encoding for generating all partitions.

Generation efficiency is not the only advantage of encoding partitions 
as ascending compositions. As part of our analysis of the succession rule for 
ascending compositions in \sectionRef{sec-gap-succession-rules} we proved a new 
formula for computing the number of partitions of $n$ in terms of the largest 
and second largest parts. In 
\sectionRef{sec-gap-accelerated-algorithms-ascending-compositions} we developed 
a new proof for a combinatorial identity, showing that the number of partitions 
of $n$ where the largest part is less that twice the second largest part is equal
to the number of partitions of $n - 2$. These mathematical results were motivated
by studying algorithms to generate ascending compositions.

Another advantage of using ascending compositions to encode partitions, not 
mentioned here, is the possibility of developing algorithms to generate 
a very flexible class of restricted partition. By generalising the algorithms 
developed in this paper it is possible to generate (and enumerate) combinatorially
important classes of partition such as the partitions into distinct 
parts~\cite[\sectionNumber{2}]{bjorner-combinatorial}, Rogers-Ramanujan 
partitions~\cite{fulman-rogers} and \Gollnitz-Gordon 
partitions~\cite{alladi-gollnitz}. The framework for describing these restrictions 
and developing generation and enumeration algorithms is described by 
Kelleher~\cite[\chapterNumber{3--4}]{kelleher-encoding}.


\ifarxiv
    \bibliographystyle{plain}
    \bibliography{gapacote}
\else
    \bibliographystyle{acmtrans}
    \bibliography{gapacote}

\begin{thebibliography}{10}

\bibitem{actor-infinite}
Alfred~Arthur Actor.
\newblock Infinite products, partition functions, and the {M}einardus theorem.
\newblock {\em Journal of Mathematical Physics}, 35(11):5749--5764, November
  1994.

\bibitem{ahlgren-addition}
Scott Ahlgren and Ken Ono.
\newblock Addition and counting: The arithmetic of partitions.
\newblock {\em Notices of the AMS}, 48(9):978--984, October 2001.

\bibitem{alladi-gollnitz}
Krishnaswami Alladi and Alexander Berkovich.
\newblock {G{\"o}llnitz-Gordon} partitions with weights and parity conditions.
\newblock In T.~Aoki, S.~Kanemitsu, M.~Nakahara, and Y.~Ohno, editors, {\em
  Zeta functions topology and quantum physics}, pages 1--18. Springer Verlag,
  US, 2005.

\bibitem{andrews-theory}
George~E. Andrews.
\newblock {\em The Theory of Partitions}.
\newblock Encyclopedia of Mathematics and its Applications. Addison-Wesley,
  London, 1976.

\bibitem{andrews-partitions}
George~E. Andrews.
\newblock Partitions.
\newblock In {\em History of Combinatorics}, volume~1. To appear, 2005.

\bibitem{andrews-integer}
George~E. Andrews and Kimmo Eriksson.
\newblock {\em Integer Partitions}.
\newblock Cambridge University Press, Cambridge, 2004.

\bibitem{bivins-characters}
Robert~L. Bivins, N.~Metropolis, Paul~R. Stein, and Mark~B. Wells.
\newblock Characters of the symmetric groups of degree 15 and 16.
\newblock {\em Mathematical Tables and Other Aids to Computation},
  8(48):212--216, October 1954.

\bibitem{bjorner-combinatorial}
Anders Bj{\"o}rner and Richard~P. Stanley.
\newblock A combinatorial miscellany.
\newblock \url{http://www.math.kth.se/~bjorner/files/CUP.ps}, 2005.
\newblock To appear in \emph{L'Enseignement Math{\'e}matique}.

\bibitem{boyer-simple}
John~M. Boyer.
\newblock Simple constant amortized time generation of fixed length numeric
  partitions.
\newblock {\em Journal of Algorithms}, 54(1):31--39, January 2005.

\bibitem{comet-notations}
Stig Com{\'e}t.
\newblock Notations for partitions.
\newblock {\em Mathematical Tables and Other Aids to Computation},
  9(52):143--146, October 1955.

\bibitem{deMoivre-method}
Abraham de~Moivre.
\newblock A method of raising an infinite multinomial to any given power, or
  extracting any given root of the same.
\newblock {\em Philosophical Transactions}, 19(230):619--625, 1697.

\bibitem{desesquelles-calculation}
P.~D{\'e}sesquelles.
\newblock Calculation of the number of partitions with constraints on the
  fragment size.
\newblock {\em Physical Review C (Nuclear Physics)}, 65(3):034603, March 2002.

\bibitem{dickson-history}
Leonard~E. Dickson.
\newblock {\em History of the theory of numbers}, volume II, Diophantine
  Analysis, chapter~3, pages 101--164.
\newblock Chelsea, New York, 1952.

\bibitem{ehrlich-loopless}
Gideon Ehrlich.
\newblock Loopless algorithms for generating permutations, combinations, and
  other combinatorial configurations.
\newblock {\em Journal of the ACM}, 20(3):500--513, July 1973.

\bibitem{fenner-binary}
Trevor~I. Fenner and Georghois Loizou.
\newblock A binary tree representation and related algorithms for generating
  integer partitions.
\newblock {\em The Computer Journal}, 23(4):332--337, 1980.

\bibitem{fenner-analysis}
Trevor~I. Fenner and Georghois Loizou.
\newblock An analysis of two related loop-free algorithms for generating
  integer partitions.
\newblock {\em Acta Informatica}, 16:237--252, 1981.

\bibitem{fulman-rogers}
Jason Fulman.
\newblock The {R}ogers-{R}amanujan identities, the finite general linear
  groups, and the {H}all-{L}ittlewood polynomials.
\newblock {\em Proceedings of the American Mathematical Society},
  128(1):17--25, 2000.

\bibitem{grossmann-from}
Siegfried Grossmann and Martin Holthaus.
\newblock From number theory to statistical mechanics: {B}ose--{E}instein
  condensation in isolated traps.
\newblock {\em Chaos, Solitons \& Fractals}, 10(4--5):795--804, April 1999.

\bibitem{gupta-ranking-B}
Udai~I. Gupta, D.~T. Lee, and C.~K. Wong.
\newblock Ranking and unranking of {B}-trees.
\newblock {\em Journal of Algorithms}, 4(1):51--60, March 1983.

\bibitem{hardy-asymptotic}
Godfrey~H. Hardy.
\newblock Asymptotic formulae in combinatory analysis.
\newblock In {\em Collected papers of {G}.{H}. {H}ardy: including joint papers
  with {J}.{E}. {L}ittlewood and others edited by a committtee appointed by the
  London Mathematical Society}, volume~1, pages 265--273. Clarendon Press,
  Oxford, 1966.

\bibitem{honsberger-mathematical}
Ross Honsberger.
\newblock {\em Mathematical Gems III}.
\newblock Number~9 in Dolciani Mathematical Expositions. Mathematical
  Association of America, 1985.

\bibitem{kelleher-encoding}
Jerome Kelleher.
\newblock {\em Encoding Partitions as Ascending Compositions}.
\newblock PhD thesis, University College Cork, 2006.

\bibitem{kemp-generating}
Rainer Kemp.
\newblock Generating words lexicographically: {A}n average-case analysis.
\newblock {\em Acta Informatica}, 35(1):17--89, January 1998.

\bibitem{klimko-algorithm}
Eugene~M. Klimko.
\newblock An algorithm for calculating indices in {\FaadiBruno}'s formula.
\newblock {\em BIT}, 13(1):38--49, 1973.

\bibitem{knuth-stanford}
Donald~E. Knuth.
\newblock {\em The Stanford GraphBase: a platform for combinatorial computing}.
\newblock Addison-Wesley, 1994.

\bibitem{knuth-generating-all-n-tuples}
Donald~E. Knuth.
\newblock Generating all n-tuples, 2004.
\newblock Pre-fascicle 2A of \emph{The Art of Computer Programming} {A} draft
  of section 7.2.1.1.
  \url{http://www-cs-faculty.stanford.edu/~knuth/fasc2a.ps.gz}.

\bibitem{knuth-generating-all-partitions}
Donald~E. Knuth.
\newblock Generating all partitions, 2004.
\newblock Pre-fascicle 3B of \emph{The Art of Computer Programming}, {A} draft
  of sections 7.2.1.4--5
  \url{http://www-cs-faculty.stanford.edu/~knuth/fasc3b.ps.gz}.

\bibitem{knuth-history}
Donald~E. Knuth.
\newblock History of combinatorial generation, 2004.
\newblock Pre-fascicle 4B of \emph{The Art of Computer Programming}, {A} draft
  of section 7.2.1.7.
  \url{http://www-cs-faculty.stanford.edu/~knuth/fasc4b.ps.gz}.

\bibitem{kreher-combinatorial}
Donald~L. Kreher and Douglas~R. Stinson.
\newblock {\em Combinatorial Algorithms: Generation, Enumeration and Search}.
\newblock CRC press LTC, Boca Raton, Florida, 1998.

\bibitem{kubasiak-fermi}
Anna Kubasiak, Jaros{\l}aw~K. Korbicz, Jakub Zakrzewski, and Maciej Lewenstein.
\newblock Fermi-{D}irac statistics and the number theory.
\newblock {\em Europhysics Letters}, 72(4):506--512, 2005.

\bibitem{lehmer-machine}
Derrick~H. Lehmer.
\newblock The machine tools of combinatorics.
\newblock In Edwin~F. Beckenbach, editor, {\em Applied Combinatorial
  Mathematics}, chapter~1, pages 5--31. Wiley, New York, 1964.

\bibitem{macmahon-combinatory}
Percy~A. MacMahon.
\newblock {\em Combinatory Analysis}.
\newblock Cambridge University Press, 1915.

\bibitem{martin-schur}
Stuart Martin.
\newblock {\em Schur Algebras and Representation Theory}.
\newblock Cambridge University Press, 1999.

\bibitem{mckay-partitions}
J.~K.~S. McKay.
\newblock Algorithm 371: {P}artitions in natural order.
\newblock {\em Communications of the ACM}, 13(1):52, January 1970.

\bibitem{narayana-algorithm}
T.~V. Narayana, R.~M. Mathsen, and J.~Saranji.
\newblock An algorithm for generating partitions and its applications.
\newblock {\em Journal of Combinatorial Theory, Series A}, 11(1):54--61, July
  1971.

\bibitem{nijenhuis-combinatorial}
Albert Nijenhuis and Herbert~S. Wilf.
\newblock {\em Combinatorial Algorithms for Computers and Calculators}.
\newblock Academic Press, New York, second edition, 1978.

\bibitem{odlyzko-asymptotic}
Andrew~M. Odlyzko.
\newblock Asymptotic enumeration methods.
\newblock In R.~L. Graham, M.~Gr{\"o}tschel, and L.~Lov{\'a}sz, editors, {\em
  Handbook of combinatorics}, volume~II, pages 1063--1229. MIT Press,
  Cambridge, MA, USA, 1996.

\bibitem{page-introduction}
E.~S. Page and L.~B. Wilson.
\newblock {\em An Introduction to Computational Combinatorics}.
\newblock Cambridge University Press, Cambridge, 1979.

\bibitem{pak-partition}
Igor Pak.
\newblock Partition bijections, a survey.
\newblock {\em The Ramanujan Journal}, 12(1):5--75, 2006.

\bibitem{pemmaraju-computational}
Sriram Pemmaraju and Steven~S. Skiena.
\newblock {\em Computational Discrete Mathematics: Combinatorics and Graph
  Theory With Mathematica}.
\newblock Cambridge University Press, 2003.

\bibitem{planat-thermal}
Michel Planat.
\newblock Thermal $1/f$ noise from the theory of partitions: application to a
  quartz resonator.
\newblock {\em Physica A: Statistical Mechanics and its Applications},
  318(3--4):371--386, February 2003.

\bibitem{reingold-combinatorial}
Edward~M. Reingold, Jurg Nievergelt, and Narsingh Deo.
\newblock {\em Combinatorial algorithms: theory and practice}.
\newblock Ridge Press/Random House, 1977.

\bibitem{riha-efficient}
W.~Riha and K.~R. James.
\newblock Algorithm 29: {E}fficient algorithms for doubly and multiply
  restricted partitions.
\newblock {\em Computing}, 16:163--168, 1976.

\bibitem{ruskey-combinatorial}
Frank Ruskey.
\newblock {\em Combinatorial Generation}.
\newblock Working version 1i \url{http://www.cs.usyd.edu.au/~algo4301/Book.ps},
  2001.

\bibitem{savage-gray}
Carla~D. Savage.
\newblock Gray code sequences of partitions.
\newblock {\em Journal of Algorithms}, 10(4):577--595, 1989.

\bibitem{savage-survey}
Carla~D. Savage.
\newblock A survey of combinatorial gray codes.
\newblock {\em SIAM Review}, 39(4):605--629, December 1997.

\bibitem{sawada-generating}
Joe Sawada.
\newblock Generating bracelets in constant amortized time.
\newblock {\em SIAM Journal on Computing}, 31(1):259--268, 2001.

\bibitem{schroeder-number}
Manfred~R. Schroeder.
\newblock {\em Number Theory in Science and Communication with Applications in
  Cryptography, Physics, Digital Information, Computing, and Self-Similarity}.
\newblock Springer-Verlag, Berlin, second enlarged edition, 1986.

\bibitem{sedgewick-permutation}
Robert Sedgewick.
\newblock Permutation generation methods.
\newblock {\em ACM Computing Surveys}, 9(2):137--164, June 1977.

\bibitem{skiena-implementing}
Steven~S. Skiena.
\newblock {\em Implementing discrete mathematics: combinatorics and graph
  theory with mathematica}.
\newblock Addison-Wesley, Redwood City, California, 1990.

\bibitem{sloane-online}
Neil J.~A. Sloane.
\newblock The on-line encyclopedia of integer sequences.
\newblock \url{http://www.research.att.com/~njas/sequences/}, 2009.

\bibitem{stanley-enumerative}
Richard~P. Stanley.
\newblock {\em Enumerative Combinatorics}.
\newblock Wadsworth, Belmont, California, 1986.

\bibitem{stanton-constructive}
Dennis Stanton and Dennis White.
\newblock {\em Constructive combinatorics}.
\newblock Springer-Verlag, Berlin, 1986.

\bibitem{stockmal-generation-constraints}
Frank Stockmal.
\newblock Algorithm 114: {G}eneration of partitions with constraints.
\newblock {\em Communications of the {ACM}}, 5(8):434, August 1962.

\bibitem{stockmal-generation}
Frank Stockmal.
\newblock Algorithm 95: {G}eneration of partitions in part-count form.
\newblock {\em Communications of the ACM}, 5(6):344, June 1962.

\bibitem{sylvester-constructive}
James~J. Sylvester.
\newblock A constructive theory of partitions, arranged in three acts, an
  interact and an exodion.
\newblock {\em American Journal of Mathematics}, 5(1/4):251--330, 1882.

\bibitem{temperley-statistical}
H.~N.~V. Temperley.
\newblock Statistical mechanics and the partition of numbers. {I}. {T}he
  transition of liquid helium.
\newblock {\em Proceedings of the Royal Society of London. Series A,
  Mathematical and Physical Sciences}, 199(1058):361--375, November 1949.

\bibitem{tomasi-two}
C.~Tomasi.
\newblock Two simple algorithms for the generation of partitions of an integer.
\newblock {\em Alta Frequenza}, 51(6):352--356, 1982.

\bibitem{tran-on}
Muoi~N. Tran, M.~V.~N. Murthy, and Rajat~K. Bhaduri.
\newblock On the quantum density of states and partitioning an integer.
\newblock {\em Annals of Physics}, 311(1):204--219, May 2004.

\bibitem{wells-elements}
Mark~B. Wells.
\newblock {\em Elements of Combinatorial Computing}.
\newblock Pergamon Press, Oxford, 1971.

\bibitem{yang-derivatives}
Winston~C. Yang.
\newblock Derivatives are essentially integer partitions.
\newblock {\em Discrete Mathematics}, 222(1--3):235--245, July 2000.

\bibitem{zoghbi-fast}
Antoine Zoghbi and Ivan Stojmenovi{\'c}.
\newblock Fast algorithms for generating integer partitions.
\newblock {\em International Journal of Computer Math}, 70:319--332, 1998.

\end{thebibliography}
    \endreceived
\fi
\end{document}